\documentclass{emulateapj}
\usepackage{natbib}
\usepackage{epsfig}
\usepackage{graphicx}
\newcommand{\acounits}{\mbox{M$_\odot$ pc$^{-2}$} \mbox{(K km s$^{-1}$)$^{-1}$}}

\shorttitle{High Resolution ISM Structure in Galaxies}
\shortauthors{Leroy et al.}

\begin{document}

\slugcomment{Accepted for Publication in the Astrophysical Journal} 
\title{A Portrait of Cold Gas in Galaxies at 60~pc Resolution and a Simple Method to Test Hypotheses That Link Small-Scale ISM Structure to Galaxy-Scale Processes}

\author{
Adam K. Leroy\altaffilmark{1},
Annie Hughes\altaffilmark{2,3},
Andreas Schruba\altaffilmark{4},
Erik Rosolowsky\altaffilmark{5},
Guillermo Blanc\altaffilmark{6},
Alberto D. Bolatto\altaffilmark{7},
Dario Colombo\altaffilmark{8},
Andres Escala\altaffilmark{6},
Carsten Kramer\altaffilmark{9},
J.~M.~Diederik Kruijssen\altaffilmark{10},
Sharon Meidt\altaffilmark{11},
Jerome Pety\altaffilmark{12,13},
Miguel Querejeta\altaffilmark{11},
Karin Sandstrom\altaffilmark{14},
Eva Schinnerer\altaffilmark{11},
Kazimierz Sliwa\altaffilmark{11},
Antonio Usero\altaffilmark{15}
}
\altaffiltext{1}{Department of Astronomy, The Ohio State University, 140 West 18th Avenue, Columbus, OH 43210}
\altaffiltext{2}{CNRS, IRAP, 9 av. du Colonel Roche, BP 44346, F-31028 Toulouse cedex 4, France}
\altaffiltext{3}{Universit\'{e} de Toulouse, UPS-OMP, IRAP, F-31028 Toulouse cedex 4, France}
\altaffiltext{4}{Max-Planck-Institut f\"ur extraterrestrische Physik, Giessenbachstra{\ss}e 1, 85748 Garching, Germany}
\altaffiltext{5}{Department of Physics, University of Alberta, Edmonton, AB, Canada}
\altaffiltext{6}{Departamento de Astronom\'{\i}a, Universidad de Chile, Casilla 36-D, Santiago, Chile.}
\altaffiltext{7}{Department of Astronomy, Laboratory for Millimeter-wave Astronomy, and Joint Space Institute, University of Maryland, College Park, Maryland 20742, USA}
\altaffiltext{8}{Max-Planck-Institut f\"ur Radioastronomie, Auf dem H\"ugel 69, D-53121 Bonn, Germany}
\altaffiltext{9}{Instituto Radioastronom\'{\i}a Milim\'{e}trica (IRAM), Av. Divina Pastora 7, Nucleo Central, E-18012 Granada, Spain}
\altaffiltext{10}{Astronomisches Rechen-Institut, Zentrum f\"{u}r Astronomie der Universit\"{a}t Heidelberg, M\"{o}nchhofstra\ss e 12-14, 69120 Heidelberg, Germany}
\altaffiltext{11}{Max Planck Institute f\"ur Astronomie, K\"onigstuhl 17, 69117, Heidelberg, Germany}
\altaffiltext{12}{Institut de Radioastronomie Millim�trique (IRAM), 300 Rue de la Piscine, F-38406 Saint Martin d?H�res, France}
\altaffiltext{13}{Observatoire de Paris, 61 Avenue de l?Observatoire, F-75014 Paris, France}
\altaffiltext{14}{Center for Astrophysics and Space Sciences, Department of Physics, University of California, San Diego, 9500 Gilman Drive, La Jolla, CA 92093, USA}
\altaffiltext{15}{Observatorio Astron�mico Nacional (IGN), C/ Alfonso XII, 3, 28014 Madrid, Spain}

\begin{abstract}
The cloud-scale density, velocity dispersion, and gravitational boundedness of the interstellar medium (ISM) vary within and among galaxies. In turbulent models, these properties play key roles in the ability of gas to form stars. New high fidelity, high resolution surveys offer the prospect to measure these quantities across galaxies. We present a simple approach to make such measurements and to test hypotheses that link small-scale gas structure to star formation and galactic environment. Our calculations capture the key physics of the Larson scaling relations, and we show good correspondence between our approach and a traditional ``cloud properties'' treatment. However, we argue that our method is preferable in many cases because of its simple, reproducible characterization of all emission. Using, low-$J$ $^{12}$CO data from recent surveys, we characterize the molecular ISM at $60$~pc resolution in the Antennae, the Large Magellanic Cloud, M31, M33, M51, and M74. We report the distributions of surface density, velocity dispersion, and gravitational boundedness at $60$~pc scales and show galaxy-to-galaxy and intra-galaxy variations in each. The distribution of flux as a function of surface density appears roughly lognormal with a $1\sigma$ width of ${\sim}0.3$~dex, though the center of this distribution varies from galaxy to galaxy. The $60$~pc resolution line width and molecular gas surface density correlate well, which is a fundamental behavior expected for virialized or free-falling gas. Varying the measurement scale for the LMC and M31, we show that the molecular ISM has higher surface densities, lower line widths, and more self-gravity at smaller scales.
\end{abstract}


\section{Introduction}
\label{sec:intro}

In a turbulent view of star formation, the physical state of the interstellar medium (ISM) on the scale of individual gravitationally bound clouds regulates the ability of gas to form stars. Turbulent theories link the star formation per unit gas to the mean density, gravitational boundedness and turbulent Mach number of star-forming molecular clouds \citep[see][]{KRUMHOLZ05,MCKEE07,PADOAN11,KRUIJSSEN12,FEDERRATH12,KRUMHOLZ12B,HENNEBELLE13}.

Telescopes including ALMA and NOEMA can measure the structure of the molecular ISM at the scale of individual clouds (a few 10s of pc) across large parts of galaxies \citep[e.g.,][]{SCHINNERER13}. Such observations capture the surface density and velocity dispersion of the gas. These are close cognates of the mean density and Mach number, while their ratio probes gravitational boundedness. Thus, observations directly access these cloud-scale quantities crucial to the ability of gas to collapse and form stars. 

These observations show that cloud-scale gas structure does vary within and among galaxies. For example, the surface density, volume density, and line width of molecular clouds in local starburst galaxies far exceed those of giant molecular clouds (GMCs) in the Milky Way \citep[e.g.,][]{LEROY15A,JOHNSON15}. Real, if subtler, differences are also evident among the GMC populations of more quiescent galaxies \citep[see][]{ROSOLOWSKY05B,HUGHES13B}.  Within individual galaxies, cloud properties correlate with environment, varying between arm and interarm regions, and with radius in the galaxy \citep[e.g.,][]{KODA09,KRUIJSSEN13,COLOMBO14A,HEYER15}.

A major goal for the next years will be to measure how these variations in cloud-scale properties depend on large-scale galactic environment and how they drive the behavior of the gas on smaller scales. In this paper, we present the natural framework to carry out such tests. We then apply it, in limited form, to a suite of the best available set of high physical resolution ($\theta = 60$~pc) data for nearby galaxies.

Linking cloud-scale structure to galactic structure and star formation is a multi-scale problem. The relevant gas structural properties must be measured on scales of 10s of pc, but galaxy structure and quantities like the star formation rate (SFR) and star formation per unit gas require measurements over larger scales. To measure the time-average rate or efficiency of star formation, one must marginalize over the evolution of individual regions \citep[e.g., see][]{KAWAMURA09,SCHRUBA10,KRUIJSSEN14}. Meanwhile, many key aspects of galaxy structure are large-scale quantities, so that comparing mean cloud structure to, e.g., stellar structure or galactic rotation makes sense at larger scales. Practical considerations also make this a multi-scale problem; many key probes of ISM state, including dust properties \citep[e.g.,][]{SANDSTROM13} and sensitive spectroscopy of faint mm-wave lines \citep[e.g.,][]{USERO15,BIGIEL16}, are mainly available at coarse resolution.

Our approach is to use wide-area, high resolution spectroscopic maps of gas (in this paper, molecular gas) to measure the surface density, velocity dispersion, gravitational boundedness, and intensity distribution at the scale of individual clouds. We do this in a way that eschews the arbitrary decomposition choices involved in traditional cloud property measurements. Nevertheless, we demonstrate below that we recover the results from cloud property calculations using our simpler approach. 

We combine measurements for many lines of sight using an intensity weighting scheme to measure the intensity-weighted surface density, line width, and velocity dispersion over a region of interest in a galaxy. In this paper, our regions of interest are Gaussian beams with FWHM 500~pc, though this choice is somewhat arbitrary. Over this larger area, we measure quantities like the ratio of total gas mass to star formation rate, the structure of the galaxy, or the hydrostatic ISM pressure.

As a practical example of this approach, consider a test of the hypothesis of \citet{KRUMHOLZ12B}. They posit that across a wide range of environments ${\sim}1\%$ of the gas turns to stars over each free-fall time. In this theory, the relevant size scale is the outer scale of turbulence, $\sim 50{-}100$~pc, and the free-fall time depends via $\tau_{\rm ff} \propto \rho^{-0.5}$ on the density at this scale. The mass volume density, $\rho$, in turn, relates closely to the surface density, $\Sigma$, so that for purposes of this example, $\tau_{\rm ff} \propto \Sigma^{-0.5}$. The hypothesis predicts that the ratio ${\rm SFR}/M_{\rm H2} \propto \tau_{\rm ff}^{-1} \propto \Sigma^{0.5}$.  In this case, $\Sigma$ must be measured at approximately the cloud scale. However, the time-averaged ratio of SFR-to-H$_2$ is only accessible averaging over an ensemble of regions in different evolutionary states to capture both the time evolution of the star formation process and to render the SFR estimate reliable. Our method to test the \citet{KRUMHOLZ12B} hypothesis, then, is to measure $\Sigma$ at high resolution, weight the local $\Sigma$ by intensity, and average to ${\sim}500$~pc (or similar) scales, where we expect a measurement of SFR-to-H$_2$ to be reliable.

Beyond multiple scales and intensity-weighting, the other key aspect of our proposed methodology is to treat the intensity of a mass-tracing line (here CO) beam-by-beam (``beamwise'') as the key parameter. That is, we estimate surface density, line width, and gravitational boundedness point-by-point, and avoid decomposition into clouds and peak finding. Instead, we focus on a statistical characterization of the ensemble of intensity measurements at the native resolution of the data. This approach is simple, with minimal tuning parameters, and characterizes the whole ISM.  The usefulness of such calculations has already been demonstrated in studies by \citet{SAWADA12}, \citet{HUGHES13A} and \citet{LEROY13B}, each of which deployed variants of some of the techniques described here.

Many studies have characterized the gas in galaxies at cloud-scale using an approach that identifies individual molecular clouds and then measures their properties \citep[][]{BOLATTO08,DONOVANMEYER12,HUGHES13B,COLOMBO14A,LEROY15A}. These calculations often have a number of tuning parameters and assumptions that are embedded in the segmentation and property measurement algorithms \citep[see][]{WILLIAMS94,ROSOLOWSKY06,ROSOLOWSKY08,COLOMBO16}. Because they focus on compact objects -- and sometimes only on apparently bound structures -- cloud property studies typically do not characterize the full content of the ISM. They are also often forced to adopt aggressive assumptions and/or extrapolations in their treatment of marginally resolved objects.

The cloud properties treatment still has large value, including a direct link to studies of individual molecular clouds in the Milky Way. Indeed, in this paper we show that our treatment and a cloud property treatment generally show good agreement. We demonstrate this by implementing a ``gridding kernel'' treatment of cloud properties that allows straightforward comparison of the two approaches. This gridding approach allows cloud catalogs to be used in lieu of simple intensity measurements to carry out hypothesis testing over larger regions of interest. Nonetheless, we argue that due to its simplicity and more complete characterization of the flux, the intensity-based approach is often a better way to implement hypothesis tests using the latest generation of high resolution, wide-field, full-flux recovery data.

In this paper, we lay out our methods in detail (Section \ref{sec:methods} and Appendix). Then, in Section \ref{sec:data} we apply our method to six high physical resolution CO data sets spanning from dwarf spirals (the LMC, M33) to disks (M31, M51, M74), and the nearest major merger (the Antennae). In Section \ref{sec:compare}, we compare our results to those obtained from a cloud property treatment applied to the same data. In Section \ref{sec:results}, we report results for our six targets, which demonstrate how cloud-scale ISM structure varies starkly among galaxies. Section \ref{sec:discuss} summarizes our findings. This section also identifies natural next steps for this approach, which include application to {\sc Hi} and structural measurements, e.g., application to multiple scales or point-to-point correlation. In the Appendices, we present details of the methodology that are crucial to our calculations, but that are likely to interest a narrower audience. We also show an atlas of cloud-scale measurements for our six targets.

\section{Method}
\label{sec:methods}

\begin{figure*}[th]
\plotone{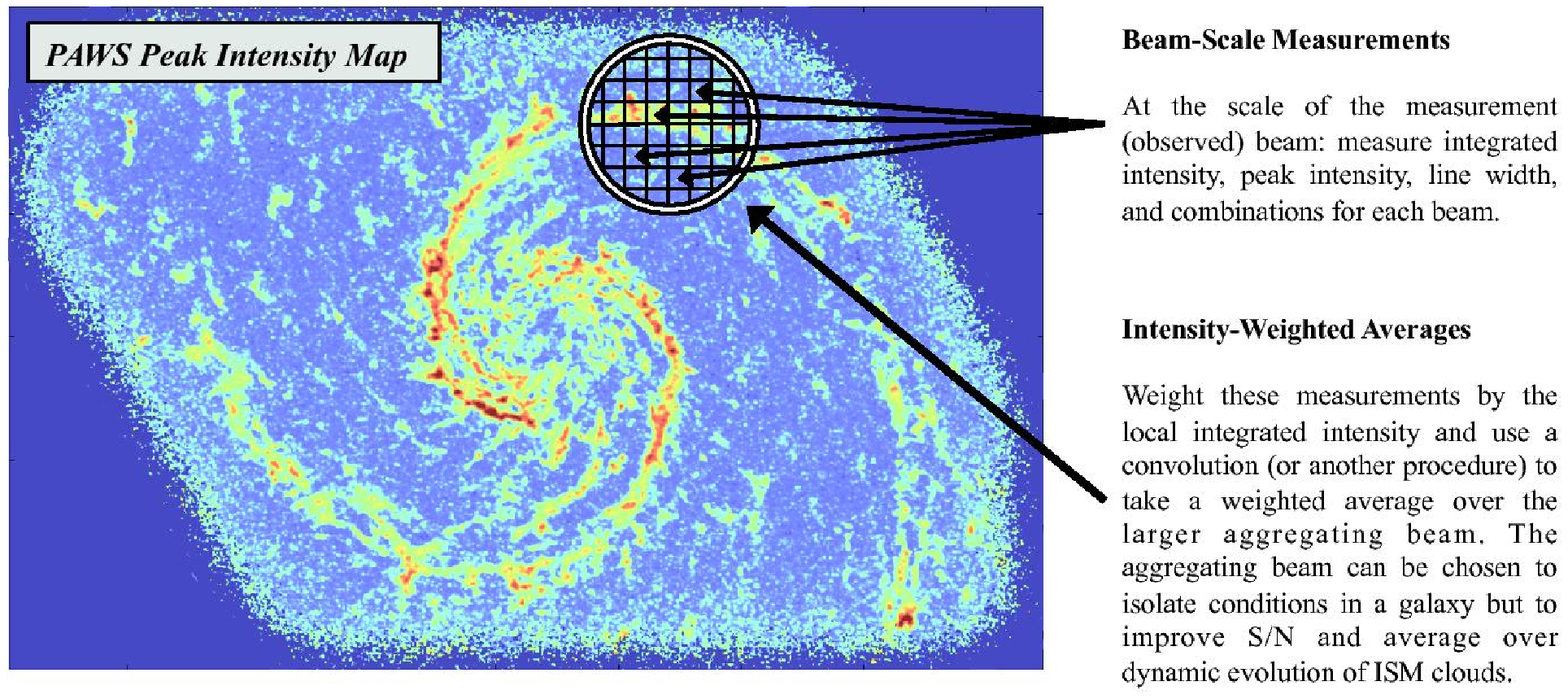}
\caption{Sketch of our method on the PAWS CO(1-0) peak intensity map of M51 \citep{SCHINNERER13,PETY13}. In each independent, high resolution beam (illustrated by the small squares), we measure the peak intensity, integrated intensity, line width, and mean velocity of the gas. With a few assumptions, these constrain the gas surface and volume density, the turbulent velocity dispersion and Mach number, and the dynamical state of the gas on the scale of an individual cloud (Section \ref{sec:props}). We then use an intensity-weighting scheme to derive ensemble averages on larger scales (the large circle), which allows us to test hypotheses linking this cloud-scale structure to the time-averaged process of star formation and the large-scale structure of the galaxy.}
\label{fig:sketch}
\end{figure*}

Our method has two steps, illustrated in Figure \ref{fig:sketch}:

\begin{enumerate}
\item Measure the properties of a mass-tracing\footnote{Here ``mass-tracing'' means a line with near uniform emissivity, so that a single value relates integrated intensity to mass surface density; for example, this is often the case for the low-$J$ $^{12}$CO lines \citep{BOLATTO13A} or the {\sc Hi} 21-cm line.} spectral line for each independent beam in a high physical resolution, wide-area data cube.
\item Characterize the ensemble of beam-scale properties across a region of interest, carrying out averages using an intensity-weighted averaging scheme. 
\end{enumerate}

\noindent We then use these ensemble averages to gauge the mean physical state of the ISM or to test hypotheses relating local ISM properties to galaxy structure and processes like star formation.

In each beam (each small square in Figure \ref{fig:sketch}), we measure the peak intensity, integrated intensity, mean velocity, and line width. With a few assumptions, these trace the gas surface and volume density averaged over the beam, the turbulent velocity dispersion (related to the Mach number), and the gravitational boundedness of the gas. Section \ref{sec:props} describes these measurements and the link to physical properties.

We use an intensity-weighted averaging scheme to aggregate cloud properties across a larger region of interest (the big circle in Figure \ref{fig:sketch}). Measurements made at the beam scale are weighted by the local integrated intensity before averaging is performed (e.g., by convolution to a coarser resolution; see Equation \ref{eq:weight} and Section \ref{sec:resolutions}). Using a related approach, we measure the distribution of intensity inside the averaging region (Section \ref{sec:distributions}). The resulting ensemble averages capture the typical cloud-scale conditions within a larger region of the galaxy.

The rest of this section lays out our methodology in detail. We describe the measurements that we use to characterize the ISM beam-by-beam (Section \ref{sec:props}), which we refer to as ``beamwise'' measurements. Then we discuss our weighted averages, including their natural extension to catalogs of point sources, e.g., cloud catalogs (Section \ref{sec:resolutions}). Finally, we discuss how we incorporate distributions into our treatment (Section \ref{sec:distributions}). We defer several more technical aspects of the methodology to the Appendices: (1) the use of spectral stacking (``shuffling'') to avoid sensitivity biases, (2) calculation of uncertainties, and (3) treatment of the finite spectral response of the data. We discuss a number of extensions and future applications of this method in Section \ref{sec:future}.

\subsection{Beam-Wise Measurements}
\label{sec:props}

We consider spectroscopic mapping observations of a line taken to be mass-tracing. In this paper, we focus on low-$J$ $^{12}$CO lines, but the method applies as well to high physical resolution observations of the {\sc Hi} 21-cm line or tracers of gas at different densities. We are interested in data sets for which the physical size of the beam, $\theta$, is comparable to the scale of individual molecular clouds ($\theta \approx 50$~pc).

Gaussian line profiles are typical for CO and {\sc Hi}  emission from the disks of star-forming galaxies \citep[e.g., see][]{PETRIC07,SCHRUBA11,CALDUPRIMO13}, though they are not universal, especially in extreme environments \citep[e.g.,][]{JOHNSON15}. Three pieces of information specify a Gaussian line profile: the peak intensity, line width, and mean velocity. 

The line width, peak intensity, and their combination, the integrated intensity, provide an observational estimate (with varying degrees of directness) of the surface density, volume density and associated free-fall time, turbulent velocity dispersion and associated Mach number, and gravitational boundedness of the gas. Their distributions contain higher order information related to the GMC mass function and density probability distribution function (PDF). Their correlations relate to cloud structure and clustering. The mean velocity is less important in this paper, though we use it for spectral stacking (Section \ref{sec:stack}).

\begin{deluxetable}{lcc}[t!]
\tabletypesize{\scriptsize}
\tablecaption{Measured or Estimated Quantities \label{tab:quantities}}
\tablewidth{0pt}
\tablehead{
\colhead{Name} & 
\colhead{Symbol} &
\colhead{Units} 
}
\startdata

\multicolumn{3}{c}{Beamwise observables} \\
\multicolumn{3}{c}{for a $\theta$pc (FWHM) beam} \\[0.5ex]
\hline
\\[-1.8ex]
Integrated intensity & $I_{\rm \theta pc}$ & K~km~s$^{-1}$ \\
Mean velocity & $v_{\rm \theta pc}$ & km~s$^{-1}$ \\
Line width\tablenotemark{a} & $\sigma_{\rm \theta pc}$ & km~s$^{-1}$ \\
Peak intensity & $I_{\rm \nu, pk, \theta pc}$ & K \\
``Boundedness''\tablenotemark{b} & $B_{\rm \theta pc} \equiv I / \sigma^2$ & $\frac{{\rm K~km~s}^{-1}}{({\rm km~s}^{-1})^2}$ \\[0.5ex]
\hline
\\[-1.8ex]
\multicolumn{3}{c}{Related quantities} \\
\multicolumn{3}{c}{accessed with additional assumptions} \\[0.5ex]
\hline
\\[-1.8ex]
Surface density\tablenotemark{b} & $\Sigma_{\rm \theta pc}$ & M$_\odot$~pc$^{-2}$ \\
Volume density\tablenotemark{c} & $\rho_{\rm \theta pc}$ & M$_\odot$~pc$^{-3}$ \\
Gravitational free-fall time & $\tau_{\rm ff}$ & yr \\
Turbulent mach number\tablenotemark{d} & $\mathcal{M}$ & \nodata \\
Virial parameter\tablenotemark{e} & $\alpha_{\rm vir} \approx \frac{2 \mathit{KE}}{\mathit{UE}}$ & \nodata \\[0.5ex]
\hline
\\[-1.8ex]
\multicolumn{3}{c}{Notation for intensity-weighted} \\
\multicolumn{3}{c}{ensemble averages} \\[0.5ex]
\hline
\\[-1.8ex]
$\left<Q_{\rm \theta pc}\right>_{A{\rm pc}}$ & \multicolumn{2}{l}{Intensity weighted average} \\
&  \multicolumn{2}{l}{... of quantity $Q$} \\
&  \multicolumn{2}{l}{... at measurement scale $\theta$} \\
&  \multicolumn{2}{l}{... over averaging beam $A$} \\[0.5ex]
\hline
\\[-1.8ex]
\multicolumn{3}{c}{Quantities measured from light} \\
\multicolumn{3}{c}{distribution in the averaging beam} \\[0.5ex]
\hline
\\[-1.8ex]
16, 50, 84$^{\rm th}$ percentile for & & \\
\ldots integrated intensity & $I_{\rm \theta pc}^{16,50,84}$ & K~km~s$^{-1}$ \\
\ldots intensity & $I_{\rm \nu, \theta pc}^{16,50,84}$ & K \\
Logarithmic distribution width & $\Delta^{84-16}$ & dex
\enddata
\tablenotetext{a}{Expressed as rms velocity dispersion about mean; measurable from fit, moments, or equivalent width. When relevant a Gaussian profile is assumed to translate expressions.}
\tablenotetext{b}{With an appropriate light-to-mass conversion factor.}
\tablenotetext{c}{With an assumed depth or scale height.}
\tablenotetext{d}{After accounting for non-turbulent contributions and for an assumed temperature.}
\tablenotetext{e}{For a fixed size scale.}
\end{deluxetable}

\subsubsection{Observables}
\label{sec:obs}

Table \ref{tab:quantities} summarizes the observables and related quantities used in our analysis.

The {\em integrated intensity}, $I$, is the sum of specific intensity over the line profile. For a mass-tracing line, $I$ will be closely related to the surface density and should also trace the volume density of the gas, though less directly.

The {\em central velocity}, $v$, defines the center of the line profile. It can be calculated in several ways. In this paper, we use the intensity-weighted mean velocity, mostly for stacking purposes.

The {\em one dimensional velocity dispersion}, $\sigma$, characterizes the width of the profile, equivalent within a numerical factor to the FWHM ($=2.35 \sigma$) or HWHM ($=1.18\sigma$) for a Gaussian profile. This quantity captures the kinetic energy of the gas. In the case where bulk and thermal motions are small, $\sigma$ relates closely to the turbulent Mach number of the gas, $\mathcal{M}$. 

Along with $\sigma$ and $v$, the {\em peak intensity} of the line, $I_{\rm \nu,pk}$, specifies a Gaussian profile. Several different physical meanings have been attributed to the peak intensity, most intriguingly a mapping to the abundance of a cold, narrow line width gas phase \citep[e.g.,][]{BRAUN97}. For a known source temperature (e.g., from multi-transition modeling) and an optically thick line, it also closely relates to the gas filling factor. Statistically, $I_{\rm \nu,pk}$ is interesting because it is independent of the line width, $\sigma$, while the integrated intensity, $I$, is covariant with $\sigma$.

In this paper, we measure $\sigma$ using the ``equivalent width'' approach because it is less sensitive to emission in the line wings than a moment-based approach, and it assumes less about the line shape than direct fitting. Following \citet{HEYER01}, we define the equivalent width, denoted as $\sigma$, as:

\begin{equation}
\label{eq:ew}
\sigma = \frac{I}{\sqrt{2 \pi} I_{\rm \nu, pk}}~.
\end{equation}

\noindent Note that the optical definition of equivalent width normalizes by the strength of the continuum, while this millimeter-wave definition uses the peak intensity of the line. Sensitivity to the peak temperature measurement represents the main drawback of this approach, because $I_{\nu, \rm pk}$ can be biased low by averaging the line within a spectral channel and biased high by the tendency to identify upward scattering noise peaks as the brightest pixel. Despite these drawbacks, the method performs well on simulated data, requires few assumptions, and is less sensitive to noise than the moment approach.

\subsubsection{Link to Physical Conditions}
\label{sec:physcond}

{\em Surface Density:} With an adopted conversion factor, $\alpha_{\rm conv}$, the integrated intensity, $I$, corresponds to the mass surface density of gas, 

\begin{equation}
\Sigma \left[ \frac{{\rm M}_\odot}{{\rm pc}^2}\right] = \alpha_{\rm conv} \left[ \frac{{\rm M}_\odot~{\rm pc}^{-2}}{{\rm K~km~s}^{-1}}\right] \times I \left[ {\rm K~km~s}^{-1} \right]~.
\end{equation}

\noindent The conversion factor, $\alpha_{\rm CO}$, between CO emission and H$_2$ has been discussed at length elsewhere \citep[see][]{BOLATTO13A,SANDSTROM13,BLANC13,LEROY11}. For CO~(1-0), we adopt a fiducial $\alpha_{\rm CO}^{1-0} = 4.35$ \acounits . 

{\em Volume Density and free-fall Time:} With an assumed physical depth, $l$, $\Sigma$ relates to the average gas mass volume density, $\rho$, and particle number density, $n$, via

\begin{eqnarray}
\label{eq:density}
\rho \left[ \frac{{\rm M}_\odot}{{\rm pc}^3} \right] &=& \frac{ \Sigma \left[{\rm M_\odot~pc}^{-2}\right]}{l \left[{\rm pc} \right]} \\
\nonumber n_{\rm H2}~\left[ {\rm cm^{-3}}\right] &=& 14.9 \times \rho \left[ \frac{{\rm M}_\odot}{{\rm pc}^3}\right]~.
\end{eqnarray}

\noindent The latter equation assumes H$_2$ and a helium mass fraction of 1.36. Both equations assume a constant density along the line of sight and correspond to the average density over a large physical area (the beam multiplied by the assumed $l$). They should thus be considered distinct from the microscopic volume density that is relevant, e.g., to excite line emission.

One can compute the gravitational free-fall time for a sphere of density $\rho$, often taken to be the controlling timescale for star formation, from

\begin{eqnarray}
\label{eq:tff}
\tau_{\rm ff} = \sqrt{\frac{3 \pi}{32 G \rho}} \approx 8.1 {\rm Myr} \left( \frac{\rho}{{\rm M}_\odot~{\rm pc}^{-3}} \right)^{-0.5}
\end{eqnarray}

{\em Mach Number:} If the observed line width is due purely to turbulent motions and the gas kinetic temperature, $T_{\rm kin}$, is known, then the three dimensional turbulent Mach number, $\mathcal{M}$, can be computed from the one dimensional velocity dispersion $\sigma$ by

\begin{equation}
\label{eq:machno}
\mathcal{M} = \frac{\sqrt{3} \sigma}{c_s} = \frac{\sqrt{3} \sigma}{0.38~{\rm km~s}^{-1}~T_{\rm 25K}^{0.5}}~.
\end{equation}

\noindent Here $T_{\rm 25K}$ is the kinetic temperature of H$_2$ divided by 25~K, a typical value for molecular gas, and we have assumed that the gas consists purely of H$_2$ molecules (these numbers would thus need to be adjusted to estimate $\mathcal{M}$ in the atomic ISM from the {\sc Hi} line). If thermal or bulk motions contribute to the line width, then Equation~\ref{eq:machno} will only hold after accounting for these effects. Note also, that many theories consider the one dimensional Mach number, which differs from Equation ~\ref{eq:machno} by a factor of $\sqrt{3}$.

{\em Gravitational Boundedness:} The balance of kinetic energy, $\mathit{KE}$, and gravitational potential energy, $\mathit{UE}$, plays a key role in the behavior of the ISM. At a basic level, strongly self-gravitating gas will collapse to form dense structures and then stars. In detail, turbulent theories treat the virial parameter, $\alpha_{\rm vir} \approx 2 \mathit{KE} / \mathit{UE}$ \citep{BERTOLDI92}, as a governing property for star formation \citep[][]{KRUMHOLZ05,FEDERRATH12,PADOAN12}. 

For a fixed size scale, $R$, the ratio $\Sigma / \sigma^2 \propto \mathit{UE}/\mathit{KE}$ is proportional to the gravitational boundedness due to self-gravity of the gas averaged over that size scale (and thus inversely proportional to the virial parameter).  We define $B \equiv I / \sigma^2$ as an observational estimate of gas boundedness at a fixed size scale $R$.

For a sphere of uniform density and radius $R$, $B$ relates to the balance of potential and kinetic energy via

\begin{eqnarray}
\label{eq:boundedness}
\frac{\mathit{UE}}{2~\mathit{KE}} &\approx& \alpha_{\rm vir}^{-1} = \frac{\frac{3 G M^2}{5 R}}{2~\frac{1}{2} M \sigma^2} = \frac{3 G M}{5 R \sigma^2} \\
\nonumber &=& \frac{3 \pi}{5}~G~R \frac{\frac{M}{\pi R^2}}{\sigma^2} = \frac{3 \pi }{5}~G~R~\frac{\Sigma}{\sigma^2} \\
\nonumber \alpha_{\rm vir}^{-1} &\approx& \frac{3 \pi }{5} G~R~\alpha_{\rm conv}~B \\
\end{eqnarray}

\noindent Here $\alpha_{\rm conv}$ is the light-to-mass conversion factor for the line of interest (e.g., $\alpha_{\rm CO}$), $G$ is the gravitational constant, and we have assumed a constant density sphere in the potential. Equation \ref{eq:boundedness} shows that $B$ maps to the ratio of potential to kinetic energy and the inverse of the virial parameter:

\begin{equation}
\label{eq:b}
B \equiv \frac{I}{\sigma^2} \propto \frac{\mathit{UE}}{\mathit{KE}} \propto \alpha_{\rm vir}^{-1} ~.
\end{equation}

\noindent This proportionality holds because $B = I/\sigma^2$ relates to the more physical ratio $\Sigma/\sigma^2$ via the mass-to-light conversion factor. As long as the size scale $R$ remains roughly fixed for a fixed resolution, $B$ probes the gravitational boundedness of the gas. In Section~\ref{sec:alpha_cld_vs_b}, we demonstrate an empirical match between $B$ and $\alpha_{\rm vir}$ measured from cloud property measurements. Here we note, for reference, the conversion from $B$ to $\alpha_{\rm vir}^{-1}$ for a uniform density sphere of radius $R$:

\begin{equation}
\label{eq:b_to_avir}
\alpha_{\rm vir}^{-1} = 1.06~R_{30}~\alpha_{4.35}^{\rm conv}~B
\end{equation}

\noindent where $R_{30}$ is $R$ divided by 30~pc, $\alpha_{\rm conv}$ is the mass-to-light ratio relative to the fiducial Milky Way value of $4.35$ \acounits , and $B$ is in units of \mbox{K~(km~s$^{-1})^{-1}$}. Other density profiles produce somewhat different proportionalities \citep[e.g., see][]{ROSOLOWSKY06}.

{\em Ratio of Dynamical Time to Free-Fall Time:} The virial parameter offers a useful way to assess the dynamical state of clouds. The same combination of parameters also traces the ratio of the cloud's crossing time, $\tau_{\rm dyn} \propto R / \sigma$, and the free-fall time, $\tau_{\rm ff} \propto \rho^{-0.5} \propto \Sigma^{-0.5}$. Then for a fixed size scale, $R$,

\begin{equation}
\label{eq:tff_to_dyn}
B \propto \frac{\Sigma}{\sigma^2} \propto \frac{\tau_{\rm ff}^{-2}}{\tau_{\rm dyn}^{-2}} = \left( \frac{\tau_{\rm dyn}}{\tau_{\rm ff}} \right)^2~.
\end{equation}

\noindent Though we defer a comparison of these gas properties to tracers of the star formation rate to a future paper, we note that \citet{PADOAN12} highlight this ratio, $\tau_{\rm dyn}/\tau_{\rm ff}$, as a controlling parameter that sets the rate of star formation in galaxies because it relates closely to the virial parameter. They predict that the efficiency of star formation per free-fall time varies as $\exp (- 1.6 \tau_{\rm ff} / \tau_{\rm dyn})$ and so should depend on $B$ within geometrical factors.

\subsection{Two Scales: Measurement and Averaging}
\label{sec:resolutions}

We consider two scales, the measurement scale and the averaging scale. The {\em measurement} scale is the physical resolution of the original data. This is the scale over which we measure $I$, $I_{\rm \nu, pk}$, $\sigma$, and $B$. We denote the measurement scale for a quantity by appending the FWHM of the beam as a subscript, e.g., $I_{\rm \theta pc}$.

The {\em averaging} scale is the scale over which we aggregate measurements using an intensity-weighted average. As discussed in Section \ref{sec:intro}, this averaging is necessary for hypothesis testing because many quantities of interest for physical theories emerge only with averaging over space or (via space) time. These include the star formation rate and star formation efficiency, which become ill-defined at small scales due to the cycling of individual regions between different stages of the star formation process \citep{KAWAMURA09,SCHRUBA10,KRUIJSSEN14}, and large-scale quantities like galactic structure. In general, any individual parcel of ISM can be expected to evolve on its local dynamical timescale, and the evolution at the scale of individual clouds can be dramatic and destructive. Averaging over a large region of emission in a common environment therefore represents the best practical way to access the time-averaged behavior of gas and star formation.

We translate between the two scales using an intensity weighted average. We weight each beam-wise measurement by $I$, the integrated intensity at the measurement scale. Next, we convolve these weighted measurements to the averaging scale and divide by the total weights in the beam. This formulation is designed to answer questions like: ``What is the 60~pc resolution integrated intensity from which the average CO photon arises within this 500~pc part of the galaxy?'' In this case, $60$~pc is the measurement scale, 500~pc is the averaging scale, and the quantity of interest is the integrated intensity. 

Formally, when using a Gaussian beam for the averaging scale, we measure the intensity-weighted average of some quantity $Q$ -- which may be $I$, $\sigma$, $B$, or something else -- via

\begin{eqnarray}
\label{eq:weight}
\left< Q_{\rm \theta pc} \right>_{A {\rm pc}} \left( x_0, y_0 \right) &=& \frac{\sum w \left( x, y \right)~I_{\rm \theta pc} (x, y)~Q_{\rm \theta pc} (x, y)}{\sum w \left( x, y, \right)~I_{\rm \theta pc} (x, y)} \nonumber \,,\\
{\rm where}\quad w (x, y) &=& \exp \left( \frac{-\left(\theta (x,y,x_0,y_0)\right)^2}{2~\sigma_{A \rm pc}^2} \right)\,.
\end{eqnarray}

\noindent Here $\theta (x, y, x_0, y_0)$ is the angular distance from the measurement point $(x_0, y_0)$, $\sigma_{A \rm pc}$ is the 1$\sigma$ width of the Gaussian averaging beam, and $A {\rm pc}$ is a shorthand subscript reporting the FWHM of the averaging beam\footnote{As written, the averaging beam does not account for inclination, but could be modified to do so and hence yield a circular beam in the plane of the galaxy.}. The sums in Equation \ref{eq:weight} run over the whole map. In practice, only pixels within a few times $\sigma_{A \rm pc}$ of $(x_0, y_0)$ contribute to ${\langle Q_{\rm \theta pc} \rangle}_{A {\rm pc}}$.

Our shorthand describing this operation thus reads ${\langle Q_{\rm \theta pc} \rangle}_{A {\rm pc}}$, which should be read as ``the intensity-weighted average of $Q$ measured at $\theta$~pc (FWHM) resolution and averaged over an $A {\rm pc}$ (FWHM) sized area.'' In this paper, we examine ${\langle I_{\rm 60pc}\rangle}_{\rm 500pc}$ (among other quantities), which is the integrated intensity measured at $60$~pc (FWHM) resolution and averaged over a 500~pc (FWHM) beam.

For any meaningful comparison of surface or volume densities, this measurement scale should be matched between data sets. This can be an absolute match; e.g., a comparison of surface densities measured with $\theta \approx 60$~pc can be meaningful. Alternatively, tailoring the measurement scale to the object in question makes sense; e.g., comparing surface densities within the effective radius of a galaxy can make sense. A gross mismatch in the measurement scale seriously undermines the meaning of any comparison. Because of the dramatic differences in resolution and distance, this issue has plagued previous efforts to unify our view of Galactic and extragalactic star formation. It has also frequently prevented meaningful comparisons between simulations and observations.

There is no absolute correct measurement scale. In theory, this formalism can be useful to merge many pairs of scales, for example relating $\sim$kpc resolution surface densities to averages over whole galaxies. In practice, our main goal is to test theories that link cloud-scale gas structure to galaxy-scale conditions. For this application, the measurement scale should approach the scale of individual clouds or the outer scale of turbulence, which are often taken to be roughly equivalent \citep[e.g.][]{KRITSUK13}. 

Similarly, there is no single correct averaging scale, but previous studies have shown that familiar scaling relations that reflect time-averaged behavior emerge at scales of a few hundred pc to a kpc \citep[see][]{SCHRUBA10,ONODERA10,LEROY13}. A Gaussian averaging beam is also not required. One could consider, e.g., dynamically distinct zones in the galaxy or radial bins instead. A round top hat could also be used to address many of the topics in this paper; we prefer the Gaussian averaging kernel for mainly aesthetic reasons: the lack of sharp edges and the ability to create smooth maps of intensity-weighted properties from irregularly sampled data.

Note that because they contain no intensity, missing regions in the map or areas outside the edge of the map will not contribute to this sum. This can lead to the case where only a small amount of emission is encompassed in the average. A similar situation can occur if the averaging beam is too small and the region is deficient in gas. In practice, this concern can be addressed by considering only regions above a minimum integrated flux, requiring some minimum covering fraction of bright molecular gas, or increasing the size of the averaging beam. 

\subsubsection{Gridding Cloud Property Measurements or Other Point Sources}
\label{sec:gridclouds}

This approach can also be applied to populations of point sources, weighting by luminosity instead of intensity. In this paper, we apply such treatment to catalogs of molecular clouds, comparing the results of our beamwise calculations to the cloud properties to benchmark our approach. We calculate ${\langle Q_{\rm GMC} \rangle}_{A \rm pc}$, the luminosity-weighted average GMC property in a Gaussian averaging beam with FWHM $A {\rm pc}$. To do so, we calculate

\begin{eqnarray}
\label{eq:weight_gmc}
\left< Q_{\rm GMC} \right>_{A {\rm pc}} \left( x_0, y_0 \right) &=& \frac{\sum^{\rm GMCs}_i w \left( x_i, y_i \right)~L_{i}~Q_{i}}{\sum^{\rm GMCs}_i w \left( x_{i}, y_{i}, \right)~L_{i}}\,,
\end{eqnarray}

\noindent where $w$ is a Gaussian convolution kernel with scale $\sigma_{A \rm pc}$ (and FWHM $A \rm pc$), defined as in Equation \ref{eq:weight}, and the sum runs over all GMCs. As before, only GMCs located within a few $\sigma_{A \rm pc}$ of the point of interest contribute. Of course, Equation \ref{eq:weight_gmc} works for any set of cataloged point sources with properties of interest, not only GMCs.

\subsection{Incorporating Distributions}
\label{sec:distributions}

\begin{figure}
\plotone{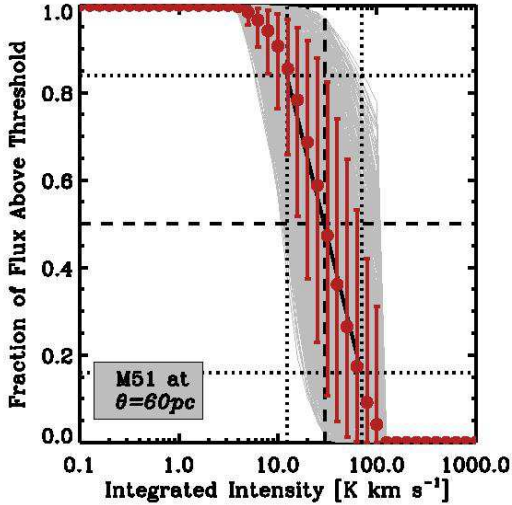}
\caption{Illustration of our approach to distributions, here applied to M51. We measure the fraction of the flux ($y$-axis) above a given integrated intensity threshold ($x$-axis), progressively varying the threshold used. Each gray line shows the result for an individual 500~pc averaging beam. Dashed and dotted horizontal lines illustrate the $16^{\rm th}$, $50^{\rm th}$, and $84^{\rm th}$ percentiles. We define the logarithm of the ratio between the integrated intensities that yield the $16^{\rm th}$ and $84^{\rm th}$ percentiles as $\Delta^{84-16}$. The solid diagonal line illustrates the implied slope in the distribution function. Red points and error bars show the median and $1\sigma$ rms scatter across all measurements for M51.}
\label{fig:dist_sketch}
\end{figure}

Beyond ensemble averages, the distributions of $I$, $\sigma$, $B$, and specific intensity, $I_\nu$, are of physical interest. These distributions roughly map to population statistics of clouds, e.g., the mass function, which show environmental correlations \citep[][Hughes et al. subm.]{ROSOLOWSKY05B,COLOMBO14A} and play important roles in many theories for galactic-scale star formation \citep[e.g.][]{TAN00,KRUMHOLZ05,MEIDT13}. 

Similar to \citet{SAWADA12} and \citet{HUGHES13A}, we gauge the distribution of emission within the averaging beam. Within each beam, we calculate the sum of integrated intensity from lines of sight of sight above a succession of threshold intensities, $t$. Varying $t$ from a low to high value, we calculate the total flux in the beam above each threshold. By dividing the total flux obtained for each value of $t$ by the total flux, we construct an analog to the integrated intensity cumulative distribution function (CDF). The process can be applied to the cube itself, thresholding on specific intensity and summing voxels, or to the maps, thresholding on integrated intensity. The volumetric implementation resembles the ``brightness distribution index'' of \citet{SAWADA12}. The main difference is that their approach is differential, and involves contrasting $I_{\nu}$ over two narrow specific ranges of $I_{\nu}$.

Our approach also differs from that of \citet{SAWADA12} and \citet{HUGHES13A} in that we consider the distribution of intensity rather than number counts of pixels. The distinction is somewhat arbitrary; the two are readily related by an integral or derivative. We focus on the distribution of intensity because we are primarily interested in how the bulk of the gas behaves. Based on the distributions measured by \citet{HUGHES13A} and the GMC mass spectra measured by \citep{ROSOLOWSKY05B}, reasonable completeness in intensity is also easier to achieve than completeness in number counts. There are many low intensity lines of sight, but for the most part they do not contribute an overwhelming fraction of the total luminosity. To an extent, this can be checked by the convergence tests we describe below.

In practice, we implement these calculations by adding an additional masking criterion to the maps and cubes created at our measurement scale. When thresholding by integrated intensity, we:

\begin{enumerate}
\item Set all lines of sight with $I < t$ to have $I=0$ in the integrated intensity (moment-0) map at the measurement scale.
\item Convolve the integrated intensity map to the averaging scale.
\item Record the convolved, thresholded intensity at each location.
\item Repeat for a succession of values of $t$.
\end{enumerate}

\noindent This algorithm acts on the integrated intensity map and yields the distribution of flux as a function of integrated intensity. When instead considering the distribution of specific intensity, we threshold on specific intensity, $I_{\nu}$, in the data cube and set voxels with $I_{\nu} < t$ to 0. We then sum the flux over the averaging beam and proceed as above.

The result of the above procedure is a measurement of the flux CDF as a function of specific or integrated intensity. Figure \ref{fig:dist_sketch} shows the integrated intensity distribution function that we obtain for M51 (PAWS). Each gray line represents a CDF for an individual 500~pc averaging beam, the red bins show the median trend and scatter. 

The CDF can be analyzed in a number of ways. In this paper, we record a few basic properties,

\begin{enumerate}
\item The intensity thresholds corresponding to $16$, $50$, and $84$ percent of the included flux.
\item The logarithmic difference between the $84^{\rm th}$ and $16^{\rm th}$ percentile, $\Delta^{84-16}$.
\end{enumerate}

\noindent These quantities are illustrated by the dotted and dashed lines in Figure \ref{fig:dist_sketch}. In Section~\ref{sec:distribution_methods}, we show that the $50^{\rm th}$ percentile value corresponds well to the intensity weighted mean that we measure above. Meanwhile, $\Delta^{84-16}$ captures the width of the mass distribution. For a linear translation of light to mass (i.e. a fixed conversion factor), it can be re-expressed as a logarithmic slope of the mass CDF. We return to this below.

\subsection{Convergence In Intensity Measurements}
\label{sec:convergence}

Our measurements will be most interesting when the data set analyzed recovers a large fraction of the total flux from the galaxy at good signal-to-noise at the measurement scale. This allows robust calculation of moments for each line-of-sight, in turn allowing shuffling and stacking to make detailed line profile measurements.

Using stacking (see Section \ref{sec:stack}), our method should be sensitive to all of the flux along each line of sight for which we find bright signal. In this paper, ``bright signal'' mean two channels at $\mathrm{S/N} > 5$; more generally, it means sufficient signal to calculate a first moment and include the emission in the stacking. To measure how much of the total flux in the cube such an analysis characterizes, we carry out the following calculation:

\begin{enumerate}
\item Mask the data at the measurement scale using the criteria used to calculate the first moment.
\item Expand this mask so that it includes all velocities along a line of sight with any bright signal.
\item Integrate the emission in the data cube that lies inside the mask and the total emission in the data cube.
\item Divide the two to estimate the fraction of flux characterized by the methods above.
\end{enumerate}

If the fraction is high, our approach offers a robust way to describe the properties of the ISM over part of a galaxy. If the fraction is low, a more sophisticated approach may be merited. One option would be to use another bright line as a prior on the local velocity of CO \citep[e.g., as done with {\sc Hi} and CO by][]{SCHRUBA11}. Another option would be to use lower resolution versions of the data or to otherwise interpolate the velocity field to predict the local velocity or to work at a coarser resolution with better flux recovery. Alternatively one could work entirely in the unmasked data, simultaneously fitting the distributions of signal and noise in three-dimensional space.

\section{Data and Application}
\label{sec:data}

\begin{figure*}
\includegraphics[width=0.34\textwidth]{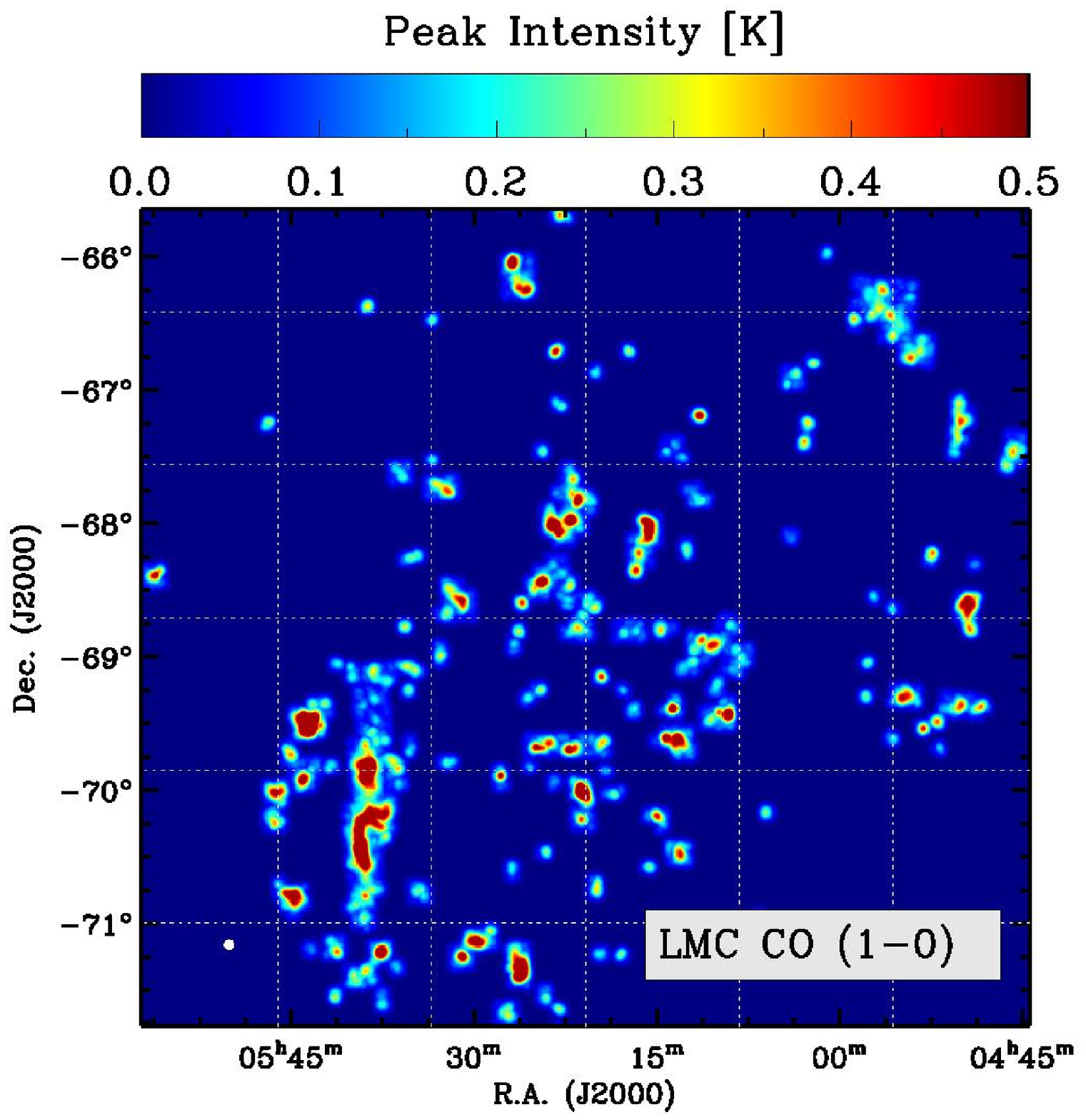}
\includegraphics[width=0.34\textwidth]{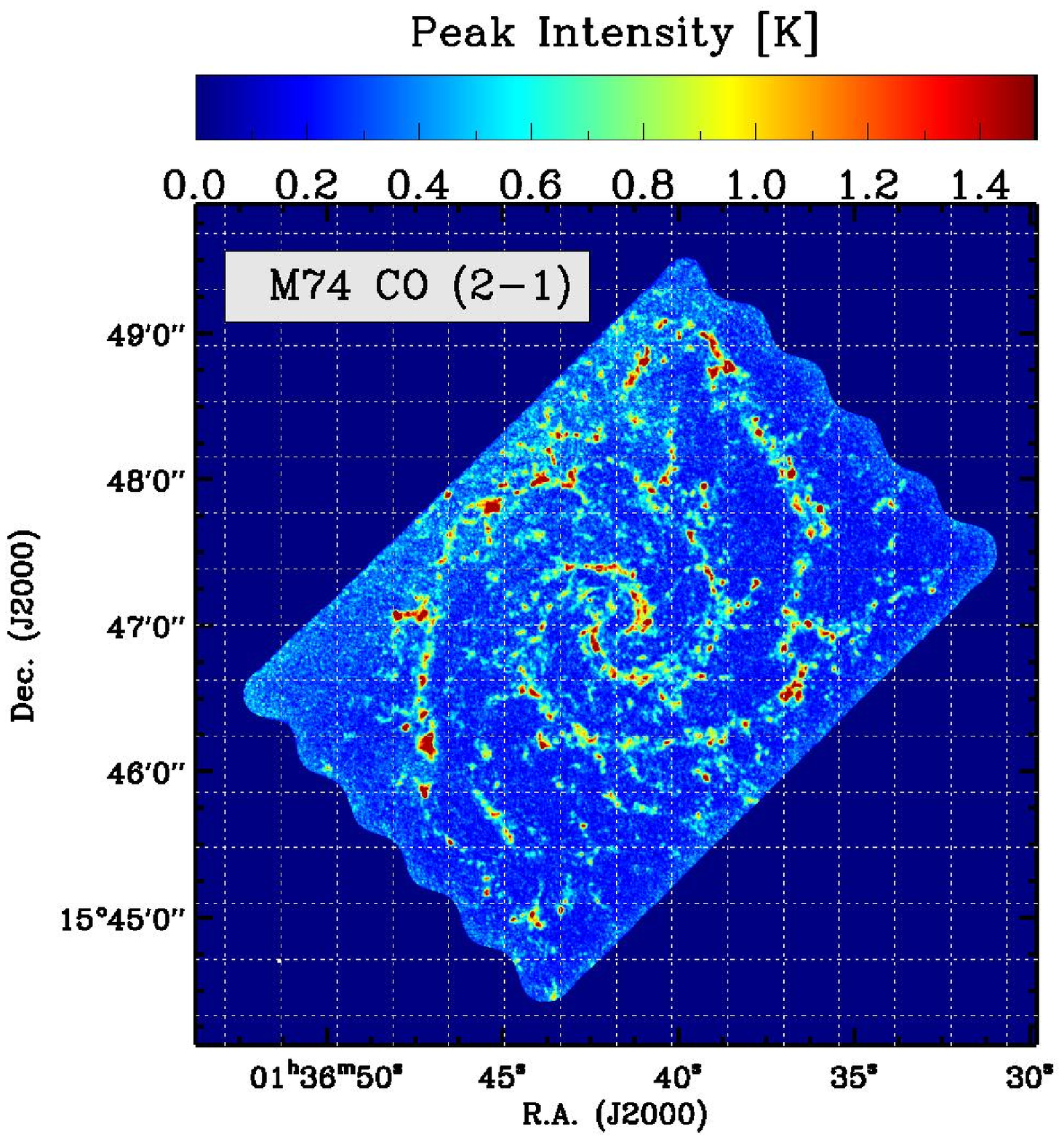}
\includegraphics[width=0.27\textwidth]{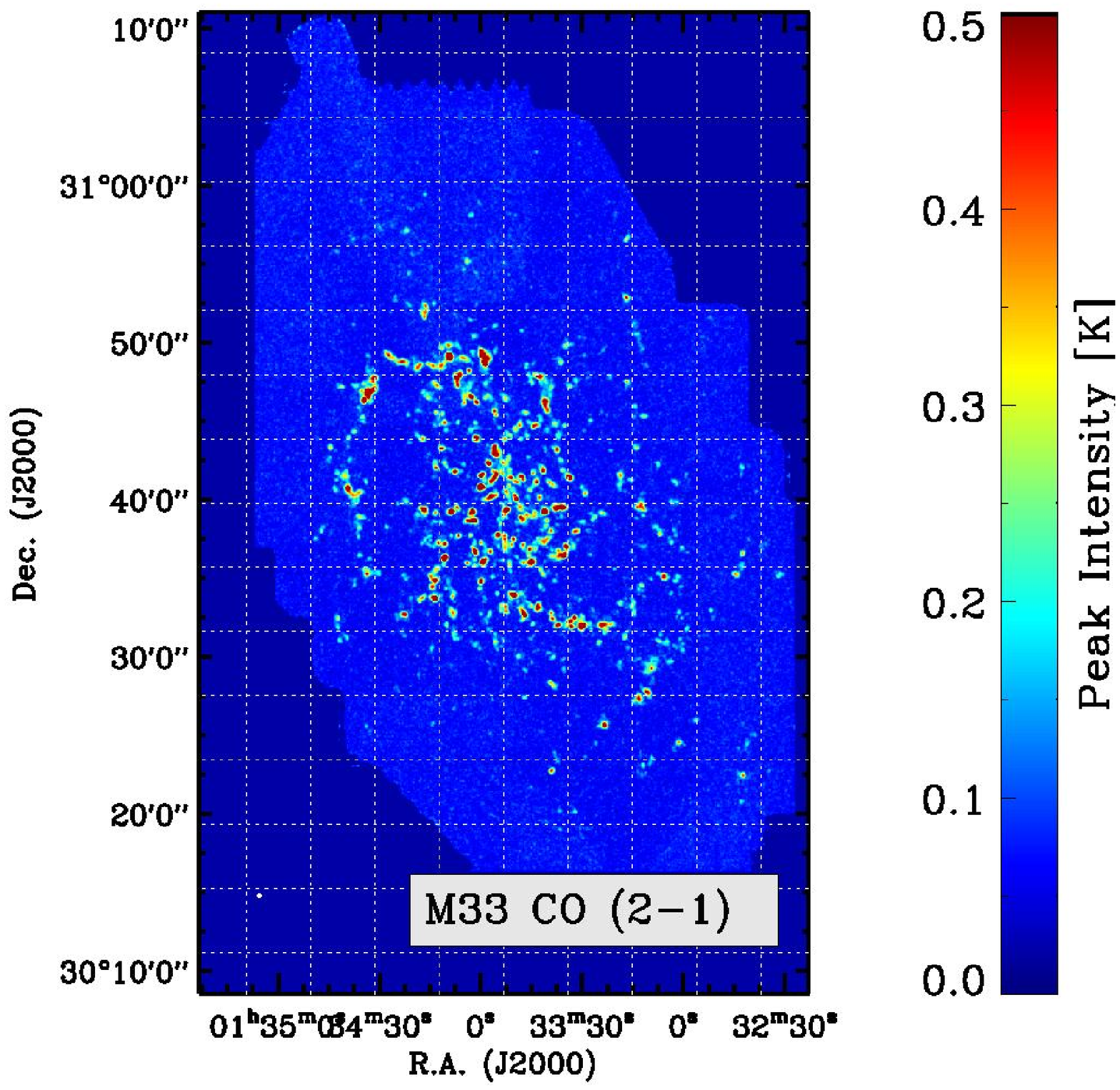}
\includegraphics[width=0.35\textwidth]{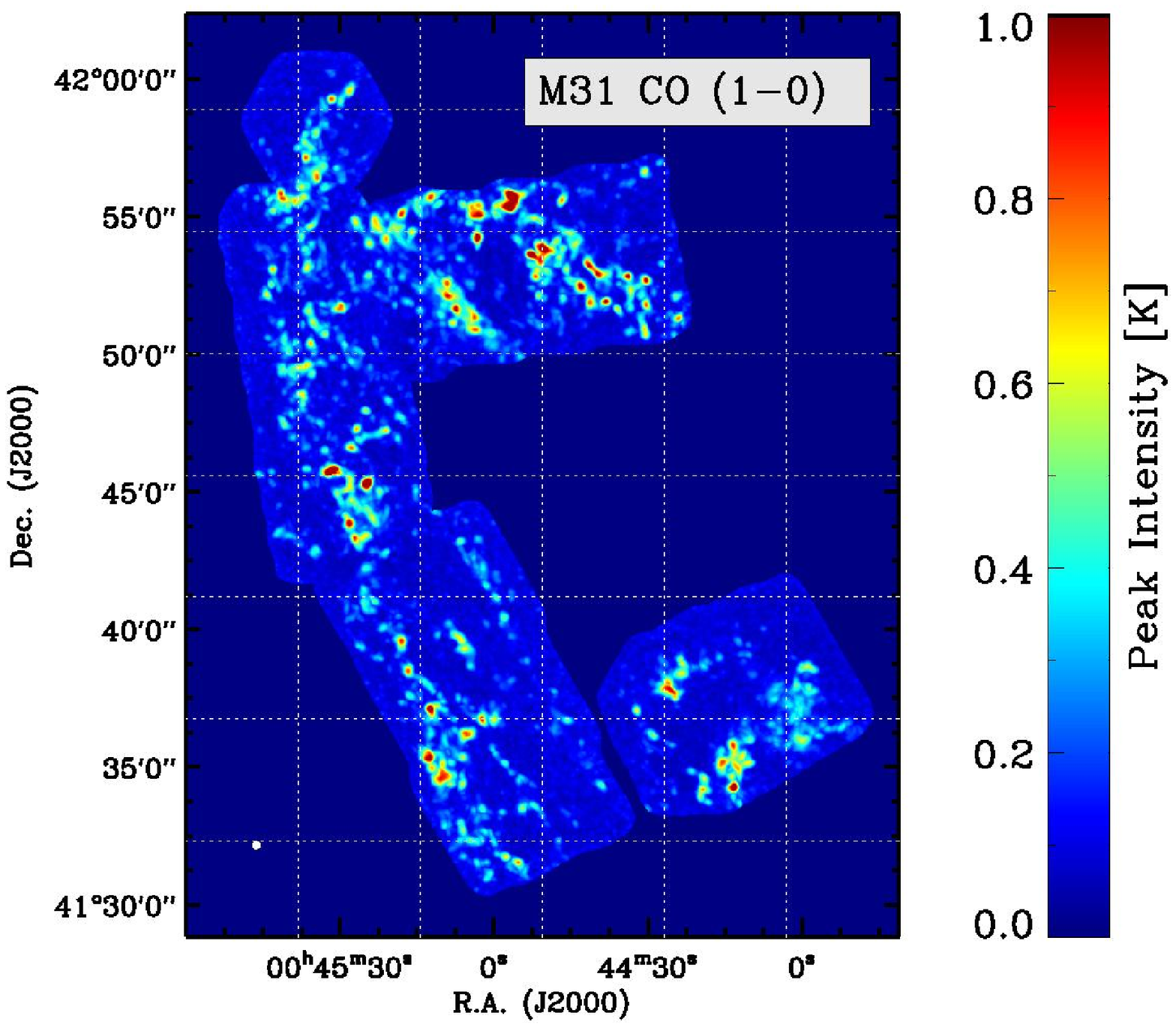}
\includegraphics[width=0.35\textwidth]{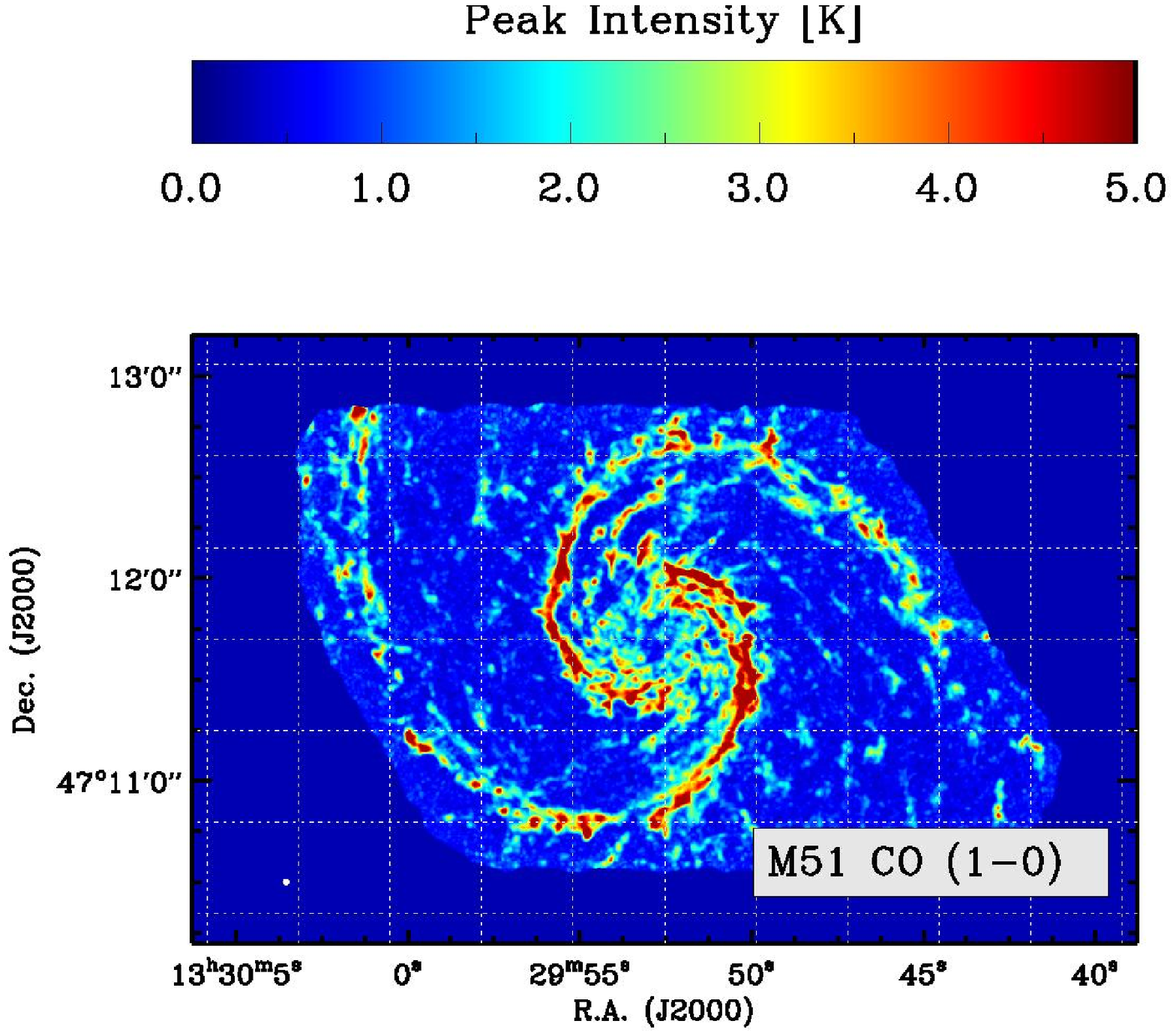}
\includegraphics[width=0.28\textwidth]{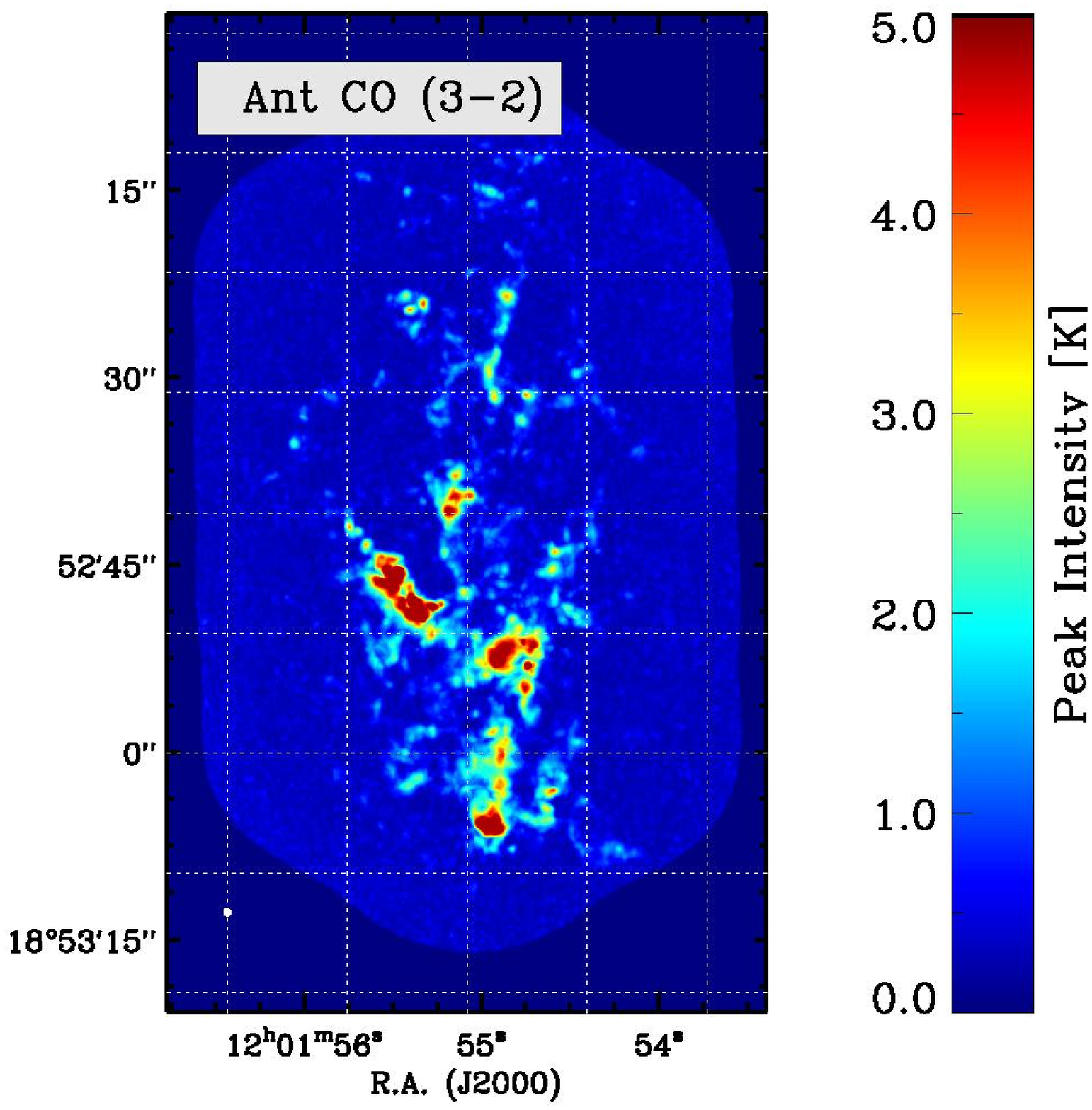}
\caption{Peak intensity maps for CO emission at $\theta = 60$~pc (FWHM) for ({\em top left}) the Large Magellanic Cloud (LMC) from \citet{WONG11}, ({\em top middle}) M74 from ALMA (Schinnerer, Sliwa, et al. in prep.), ({\em top right})  M33 from \citet{GRATIER10B} and \citet{DRUARD14} and ({\em bottom left}) M31 from CARMA (Schruba et al., in prep),  ({\em bottom middle}) M51 from \citet{SCHINNERER13} and \citet{PETY13}, and ({\em bottom right}) the overlap region of the Antennae galaxies from \citet{WHITMORE14}. A beam in the bottom left corner shows the 60~pc FWHM size of the beam; the dotted white lines are spaced by $1$~kpc at the distance to the target. We analyze these data using both our beamwise method and cloud properties software.}
\label{fig:maps}
\end{figure*}

\begin{deluxetable*}{lcccccl}
\tabletypesize{\scriptsize}
\tablecaption{Data Analyzed \label{tab:data}}
\tablewidth{0pt}
\tablehead{
\colhead{Galaxy} &
\colhead{Line} &
\colhead{Distance} &
\colhead{Resolution\tablenotemark{a}} & 
\colhead{Channel\tablenotemark{b}} &
\colhead{Adopted $\alpha_{\rm CO}$\tablenotemark{c}} &
\colhead{Reference} \\
\colhead{} & 
\colhead{} &
\colhead{(Mpc)} &
\colhead{(pc)} & 
\colhead{(km~s$^{-1}$)} &
\colhead{( $\frac{{\rm M}_\odot~{\rm pc}^{-2}}{{\rm K~km~s}^{-1}}$ )} &
\colhead{}
}
\startdata
Antennae & CO~(3-2) & 21.5 & 60 & 5.0 & 17.4 & \citet{WHITMORE14} \\
LMC & CO~(1-0) & 0.05 & 15 & 1.6 & 8.7 &  \citet{WONG11} \\
M31 & CO~(1-0) & 0.78 & 20 & 2.5 & 4.35 & Schruba et al. (in prep.) \\
M33 & CO~(2-1) & 0.84 & 50 & 2.6 & 10.8 & \citet{DRUARD14,GRATIER10B} \\
M51 & CO~(1-0) & 7.6 & 45 & 5.0 & 4.35 &  \citet{SCHINNERER13,PETY13} \\
M74 & CO~(2-1) & 9.0 & 45 & 2.0 & 7.9 & Schinnerer et al. (in prep.)
\enddata
\tablenotetext{a}{Native resolution; the analysis in this paper is carried out at a common 60~pc resolution.}
\tablenotetext{b}{Channel width used in analysis; the spectral resolution in the LMC has been degraded to this value to increase signal to noise.}
\tablenotetext{c}{Adopted mass to light ratio for the line in question, see Section \ref{sec:convfac}.}
\end{deluxetable*}

\begin{deluxetable}{lccc}
\tabletypesize{\scriptsize}
\tablecaption{Measured Data Set Properties at 60~pc Resolution \label{tab:data_props}}
\tablewidth{0pt}
\tablehead{
\colhead{Galaxy} & 
\colhead{Channel\tablenotemark{a}} &
\colhead{Noise\tablenotemark{b}} &
\colhead{Completeness\tablenotemark{c}}
\\
\colhead{} & 
\colhead{Coupling ($k$)} & 
\colhead{(K)} &
\colhead{}
}
\startdata
Antennae & 0.07 & $0.1$ & $1.1$ \\
LMC & 0.06 & $0.03$ & $0.9$ \\
M31 & 0.09 & $0.03$ & $0.9$ \\
M33 & 0.12 & $0.03$ & $0.7$ \\
M51 & 0.14 & $0.2$ & $1.0$ \\
M74 & 0.07 & $0.1$ & $0.7$
\enddata
\tablenotetext{a}{Coupling between channels expressed as $k$ (Equation \ref{eq:kern}) in the kernel that reproduced the observed channel-to-channel noise correlation.}
\tablenotetext{b}{Representative rms noise per channel at 60~pc resolution from signal-free regions. Note that the channel widths vary among data sets.}
\tablenotetext{c}{Ratio of flux included in the analysis to flux in the cube. The value above unity in the Antennae indicates that these data lack short spacing information, so that moderate ``clean bowls'' persist outside the region that we treat.}
\end{deluxetable}

We apply our methodology to characterize the ISM at $60$~pc resolution using the six high resolution, good sensitivity data sets available to us. These data sets, shown in Figure \ref{fig:maps}, are the PAWS map of CO \mbox{1-0} from M51 \citep{SCHINNERER13}; the ALMA CO \mbox{2-1} map of M74 (P.I.: Schinnerer); the CARMA CO \mbox{1-0} survey of Andromeda (Schruba et al., in prep.); the IRAM CO \mbox{2-1} survey of M33 \citep{DRUARD14,GRATIER10B}; the MAGMA CO \mbox{1-0} survey of the Large Magellanic Cloud \citep{HUGHES10,WONG11}; and the CO \mbox{3-2} survey of the overlap region of the Antennae \citep{WHITMORE14,JOHNSON15}. All of these, except the ALMA Antennae survey, include single dish data, and so they recover the full flux of the target. The interacting Antennae yield a very interesting contrast to the more quiescent spiral and dwarf galaxies, and so we include them despite the different line (CO \mbox{3-2}) and lack of short spacing data.

For each cube, we convolve the data to have a round beam, convert to units of Kelvin, and then apply the processing described above and in the Appendices. Specifically, operating on the cube at the measurement  ($\theta = 60$~pc) scale:

\begin{enumerate}
\item We identify bright signal in the data cube following the CPROPS methodology of \citet{ROSOLOWSKY06}. We begin with regions that show $\geq 2$ consecutive channels above $\mathrm{S/N} = 5$. We then expand the mask to include all contiguous regions with $\geq 2$ consecutive channels above $\mathrm{S/N} = 1.5$.
\item We calculate the integrated intensity, first and second moment, and peak intensity based on the emission inside this mask.
\item We create a ``shuffled'' version of the data cube, regridding each spectrum so that the local first moment is the new reference velocity \citep[see][and Section \ref{sec:stack}]{SCHRUBA11}. We apply this operation to the unmasked data cube.
\item We weight each measurement (moment) and each spectrum in the shuffled cube by the local integrated intensity, $I$, calculated via the zeroth moment of the masked cube.
\item We convolve these intensity-weighted measurements to the averaging scale. We also convolve the integrated intensity to this scale. We divide the weighted, convolved quantity (or spectrum) by the convolved integrated intensity map. This yields an intensity-weighted average of each measurement.
\item From the intensity-weighted shuffled spectrum at the averaging scale, we derive ${\langle I_{\rm 60pc} \rangle}_{\rm 500pc}$, ${\langle I_{\nu, \rm pk, 60pc} \rangle}_{\rm 500pc}$, ${\langle \sigma_{\rm 60pc} \rangle}_{\rm 500pc}$, and ${\langle B_{\rm 60pc} \rangle}_{\rm 500pc}$ at each location.
\item We also measure the (non-weighted) integrated intensity in each beam at the averaging scale using the progressive thresholding described in Section \ref{sec:distributions}. This diagnoses the distribution of measurement-scale intensities and integrated intensities within the averaging beam.
\end{enumerate}

Following \citet{LEROY12,LEROY13}, we build a Nyquist-spaced (for the averaging beam) hexagonal grid and sample the intensity weighted measurements. This yields a moderately oversampled database in which the intensity-weighted properties of the ISM at the measurement scale are characterized over the scale of the averaging beam at each position in each galaxy.

Following Section \ref{sec:uncertainties}, we also carry out a Monte Carlo calculation to estimate the uncertainties associated with each measurement. We realize 100 data sets with randomly generated noise and record these as an ensemble of mock databases. Analyzing their distribution, we estimate the magnitude of and covariance among statistical uncertainties in each parameter.

For each data set, we also run several cloud property analysis algorithms, which decompose the emission into discrete clouds and measure their properties. We apply the CPROPS algorithm \citep{ROSOLOWSKY06}, a seeded version of the CLUMPFIND \citep{WILLIAMS94,ROSOLOWSKY05} algorithm, and the dendrogram multiscale approach \citep{ROSOLOWSKY08}. In each case we characterize the emission in a ``cloud'' following \citet{ROSOLOWSKY06} with modifications noted in \citet{LEROY15A}.

The results of these cloud calculations are estimates of the size, line width, and luminosity for each cloud, as well as their combinations, the cloud surface density and the virial parameter (see Section \ref{sec:props}). We record these, so that we have a large set of cloud properties for each target. We aggregate these into our database using the approach to gridding cloud properties described in Section \ref{sec:gridclouds}, carrying out a luminosity-weighted Gaussian gridding using a kernel with the averaging scale.

We repeat our measurements at a series of resolutions, beginning with the native resolution listed in Table \ref{tab:data} and increasing to $300$~pc. This changes the measurement scale.

We note some processing details specific to individual data sets. In the LMC, we assume that regions outside the MAGMA field-of-view have zero intensity; MAGMA recovers ${\sim}80\%$ of the full-galaxy CO flux as gauged from the NANTEN \citep{FUKUI99} survey, so this assumption should introduce minimal bias \citep[see][]{WONG11,HUGHES13A,HUGHES13B}. For the M33 map, the noise is inhomogeneous due to the inclusion of some deep regions used in focused studies. We homogenize the noise by adding $12\arcsec$ resolution noise with appropriate spectral correlation to bring the cube to an approximately even $40$~mK ($T_{\rm mb}$) noise level. We apply the efficiency correction noted in \citet{DRUARD14}.

Table \ref{tab:data_props} reports properties of the data cubes that we use at the $\theta =60$~pc (FWHM) resolution of our analysis. We give the channel coupling, $k$, estimated from the channel-to-channel correlation following Section \ref{sec:specresponse}, and the rms noise estimated from signal-free regions. 

\subsection{Convergence}

\begin{figure}
\plotone{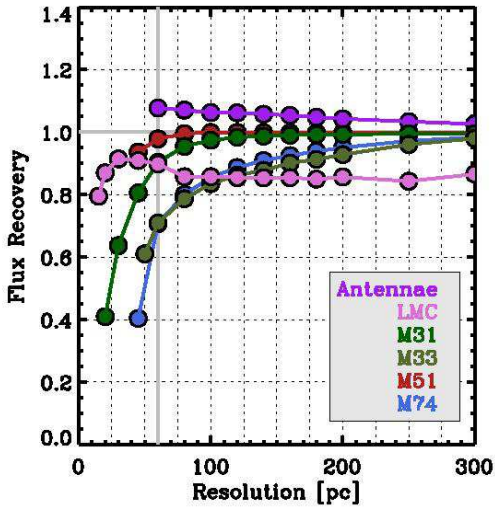}
\caption{Flux recovery in our data, defined in Section \ref{sec:convergence}, plotted as a function of resolution. The measurement scale $\theta = 60$~pc used in this paper appears as a thick vertical gray line.}
\label{fig:completeness}
\end{figure}

The last column notes of Table \ref{tab:data_props} reports the completeness of the data, calculated following Section \ref{sec:convergence}. At our resolution of $\theta=60$~pc, we recover ${\sim}70{-}100\%$ of the total flux in the cube. Note that the value larger than unity in the Antennae galaxy reflects large regions of moderately negative data, ``clean bowls,'' in that data cube, which lacks short spacing correction. Although the ${\sim}30\%$ of the flux missed by our analysis in M33 and M74 is significant, in each case our calculation characterize the bulk of the emission. The remaining flux could be included in the analysis either by using a line other than CO (likely {\sc Hi}) as a prior to stack faint regions \citep{SCHRUBA11} or by a full analysis of the intensity distribution in the data cube.

Figure \ref{fig:completeness} plots flux recovery as a function of resolution for our target data sets. For the most part, flux recovery improves as the resolution becomes coarser. This reflects that improved signal-to-noise allows more areas to be included in the mask. The effect is strongest for M31, M33, and M74, where the signal-to-noise at the native resolution is good but not sufficient to recover all of the faint emission in the cube. The decline from high to low resolution in the Antennae reflects image reconstruction artifacts.

\subsection{Adopted Conversion Factors}
\label{sec:convfac}

In Section \ref{sec:results} we will translate CO intensities into masses. To do this, we adopt a CO-to-H$_2$ conversion factor appropriate to each line and galaxy. Refining these represents its own area of research; here we only note our adopted value for each galaxy and motivate our choice.

For the Antennae, we adopted $\alpha_{\rm CO}^{3-2} = 17.4$ \acounits . This combines a Milky Way conversion factor for CO~(1-0) and a CO (3-2) / (1-0) ratio of $\approx 0.25$ \citep{UEDA12,BIGIEL15}.

For the LMC, we take $\alpha_{\rm CO} = 8.7$ \acounits , based on results from MAGMA \citep{HUGHES10,WONG11} and consistent with \citet{LEROY11} and \citet{JAMESON15}.

For M31, we adopt a Galactic $\alpha_{\rm CO}^{1-0} = 4.35$~\acounits , this is consistent with the dust modeling results of \citet{LEROY11}, but given the difficulty of modeling the {\em Spitzer} bands and M31's outlying behavior in several plots that we present in Section{~\ref{sec:results}, we underline CO-to-H$_2$ conversion factor as a possible uncertainty for M31.

For M33, we take $\alpha_{\rm CO}^{2-1} = 10.8$~\acounits , which reflects the average CO (2-1) / (1-0) ratio, $\approx 0.8$, measured by \citet{DRUARD14} and a CO~(1-0)-to-H$_2$ conversion factor twice the Milky Way value. We adopt this value based on the recommendation of \citet{DRUARD14}, but note some uncertainty given the apparent contradiction with \citet{ROSOLOWSKY03}.

For M51, we adopt a Galactic $\alpha_{\rm CO} = 4.35$~\acounits . Groves et al. (in prep.) show that several independent approaches yield results consistent with this value; although note that there is evidence to support lower values (e.g., Schirm et al. submitted).

For M74, we adopt $\alpha_{\rm CO}^{2-1} = 7.9$~\acounits , which combines a CO (2-1)/(1-0) ratio of ${\sim}0.55$ with a Galactic CO (1-0)-to-H$_2$ conversion factor. The line ratio measurement comes from two pointings by \citet{USERO15} and is consistent with the integrated flux ratio of $0.62$ between the HERACLES \citep{LEROY09} and BIMA SONG \citep[][]{HELFER03} data cubes for this galaxy.

\section{Beamwise Measurements and Cloud Properties}
\label{sec:compare}

\begin{deluxetable}{lcc}[t!]
\tabletypesize{\scriptsize}
\tablecaption{Comparison of Our Approach to GMC Properties \label{tab:props_vs_pix}}
\tablewidth{0.85\columnwidth}
\tablehead{
\colhead{Ratio} & 
\colhead{Median} &
\colhead{Scatter} \\
\colhead{} &
\colhead{} &
\colhead{(dex)} 
}
\startdata
GMC Radius to Beam & & \\
$\ldots$ CPROPS & 1.3 & 0.10 \\
$\ldots$ CLUMPFIND & 1.7 & 0.16 \\[1ex]
GMC $I$ to Beam-wise $I_{\rm 60pc}$ & & \\
$\ldots$ CPROPS & 0.8 & 0.19 \\
$\ldots$ CLUMPFIND & 0.7 & 0.22 \\[1ex]
GMC $\sigma$ to Beam-wise $\sigma_{\rm 60pc}$ & & \\
$\ldots$ CPROPS & 0.8 & 0.09 \\
$\ldots$ CLUMPFIND & 1.2 & 0.07 \\[1ex]
GMC $\alpha_{\rm vir}$ to Beam-wise $B^{-1}_{\rm 60pc}$ & & \\
$\ldots$ CPROPS & 1.2 & 0.22 \\
$\ldots$ CLUMPFIND & 2.7 & 0.30
\enddata
\tablecomments{Ratios taken across all data sets. All resolutions used for the radius-to-beam ratio. In other cases, we use a $60$~pc measurement scale. Ratios reflect luminosity weighted mean cloud property and intensity-weighted, stacked beamwise measurement over a 500~pc averaging beam.}
\end{deluxetable}

\begin{figure*}[t]
\plottwo{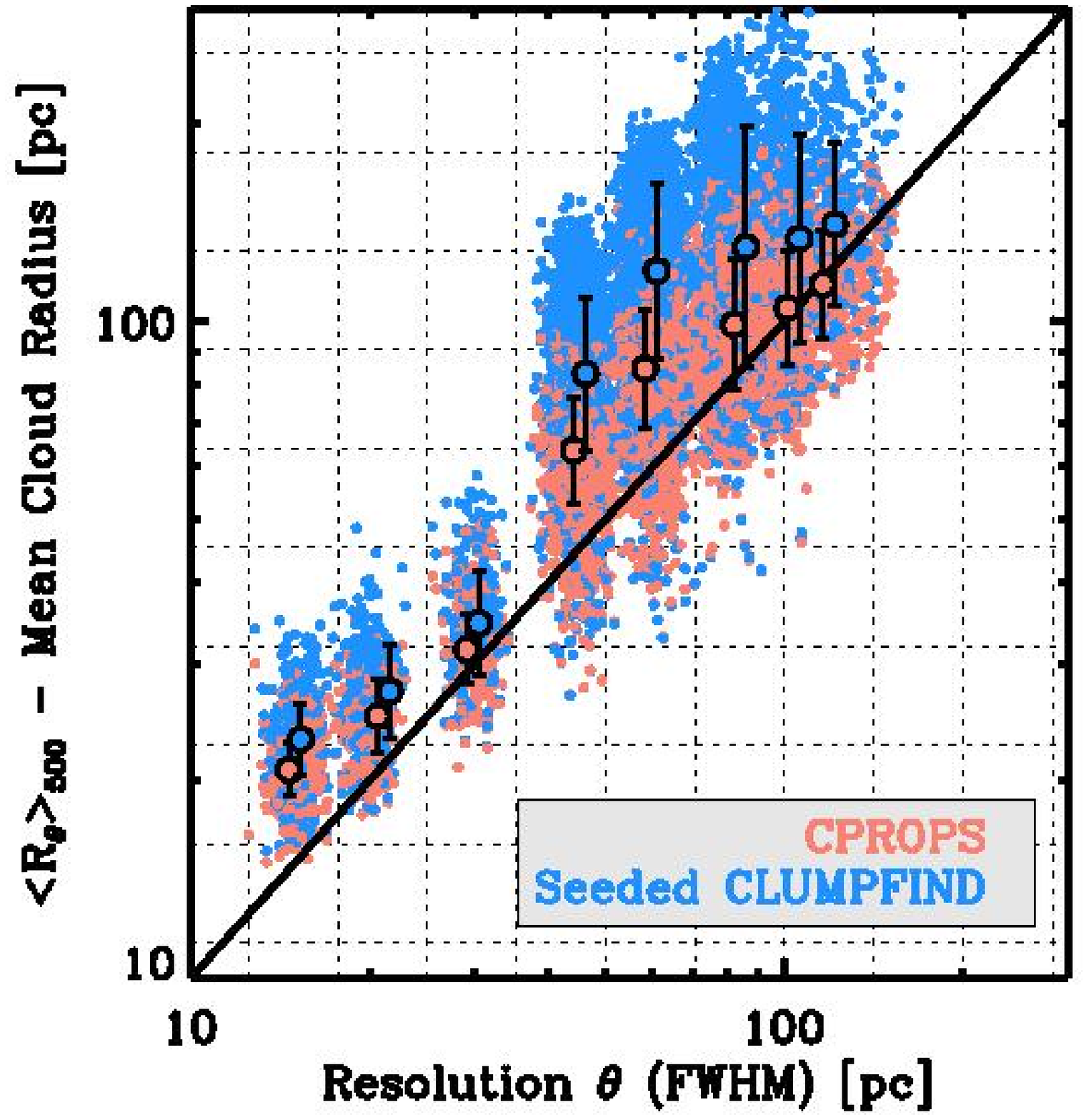}{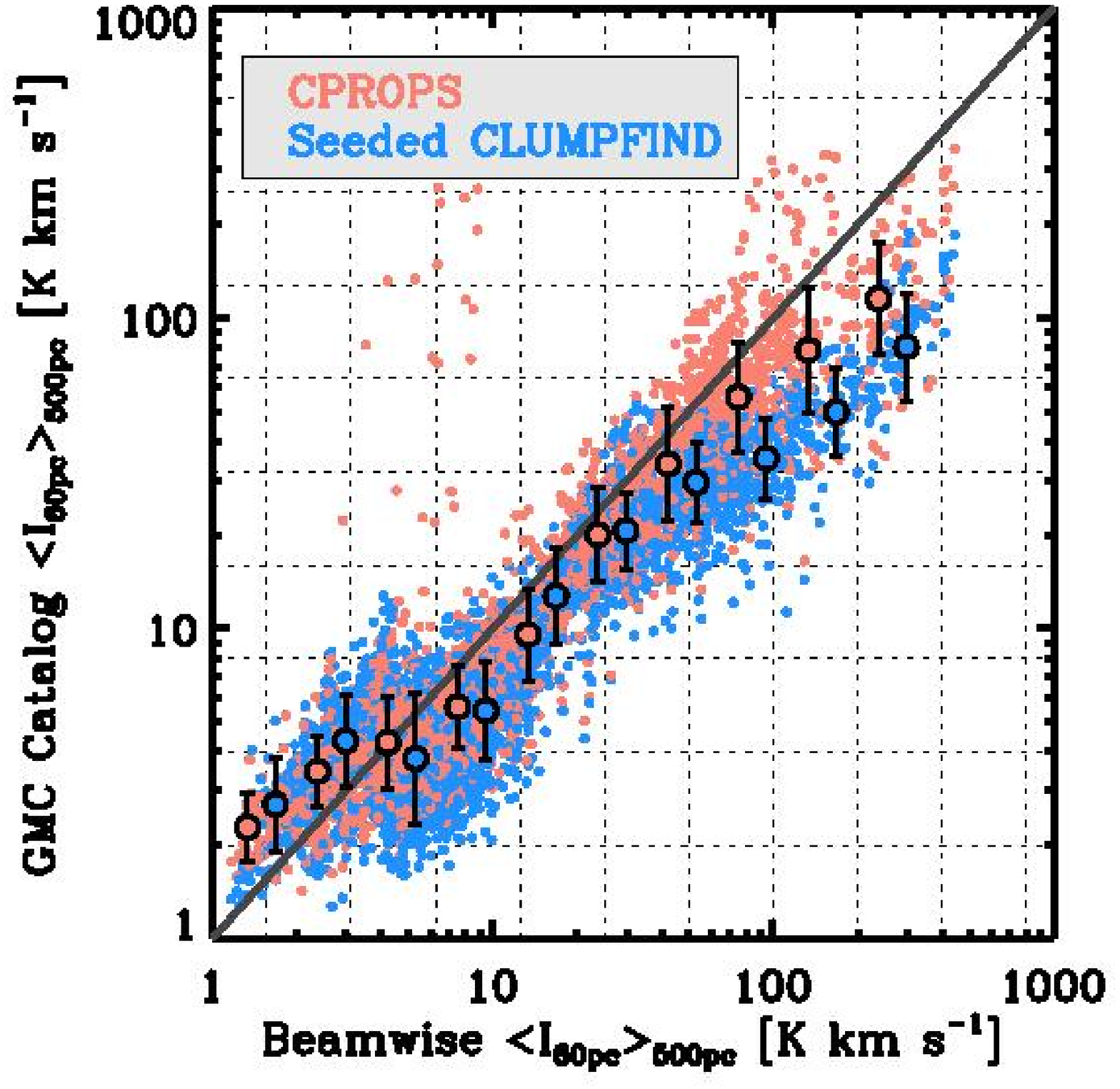}
\plottwo{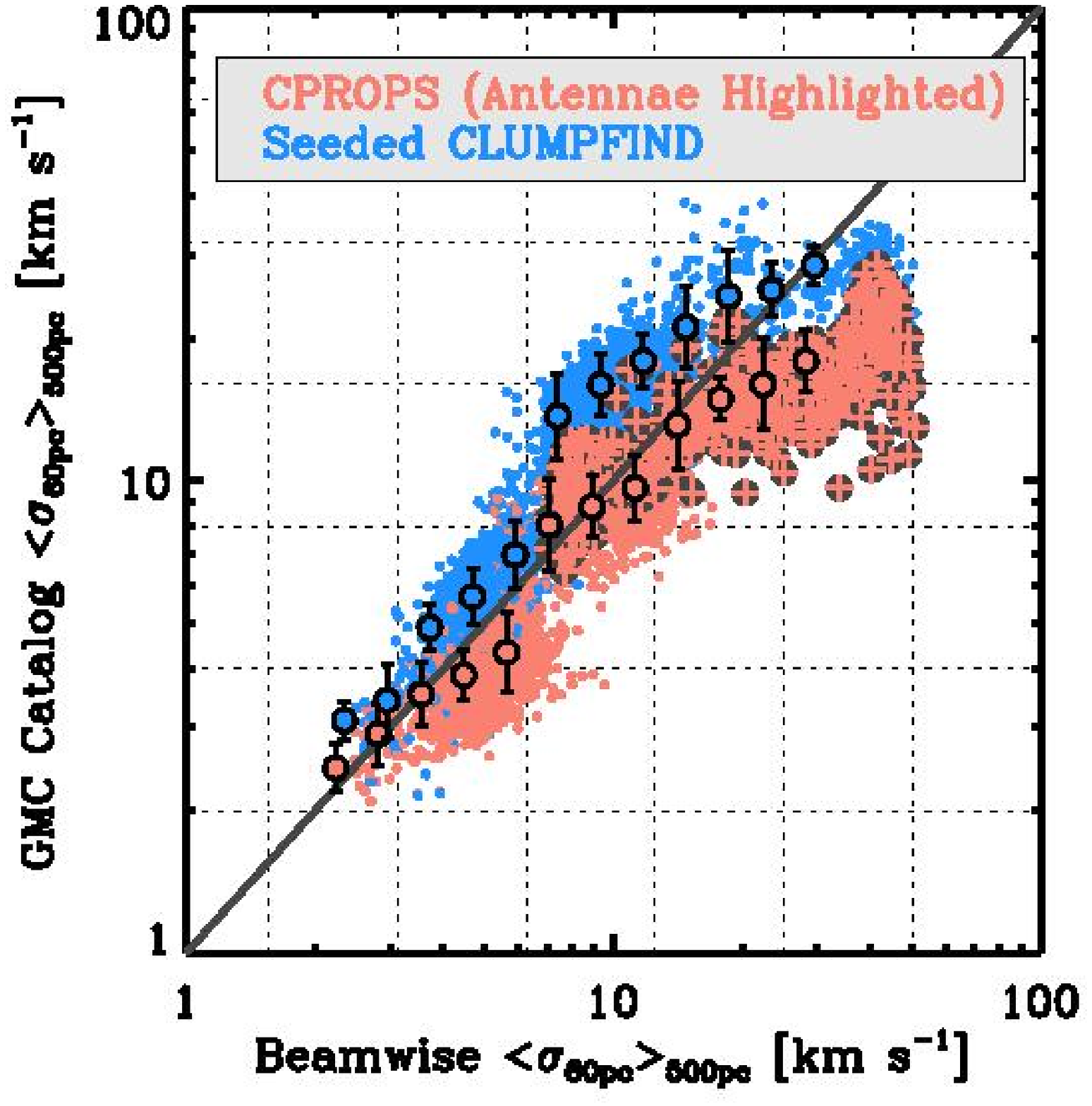}{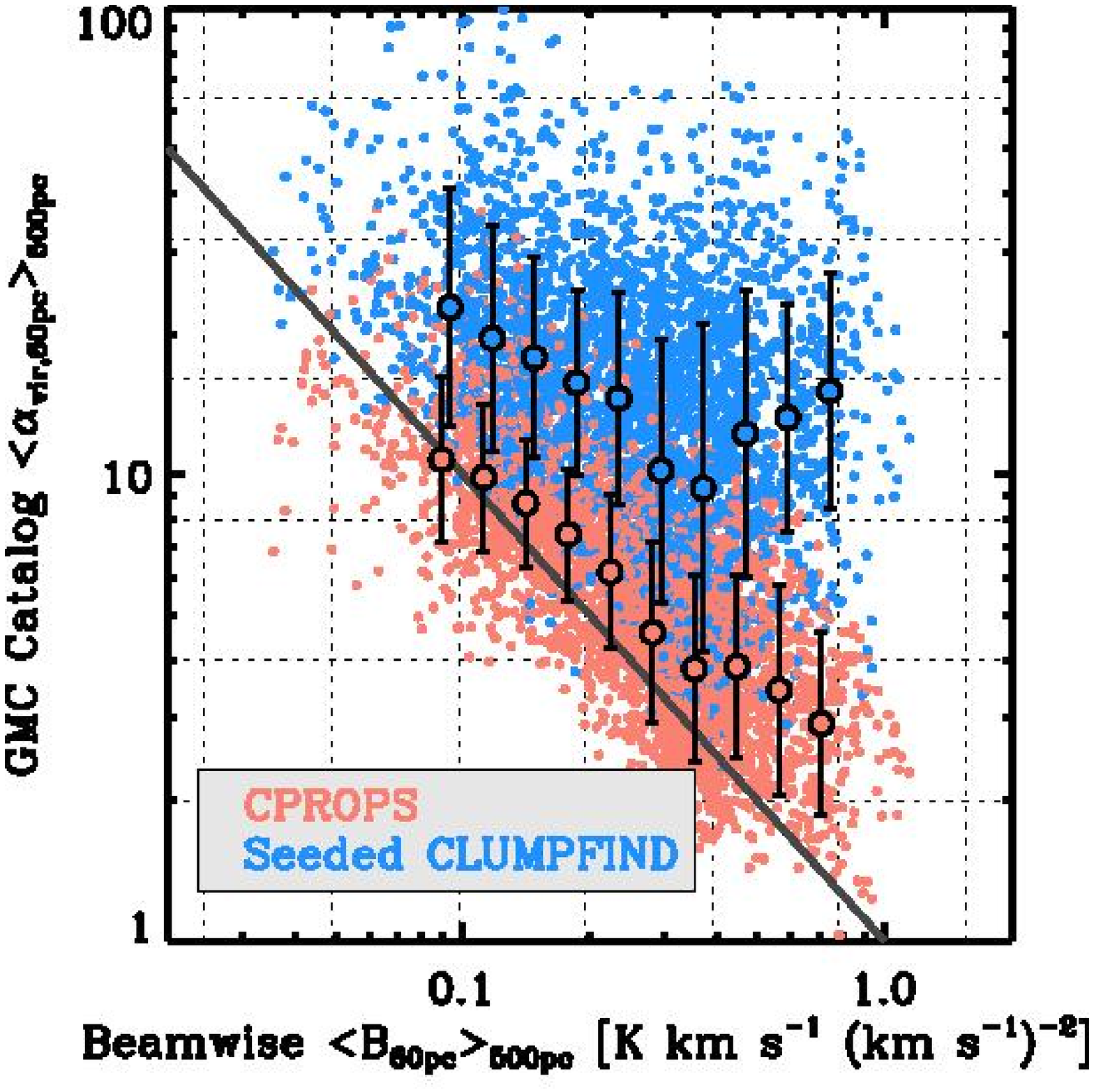}
\caption{({\em top left}) GMC radius as a function of the physical resolution (FWHM beam size) used to carry out the measurement. {\em GMCs tend to be marginally resolved, beam-scale objects}, supporting our adoption of the beam as the relevant physical scale. ({\em remaining panels}) GMC properties measured at $\theta = 60$~pc as a function of analogous beamwise quantities, also measured at $\theta = 60$~pc. For both axes, we use an averaging scale of $500$~pc (FWHM). Each panel shows all data sets using two segmentation methodologies: CPROPS (red points) and seeded CLUMPFIND (blue points). Circles with error bars show $y$ binned by $x$ with the $1\sigma$ logarithmic scatter, and so indicate the mean relationship.  The {\em top right} panel shows integrated intensity. The {\em bottom left} panel shows line width, with clouds in the overlap region of the Antennae galaxies marked by gray circles. This region shows complex, often multi-component line profiles, leading to the discrepancy between the cloud approach and our simple characterization of the line-of-sight line width. The {\em bottom right} panel shows the virial parameter, $\alpha_{\rm vir} \propto \mathit{KE} / \mathit{UE}$, as a function of our beamwise approach to ``boundedness,'' $B \equiv I / \sigma^2 \propto \mathit{UE} / \mathit{KE}$. The two correlate in the expected way, with $B \propto \alpha_{\rm vir}^{-1}$, though with substantial scatter due to cloud radius variations, particularly in CLUMPFIND.}
\label{fig:props_vs_pix}
\end{figure*}

First, we compare our $\theta = 60$~pc measurements, averaged over $500$~pc scales, to results from GMC property estimates. Figure \ref{fig:props_vs_pix} and Table \ref{tab:props_vs_pix} report the ratio of ensemble average measurements based on cloud catalogs ($y$-axis) to those calculated from our beamwise approach ($x$-axis). We highlight three important results of this comparison. First, cloud cataloging techniques always yield marginally resolved objects, meaning that our neglect of cloud sizes in the beamwise treatment in fact loses less information than one might expect. Second, the average integrated intensities, $I$, and line widths, $\sigma$, from our beamwise approach track those from cloud cataloging algorithms well. Finally, our observational ``boundedness'' parameter $B$ correlates with the estimated virial parameter, $\alpha_{\rm vir}$, in the expected way. We further derive a ratio relating $B$ to $\alpha_{\rm vir}$ and compare this to the simple geometrical estimate from Section \ref{sec:physcond}.

\subsection{Clouds are Always Marginally Resolved}

Our method takes the measurement beam as the characteristic size scale. Cloud property treatments measure the size of individual objects extracted from the data cube. Both approaches have merit and one can justify our approach simply by the desire to measure average gas properties at a fixed spatial scale. 

Based on Figure \ref{fig:props_vs_pix}, we make a stronger argument that our approach is preferable to a cloud-based treatment. The top left panel plots the average, luminosity-weighted cloud radius measured using CPROPS and CLUMPFIND as a function of the native resolution of the data. Here, we have repeated measurements for our six targets as we degrade the resolution from its native value to $100$~pc. Table \ref{tab:props_vs_pix} reports the median ratio and scatter. Clouds in our analysis have radius $1.3$ ($1.7$) times the beam size using CPROPS (CLUMPFIND), with a comparatively small scatter of $0.1$ ($0.16$) dex.

The figure shows that cloud property measurements tend to recover beam-sized objects from the data. There is some scatter, which does contain physical information, but overall it is reasonable to characterize objects in extragalactic GMC studies as marginally resolved, beam-sized objects. For a statistical characterization of a region, Figure \ref{fig:props_vs_pix} suggests that little information is lost -- and a large amount of simplicity and robustness is gained -- by using our beamwise approach.

Figure \ref{fig:props_vs_pix} reinforces and extends the point made by \citet{HUGHES13B}. They showed that the line width-size relation, one of ``Larson's Laws,'' is almost impossible to extract from cloud property measurements for a single galaxy or a set of galaxies at a common resolution. Figure \ref{fig:props_vs_pix} reveals an easy explanation: the dynamic range in size scales that are accessed by applying cloud property measurements to a fixed physical resolution is small. A much stronger relationship between line width and size emerges when combining studies with different spatial resolution \citep[e.g., see][]{BOLATTO08,FUKUI10,LEROY15A}. However, as pointed out by \citet{HUGHES13B}, leveraging different galaxies to obtain dynamic range makes it harder to measure differences among the ISM properties in different galaxies.

When a data set has intrinsically high resolution, one can always degrade the resolution. Multi-scale analysis of an individual data set would therefore seem a promising way to access the line width-size relation in a single target. This has been applied in the Milky Way, first by \citet{HEYER04} and then several times using the dendrogram approach of \citet{ROSOLOWSKY08} \citep[e.g.,][]{SHETTY12B}. With the latest generation of high resolution CO nearby galaxy surveys, multi-scale structural analysis is also a viable technique for extragalactic applications (e.g., Schruba et al., in prep.; Colombo et al., in prep.). An alternative approach using our framework is to simply re-characterize the beam-wise statistics of the image for progressively larger measurement beams. We defer such an investigation to a future paper.

\subsection{Beamwise vs.\ Cloud-Based Integrated Intensity and Line Width}

The top right and bottom left panel of Figure \ref{fig:props_vs_pix} plot average integrated intensity and line width from luminosity-weighted, gridded GMC catalogs as a function of measurements over the same area using our beamwise approach. Here, we use all six targets and a uniform measurement scale of $\theta = 60$~pc. 

The average integrated intensity and line width estimated from cloud catalogs correlate well with our beamwise measurements. The median ratios, reported in Table \ref{tab:props_vs_pix}, are not exactly unity. Particularly for the case of the integrated intensity, this should not be surprising. The integrated intensity for clouds depends on the cloud radius, which is defined assuming some fiducial geometry \citep[see][]{SOLOMON87,ROSOLOWSKY06}. The rms scatter between the different approaches is ${\sim}50\%$ for the integrated intensity and smaller, ${\sim}30\%$, for the line width.

The largest discrepancy between the two methods appears at high line width. The bottom left panel in Figure \ref{fig:props_vs_pix} shows a population of points with high beam-wise line width compared to their GMC line widths. These come from the Antennae galaxies and reflect lines-of-sight in the ``overlap region'' where the line profiles show multiple components. This is not an intrinsic failure of our beamwise approach; it would be possible to refine the calculation of the line width to deal with multiple components. It does, however, highlight that the line-of-sight line width may not map trivially to a turbulent velocity dispersion, especially in complex regions. Projection effects and local dynamics at or below the scale of the beam may likewise affect the correspondence between the line width and the turbulent velocity dispersion. The overlap region in the Antennae is the site of an ongoing collision between two galaxies; but similar issues may appear in subtler ways in galactic spiral arms with strong streaming motions \citep[e.g.,][]{MEIDT13}, or in gas flows influenced by bars \citep[e.g.,][]{SORAI00}. We defer a more detailed treatment of complex line profiles to future work. Importantly for the discussion below, we note that some of the line width that our method estimates for the Antennae galaxies arises from combining multiple velocity components. However, we emphasize that a cloud-property treatment, which does segment the different velocity components into separate clouds, also finds that GMCs in the Antennae overlap region have the highest average line width among our galaxy sample \citep[e.g.,][]{WEI12}.

\subsection{The Virial Parameter and $B \equiv I/\sigma^2$}
\label{sec:alpha_cld_vs_b}

The bottom right panel of Figure \ref{fig:props_vs_pix} plots the mean virial parameter (Equation \ref{eq:boundedness}) as a function of our $B$ parameter, defined as $B \equiv I/\sigma^2$. Following Section \ref{sec:physcond}, we expect these two quantities to anti-correlate, with $\alpha_{\rm vir} \propto \mathit{KE}/\mathit{UE}$ and $B \propto \mathit{UE}/\mathit{KE}$. As expected, we see this behavior in the figure. 

Because the physical configurations corresponding to $\mathit{UE} = \mathit{KE}$ (bound; $\alpha_{\rm vir} \approx 2$) and $\mathit{UE} = 2 \mathit{KE}$ (virialized; $\alpha_{\rm vir} \approx 1$) are of special interest, the numerical factor that relates $B$ to $\alpha_{\rm vir}$ is also important. In Section \ref{sec:physcond}, we note the prefactor expected for a Galactic CO-to-H$_2$ conversion factor and a uniform density sphere of radius $30$~pc.  Table \ref{tab:props_vs_pix} reports the ratio calculated from comparing cloud catalog estimates of $\alpha_{\rm vir}$ to $B$. This result differs depending on the GMC segmentation methodology. We prefer the CPROPS value and will take $\alpha_{\rm vir, 60pc} \approx 1.3 B_{\rm 60pc}^{-1}$, with $B_{\rm 60pc}$ in units of \mbox{K~km~s$^{-1}$} \mbox{(km~s$^{-1}$)$^{-2}$}, which resembles the theoretically expected value if clouds are modestly extended relative to the beam. The larger value from CLUMPFIND reflects the higher cloud radii found by that algorithm (see top left panel of Figure \ref{fig:props_vs_pix} and Table \ref{tab:props_vs_pix}). 

For this ratio of $B$ to $\alpha_{\rm vir}$, and if $\alpha_{\rm CO} = 4.35$ \acounits , then $B_{\rm 60pc} \approx 0.6$ \mbox{K~km~s$^{-1}$} \mbox{(km~s$^{-1}$)$^{-2}$} corresponds to marginally bound material and $B \approx 1.2$ \mbox{K~km~s$^{-1}$} \mbox{(km~s$^{-1}$)$^{-2}$} describes virialized material. These values provide a framework to interpret $B$ in an absolute sense, though we note that this is an area for future improvement in our methodology. Because it affects the estimated mass, the adopted conversion factor enters the relationship between $B$ and $\alpha_{\rm vir}^{-1}$. Given some value of $B$, a higher $\alpha_{\rm CO}$ implies more mass, and so more gravitationally bound material, and thus a higher $\alpha_{\rm vir}^{-1}$ and a lower $\alpha_{\rm vir}$. The translation is linear.

\section{Results}
\label{sec:results}

\begin{deluxetable*}{lcccccc}[t!]
\tabletypesize{\scriptsize}
\tablecaption{Properties of CO Structures at $60$~pc Scale for Six Galaxies \label{tab:results}}
\tablewidth{0pt}
\tablehead{
\colhead{Property} & 
\colhead{Antennae} &
\colhead{LMC} &
\colhead{M31} &
\colhead{M33} &
\colhead{M51} &
\colhead{M74}
}
\startdata
$\log_{10} I$ &  1.56 &  0.34 &  0.45 &  0.27 &  1.56 &  0.82 \\
(K~km~s$^{-1}$) & $(  1.27$ to $ 2.28)$ & $(  0.19$ to $ 0.54)$ & $(  0.29$ to $ 0.63)$ & $(  0.07$ to $ 0.48)$ & $(  1.29$ to $ 1.85)$ & $(  0.65$ to $ 1.00)$ \\
\\ $\log_{10} \Sigma$ &  2.81 &  1.27 &  1.09 &  1.30 &  2.20 &  1.71 \\
(M$_\odot$~pc$^{-2}$) & $(  2.51$ to $ 3.54)$ & $(  1.13$ to $ 1.48)$ & $(  0.93$ to $ 1.26)$ & $(  1.09$ to $ 1.51)$ & $(  1.92$ to $ 2.47)$ & $(  1.55$ to $ 1.90)$ \\
\\ $\log_{10} \sigma$ &  1.25 &  0.54 &  0.67 &  0.62 &  0.97 &  0.67 \\
(km~s$^{-1}$) & $(  0.91$ to $ 1.55)$ & $(  0.39$ to $ 0.65)$ & $(  0.59$ to $ 0.80)$ & $(  0.54$ to $ 0.71)$ & $(  0.86$ to $ 1.08)$ & $(  0.60$ to $ 0.74)$ \\
\\ $\log_{10} B$ & -0.85 & -0.74 & -0.94 & -0.99 & -0.39 & -0.53 \\
$\left( \frac{{\rm K~km~s}^{-1}}{({\rm km~s}^{-1})^2} \right)$ & $( -1.04$ to $-0.58)$ & $( -0.90$ to $-0.55)$ & $( -1.11$ to $-0.78)$ & $( -1.17$ to $-0.82)$ & $( -0.59$ to $-0.21)$ & $( -0.64$ to $-0.42)$ \\
\\ $\log_{10} \alpha_{\rm vir}^{-1}$ & -0.17 & -0.35 & -0.86 & -0.51 & -0.31 & -0.19 \\
 & $( -0.36$ to $ 0.10)$ & $( -0.53$ to $-0.17)$ & $( -1.03$ to $-0.71)$ & $( -0.70$ to $-0.35)$ & $( -0.50$ to $-0.13)$ & $( -0.30$ to $-0.08)$ \\
\\ $\Delta^{\rm 84-16}$ &  0.70 &  0.58 &  0.60 &  0.65 &  0.60 &  0.60 \\
(dex) & $(  0.56$ to $ 1.16)$ & $(  0.50$ to $ 0.70)$ & $(  0.52$ to $ 0.72)$ & $(  0.54$ to $ 0.78)$ & $(  0.48$ to $ 1.12)$ & $(  0.51$ to $ 0.73)$
\enddata
\tablecomments{Results using an averaging scale of 500~pc and a measurement scale of 60~pc. First line shows median of all regions with indicated weighting. Second line (in parentheses), shows the 16$^{\rm th}$ to 84$^{\rm th}$ percentile range. See Figures \ref{fig:props_hist}, \ref{fig:b_hist}, \ref{fig:mass_hist}, and \ref{fig:delta_hist}.}
\end{deluxetable*}

We apply our calculations to six galaxies with high resolution, high sensitivity mapping. The Antennae galaxies are the nearest major merger. The ALMA maps of \citet{WHITMORE14} cover the region where the two galaxies collide, including the ``super-giant molecular clouds'' (SGMCs) identified by \citet{WILSON03} at lower resolution \citep[see also][]{WEI12,JOHNSON15}. The Large Magellanic Cloud (LMC) and M33 are both Local Group dwarf spiral galaxies, which have weaker stellar potential wells than large spirals and where the molecular gas exists within a dominant atomic gas reservoir. Because of their proximity, these have been the targets of GMC studies for more than two decades \citep[e.g.,][]{WILSON90,FUKUI99,ROSOLOWSKY03,HUGHES10}. M31 is a massive early-type spiral also in the Local Group. It has fairly anemic star formation \citep[e.g., see][]{LEWIS15} and a large stellar bulge. The CARMA map by Schruba et al. targets  mostly the $10$~kpc molecular ring, which hosts most of the CO emission in the galaxy \citep{NIETEN06}; however the southwest field covers part of M31's bulge. M51 (NGC 5194) and M74 (NGC 628) are both grand design spiral galaxies within 10~Mpc distance. M51, the more massive of the two, is currently undergoing an interaction with its early type companion. It shows a higher surface density of gas in its disk, a higher molecular fraction, and more active star formation than M74.

In Section \ref{sec:data}, we applied our analysis to CO data at $\theta=60$~pc resolution for each of these targets. This yields a detailed view of the structure of molecular gas in galaxies on the scale where one resolution element corresponds to a massive GMC. Table \ref{tab:results} summarizes the results of these calculations for each target. The table reports intensity-weighted ensemble average using a measurement scale of $60$~pc and an averaging scale of $500$~pc. The table quotes the median over all 500~pc regions in each galaxy. Below this, in parentheses, we give the $16^{\rm th}$ to $84^{\rm th}$ percentile range, corresponding to $\pm 1\sigma$ for a normal distribution. Thus, this table gives our measurement of the typical properties of the molecular ISM at a resolution of 60~pc across a diverse sample of local galaxies.

Table \ref{tab:results} reports the median and range treating each 500~pc averaging beam equally. Some beams contain more flux than others. If we instead take the median value and $\pm 1\sigma$ weighting each measurement by flux, the values for most of our targets increase by ${\sim}0.1$~dex on average. That is, more flux comes from high line width, high intensity regions, but the effect is mild in most of our targets except for the Antennae. 

In the Antennae, the SGMCs identified by \citet{WILSON03} contribute a large fraction of the flux but subtend only a modest area. As Table \ref{tab:weighting} shows, the ensemble properties of these few bright regions are more extreme than those over the rest of the galaxy.  A far weaker version of this effect in the LMC distinguishes the bright molecular ridge south of 30 Doradus from the rest of that system. The effect could also be seen using our formalism by setting the averaging scale to the whole galaxy, which we would expect to approach the ``flux weighting'' case in Table \ref{tab:weighting}.

Figures \ref{fig:props_hist}, \ref{fig:delta_hist},  \ref{fig:mass_hist}, and \ref{fig:b_hist} visualize the results in Table \ref{tab:results}. These present the distributions of properties -- ${\langle I_{\rm 60pc}\rangle}_{\rm 500pc}$, ${\langle \sigma_{\rm 60pc} \rangle}_{\rm 500pc}$, ${\langle B_{\rm 60pc} \rangle}_{\rm 500pc}$, ${\langle \Sigma_{\rm 60pc} \rangle}_{\rm 500pc}$, ${\langle \alpha_{\rm vir,60pc}^{-1} \rangle}_{\rm 500pc}$, and ${\langle \Delta^{84-16}_{\rm 60pc} \rangle}_{\rm 500pc}$ -- as ``violin plots''. These are normalized histograms, where the $x$-width of each shape indicates the fraction of data at the value on the $y$-axis. In these plots, we indicate the $50^{\rm th}$ percentile value with a black dot and the $\pm 1\sigma$ range (again from the $16^{\rm th}$ to $84^{\rm th}$ percentile) as a white cavity inside the full distribution (shown in color).

\begin{deluxetable}{lcc}[t!]
\tabletypesize{\scriptsize}
\tablecaption{Flux vs.\ Equal Weighting for Antennae \label{tab:weighting}}
\tablewidth{0pt}
\tablehead{
\colhead{Property} & 
\colhead{Equal Weighting} &
\colhead{Flux Weighting}
}
\startdata

$\log_{10} I$ (K~km~s$^{-1}$) & $1.56$ & $2.33$ \\
$\log_{10} \Sigma$ (M$_\odot$~pc$^{-2}$) & $2.81$ & $3.58$ \\
$\log_{10} \sigma$ (km~s$^{-1}$) & $1.25$ & $1.55$ \\
$\log_{10} B$ $\left( \frac{{\rm K~km~s}^{-1}}{({\rm km~s}^{-1})^2} \right)$ & $-0.85$ & $-0.78$ \\
$\log_{10} \alpha_{\rm vir}^{-1}$ & $-0.17$ & $-0.10$ \\
$\Delta^{\rm 84-16}$ & $0.70$ & $0.95$
\enddata
\tablecomments{As for Table \ref{tab:results} but for two approaches to taking the median over all averaging beams in the Antennae. ``Equal Weighting'' treats each 500~pc averaging beam equally. ``Flux Weighting'' takes the median weighted by flux, so that 50\% of the flux comes from above or below the reported value. The significant difference between the two weightings indicates that a large amount of the flux in the Antennae emerges from the SGMCs at the interface of the two galaxies.}
\end{deluxetable}

\subsection{Variations Within and Among Galaxies}
\label{sec:vary}

\begin{figure*}
\plottwo{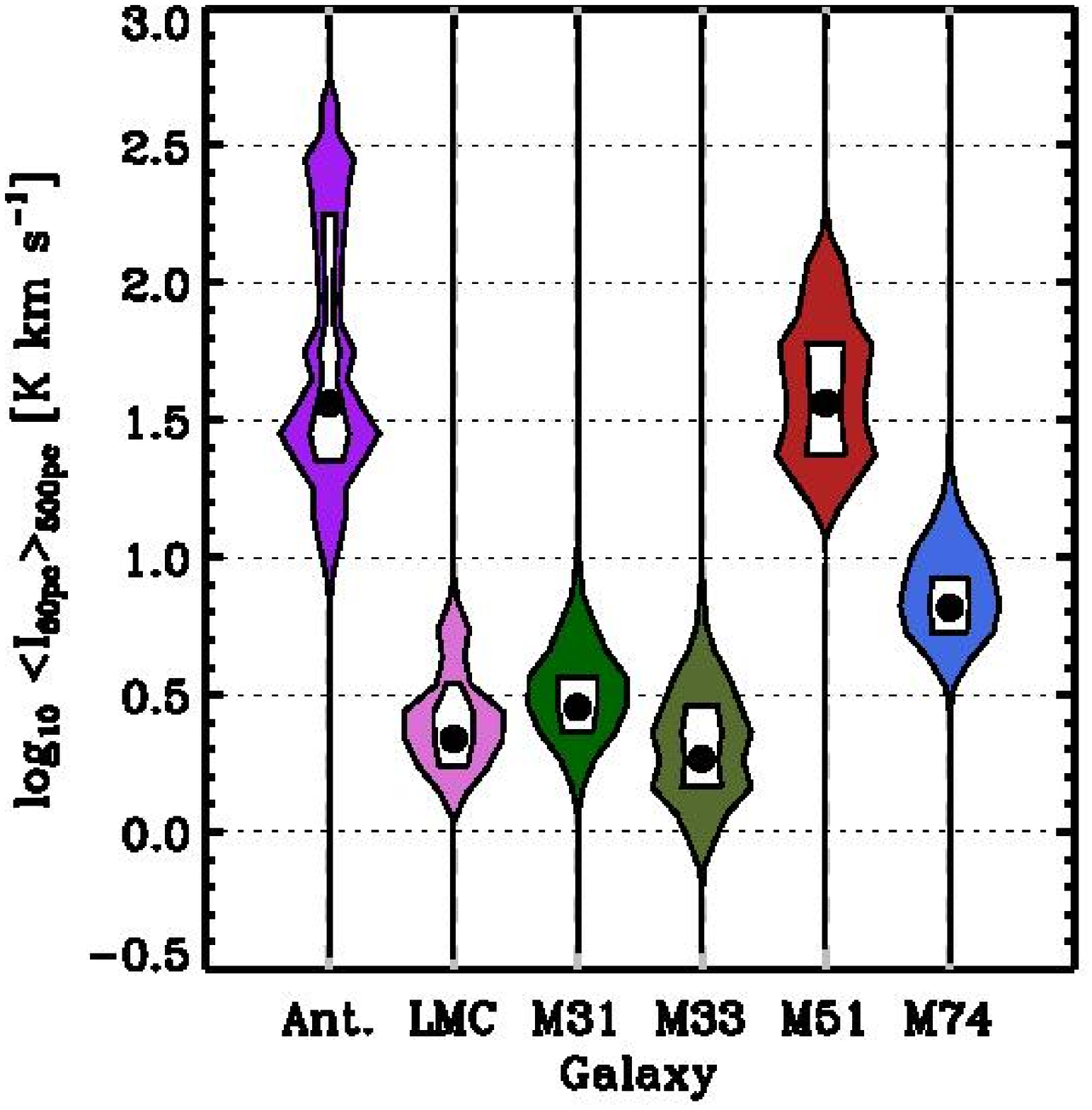}{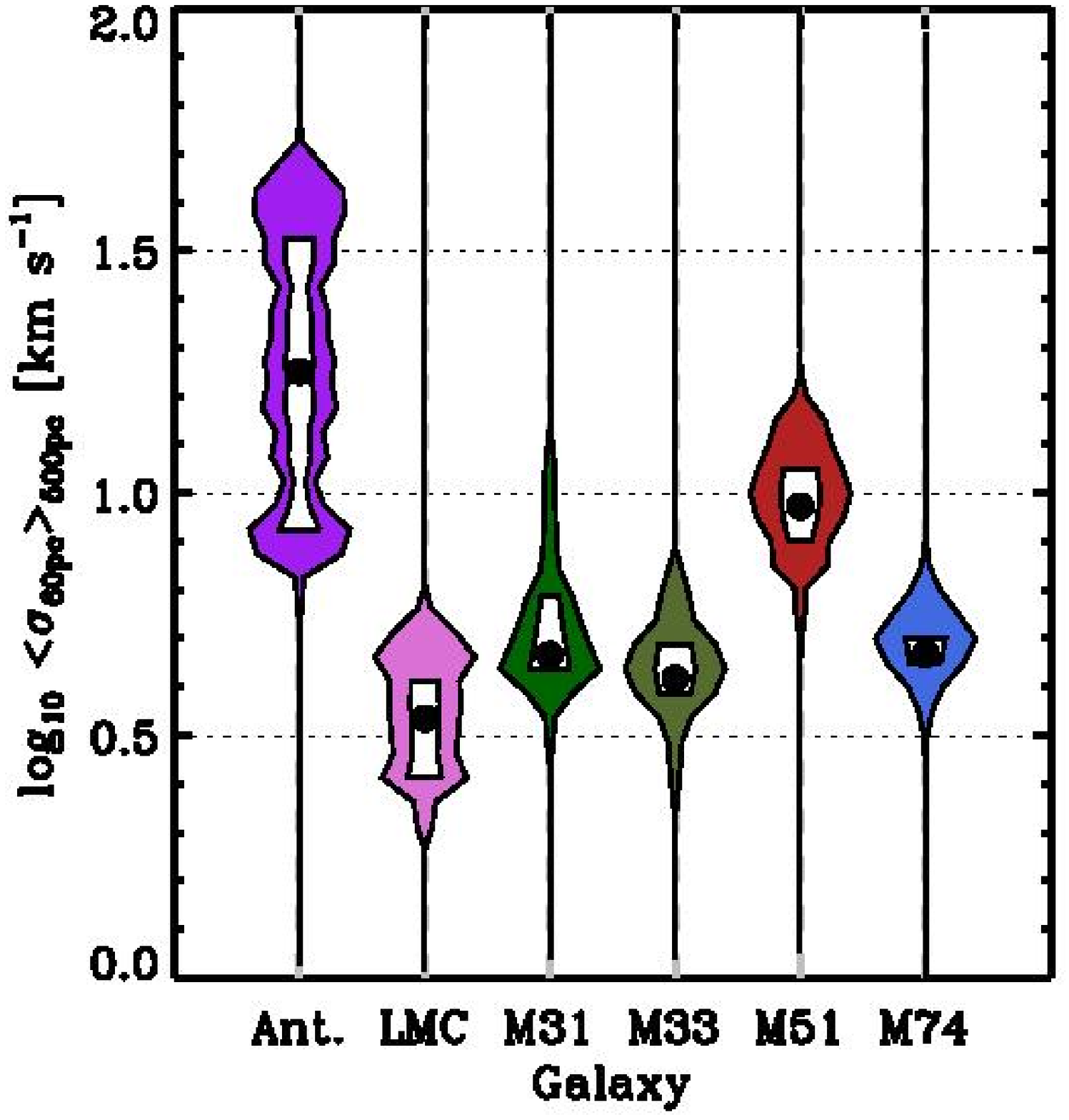}
\caption{Distribution of average $\theta=60$~pc resolution properties over $A=500$~pc regions in our six targets. ({\em left}) Integrated intensity, ${\langle I_{\rm 60pc} \rangle}_{\rm 500pc}$. The data span two decades in CO surface brightness at fixed resolution with variation both among and within galaxies. ({\em right}) Line width, ${\langle \sigma_{\rm 60pc} \rangle}_{\rm 500pc}$. Line width also varies within and among galaxies. The galaxies follow a similar order in the two plots, with the quiescent LMC showing low surface brightness and narrow lines, followed by M33, M31, and M74. In both plots, the Antennae galaxies show a wide range of conditions, overlapping M51 in many areas but showing regions of extreme surface density and line width in the SGMCs at the interface of the two galaxies.}
\label{fig:props_hist}
\end{figure*}

Figure~\ref{fig:props_hist} shows the distributions of integrated intensity, $I$, and line width, $\sigma$, for our sample (see also Figure~\ref{fig:histograms}). Variations are immediately apparent within our sample. 

To first order, the three Local Group galaxies -- LMC, M31, M33 -- appear similar to one another and distinct from the three more active, more distant systems -- the Antennae, M51, and M74. Because of their proximity, the Local Group galaxies served as the initial targets for GMC property studies; these represent the only galaxies where early mm-wave telescopes could resolve and detect GMCs \citep[see review in][]{FUKUI10}. The similarity of molecular gas structures in these systems, visible in the matched distributions of line width and integrated intensity, helped fuel the idea of an approximately universal population of GMCs \citep[see][]{BOLATTO08}.

Figure \ref{fig:props_hist} also makes clear that the galaxies of the Local Group offer a biased view of molecular gas properties in galaxies. The Antennae and M51 show markedly higher $I$ and $\sigma$ than the Local Group targets. M74 represents an intermediate case between the Local Group targets and M51. At $\theta = 60$~pc scales -- that is, at scales where a beam matches the size of an individual large GMC -- the sense of the variation is that the more gas rich, higher star formation rate systems show higher CO intensities. The differences in Figure \ref{fig:props_hist} became particularly apparent with PAWS, which \citet{HUGHES13A, HUGHES13B} used to demonstrate stark differences between M51, M33, and the LMC.

Galaxies also differ in their distributions of line widths, $\sigma$, plotted in the right panel of Figure \ref{fig:props_hist}. These vary from a few km~s$^{-1}$, pushing up against the resolution of the data in M33, to more than $30$~km~s$^{-1}$ in the bright regions of the Antennae. As mentioned above, some of the most extreme Antennae line widths can come from emission profiles with multiple components. Though the line width does reflect the dispersion of velocities along the line-of-sight, some of these values should not be interpreted as purely turbulent \citep[see][for an example using these same data]{JOHNSON15}. But, as the bottom left panel of Figure \ref{fig:props_vs_pix} shows, even the single component line widths (reflected in the GMC values) are $15{-}25$ \mbox{km~s$^{-1}$} in these regions.

The line width variations in Figure \ref{fig:props_hist} reinforce the existence of real physical differences in ISM properties among galaxies. The sense of the variations match those expected if internal pressure in the molecular gas tracks the hydrostatic pressure needed to support the ISM \citep{HUGHES13B}. Galaxies with higher mass and deeper potential wells require higher hydrostatic pressure to support the gas \citep[in the Antennae, the interaction further raises the pressure; e.g.,][]{RENAUD14}. The line width, $\sigma$, relates to the internal pressure in the molecular gas, $P \propto \rho \sigma^2$. The more massive M51 lies at higher values than the more quiescent, gas-poor systems. M31 and M74 lie at intermediate values, with the low mass dwarf spirals, M33 and the LMC, showing the smallest line widths. We return to this topic below in the discussions of boundedness (Section \ref{sec:cond}) and the line width-surface density relation (Section \ref{sec:sd_lw}). In an upcoming paper, we present a direct region-by-region correlation of the internal and hydrostatic pressure \citep[extending][]{HUGHES13B}.

Variations exist not only among galaxies but within galaxies. The tail of high $I$ values for the Antennae in Figure~\ref{fig:props_hist} reflects the high CO intensities in the overlap region. There, a few bright complexes (the SGMCs) dominate the light, but a large amount of more quiescent gas still extends across the overlap region \citep[see][]{WHITMORE14}. In the LMC, the extension to high $I$ is due to the molecular ridge on the eastern edge of the galaxy (south of 30 Doradus), which contains the brightest emission in the galaxy. The high $I$ regions in M51 are the inner parts of the spiral arms. Meanwhile, the high $I$ and high $\sigma$ tails in the M31 distribution come from the inner part of the galaxy where the stellar surface density is high. In short, though we plot one-dimensional distributions, the variations in Figure \ref{fig:props_hist} are real, physical, and directly map to distinct environments in the sample galaxies.

\subsection{Distributions}
\label{sec:distribution_results}

The distribution of cloud-scale ISM properties within an averaging beam, in addition to the average properties, is of physical interest. Following Section \ref{sec:distributions}, we calculate cumulative mass distributions for each averaging beam. As illustrated in Figure \ref{fig:dist_sketch}, we parameterize these distributions in a simple way, recording the integrated intensity at the $84^{\rm th}$, $50^{\rm th}$, and $16^{\rm th}$ percentile. 

Before inspecting the detailed results, we consider the distribution of integrated intensities at $\theta = 60$~pc resolution across the whole map for each of our targets. We show that the flux in our maps is distributed over a relatively narrow range of $I$, following a roughly, though not exactly, lognormal distribution in most targets. In a lognormal framework, $\Delta^{84-16}$ maps straightforwardly to the width of the distribution and ${\langle I_{\rm 60pc} \rangle}_{\rm 500pc}$ to the median, modulo a factor to convert mean to median. Bearing these results in mind, we then investigate how the distribution varies from point to point and galaxy to galaxy in our sample.

\subsection{Overall Distributions of Integrated Intensity}
\label{sec:distribution_methods}

\begin{figure*}[t]
\plotone{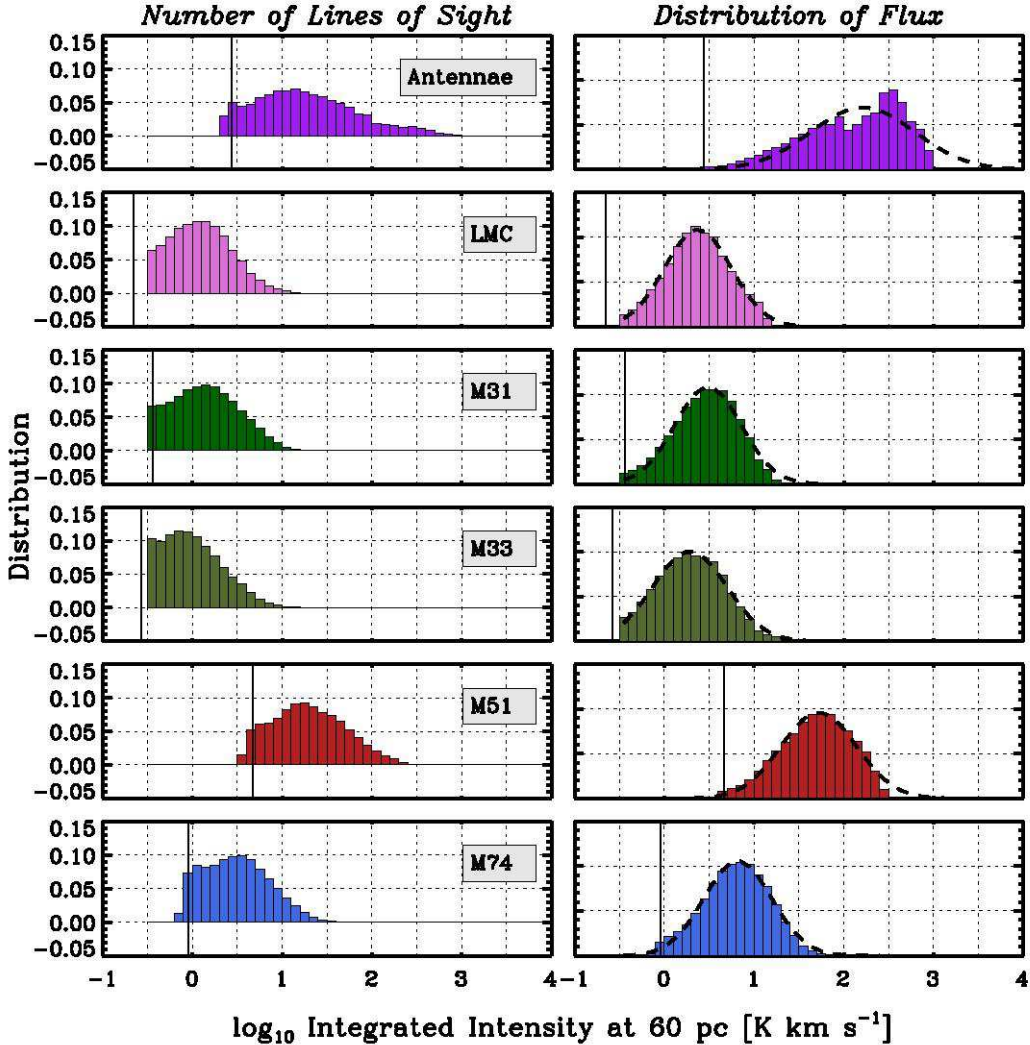}
\caption{Histograms showing ({\em left}) the distribution of integrated intensity, $I$, at $60$~pc resolution in each target and ({\em right}) the distribution of flux as a function of $I_{\rm 60pc}$ for the same data sets. The solid vertical lines show the 5$^{th}$ percentile of integrated intensities in the mask. Black dashed lines in the right column show Gaussian fits to the histograms. While many lines-of-sight exist near the threshold, most of the flux tends to lie well above our threshold, following an approximately lognormal distribution. The most notable exception is the Antennae, which shows a slightly bimodal flux distribution.}
\label{fig:histograms}
\end{figure*}

\begin{figure*}[t]
\plottwo{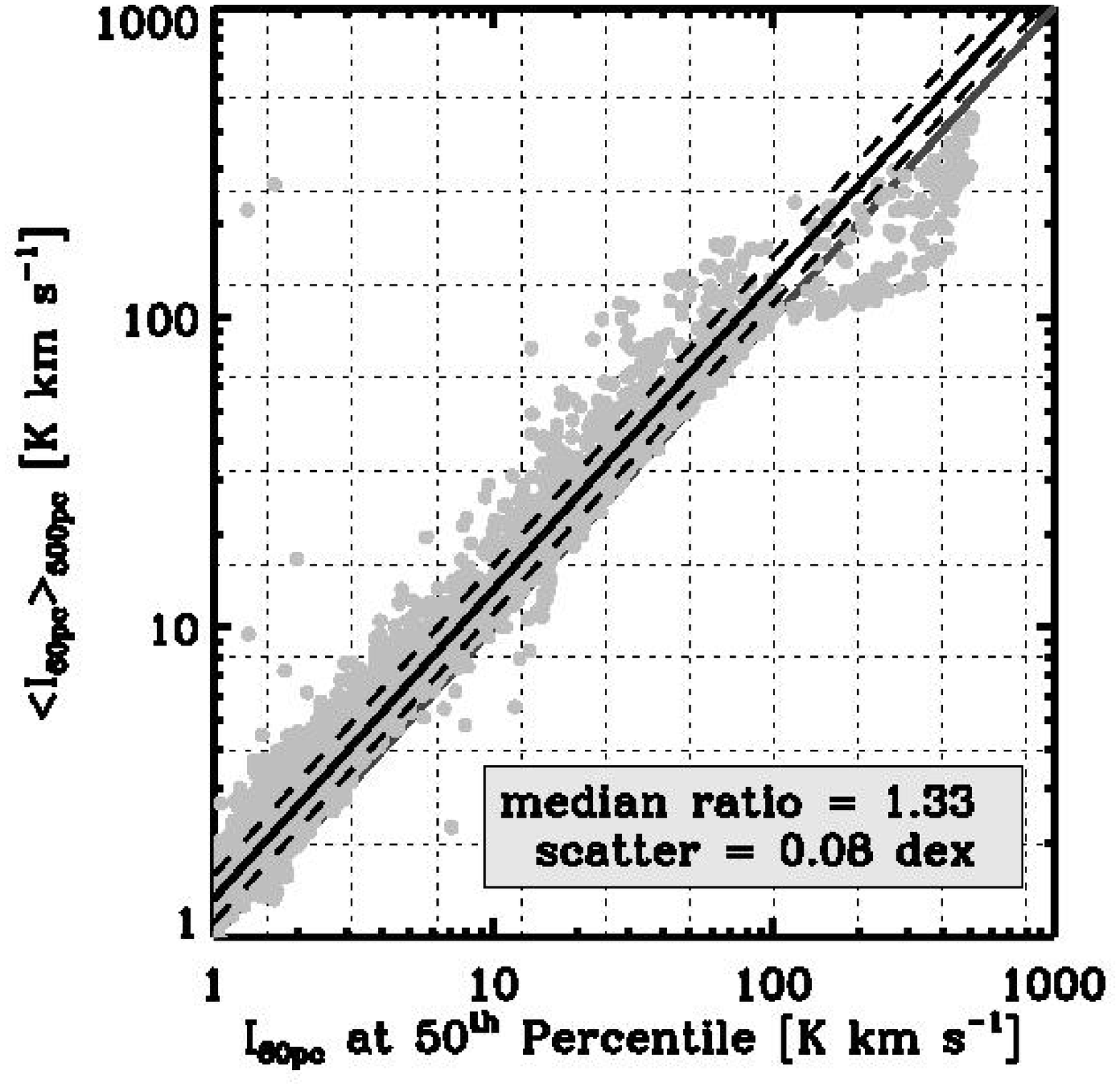}{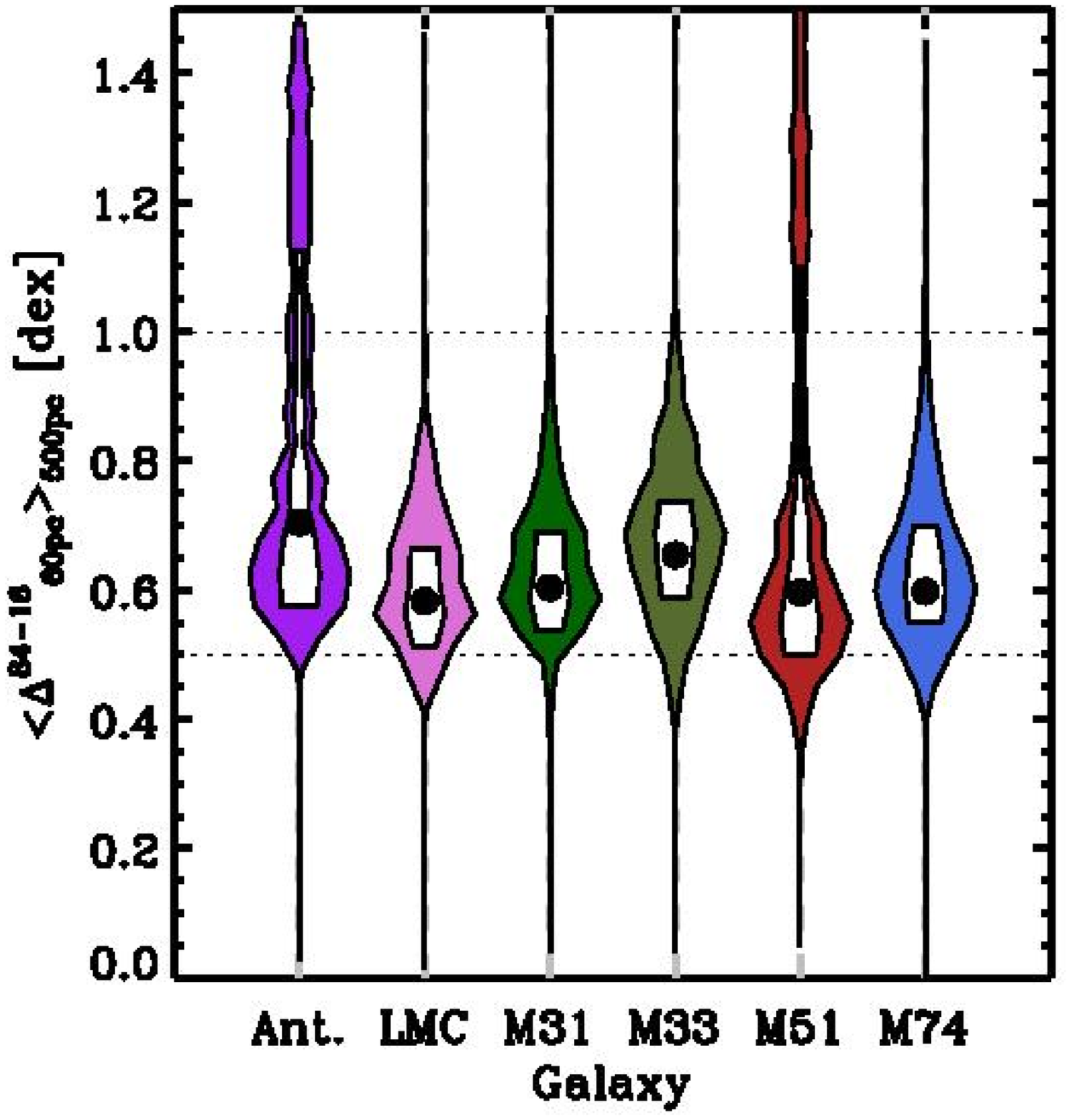}
\caption{({\em left}) Intensity-weighted integrated intensity, $<I_{\rm 60pc}>_{\rm 500pc}$ ($y$-axis) as a function of the $50^{\rm th}$ percentile integrated intensity within the averaging beam ($x$-axis). The two quantities track one another well. Some offset is expected based on the definitions of the two quantities, with larger offsets for wider distributions (see text). The intensity-weighted integrated intensity is thus closely related to the peak of the integrated intensity distribution function at 60~pc scales (see Figure \ref{fig:histograms}). ({\em right}) Width of the cumulative integrated intensity distribution function (CDF), parameterized as $\Delta^{84-16}$, for our targets. For a lognormal distribution of light (Figure \ref{fig:histograms}) $\Delta^{84-16}/2$ is the $1\sigma$ width. Thus, most of the regions in our analysis show narrow distributions, with most flux within $\pm 0.3$~dex (a factor of two) of the median. Bright regions in M51 (spiral arms) and the Antennae (the SGMCs) contain a large amount of light concentrated by dynamical effects on small scales, creating higher contrast in these regions.}
\label{fig:delta_hist}
\end{figure*}

Figure \ref{fig:histograms} plots ({\em left}) the distribution of integrated intensity, $I_{\rm 60pc}$, and ({\em right}) the distribution of flux as a function of integrated intensity for each of our targets. Note that this figure does not include any ensemble averaging. It shows the distribution of $I_{\rm 60pc}$ for the entire area within our masks at $\theta=60$~pc resolution for each target. On the right, the histogram shows the sum of the flux associated with pixels that have $I_{\rm 60pc}$ near the indicated value.

The right panel of Figure \ref{fig:histograms} may be surprising. The number distribution of molecular clouds is usually treated using a truncated power law \citep{ROSOLOWSKY05B}. Previous studies of the number distribution of CO intensities (the left column of Figure \ref{fig:histograms}) have found a mixture of lognormal and power law distributions \citep{HUGHES13A}. In both cases, many clouds and may pixels live near the sensitivity threshold. Here we show that the distribution of flux as a function of $I$ appears approximately lognormal and lies well above the sensitivity threshold in each of our sources. The large numbers of low mass clouds or low intensity pixels do not contribute overwhelming amounts of flux. We checked the robustness of this result by integrating the cubes directly, with no masking, and subtracting the flux distribution for negative $I$ from that for positive $I$; the result shows small shifts from what we observe in Figure \ref{fig:histograms} but appear qualitatively the same.

The distributions in Figure \ref{fig:histograms} are not perfectly lognormal. The fits, which are only approximate, somewhat overpredict the data at high values, suggestive of a truncation; for example, see the cases of M31 and M51 \citep[and][]{ROSOLOWSKY05B,HUGHES13A}. The Antennae appear to be better described by two lognormal distributions than one, again reflecting the difference between the SGMCs and the rest of the overlap region. We see some point-to-point variation in the width of the distribution (see next subsection), as well as real internal scatter in the center of the distribution. The full-galaxy distributions presented in Figure \ref{fig:histograms} therefore represent the combination of a number of discrete local distributions  \citep[as previously noted by][ using kinematically-defined environments in M51 PAWS]{HUGHES13A}. 

For a lognormal distribution, the $50^{\rm th}$ percentile in the CDF captures the peak of the distribution. This is closely related, but not identical, to the intensity-weighted $I$ that we measure. We compare the two quantities measured for our data in Figure \ref{fig:delta_hist}. They track one another well but the intensity-weighted average value is higher. The intensity-weighted $I$ is a {\em linear} weighted average; the logarithmic axis for the lognormal distribution means that the intensity-weighted $I$ will tend to be above the $50^{\rm th}$ percentile by an amount related to the width of the distribution (essentially just the difference between geometric and linear averaging).

\subsubsection{Variations in the Cumulative Distribution Width}

Figure \ref{fig:histograms} shows that when all lines-of-sight are considered, the CO flux appears to be approximately lognormally distributed as a function of $I$ in our targets. Figure \ref{fig:delta_hist} shows the width of the distributions within individual 500~pc averaging beams. $\Delta^{84-16}/2$, which will be the rms dispersion for a lognormal distribution, is ${\sim}0.3$~dex over most of the area in all of our targets (i.e., $\Delta^{84-16} \approx 0.6$). This is modestly narrower than the full distributions in Figure~\ref{fig:histograms}, which have $1\sigma$ dispersion ${\sim}0.4$. The difference reflects the blending of all regions together into Figure \ref{fig:histograms}.

This narrow distribution means that $\sim 70\%$ of the CO flux, and so presumably ${\sim}70\%$ of the molecular mass, is distributed over only a narrow range of factor ${\sim}4$ in $I$ (or $\Sigma$ for a fixed $\alpha_{\rm CO}$). This modest $\Delta^{84-16}$ at $\theta = 60$~pc resembles the idea of a fixed GMC surface density, one expression of the \citet{LARSON81} relations \citep[see][]{BOLATTO08,HEYER15}. However, this is {\em not} what is shown here. Instead, we observe the center of the distribution, tracked by $I$, to vary within and among targets (Figure \ref{fig:props_hist}). However, in individual 500~pc regions (Figure \ref{fig:delta_hist}), and even over large parts of galaxies (Figure \ref{fig:histograms}), much of the flux lies within a factor of $\pm 2$ of the median value. Our result supports the idea that the properties of molecular gas on cloud-scales achieve an equilibrium configuration that depends on larger scale conditions, rather than an absolute value for the molecular gas surface density that applies across all galaxies and galactic environments.

In the LMC, M31, M33, and M74, this narrow distribution also describes most of the light (right panel of Figure \ref{fig:delta_hist}), while broader distributions are present for a small but bright subset of regions in the Antennae and M51. The high-$\Delta$ regions in M51 come from the two bright spiral arms, while those in the Antennae galaxies come from near the SGMCs in the overlap region. In both cases, gas is concentrated to high surface densities by dynamical effects. The bright, dense structures are small compared to our beam, creating a large contrast with the surrounding medium inside our 500~pc averaging beam.

\subsection{Physical State of the Gas}
\label{sec:cond}

\begin{figure}
\plotone{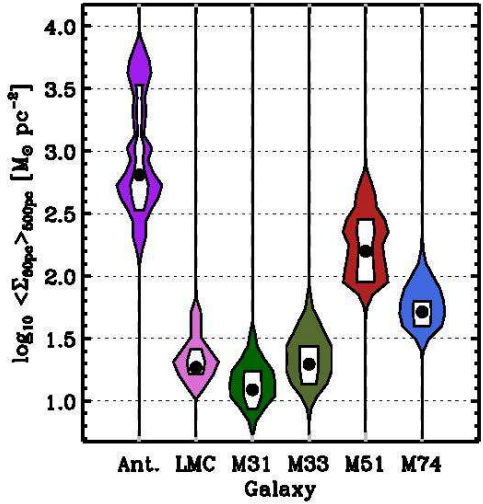}
\caption{Distribution of mass surface density for our six data sets at 60~pc resolution averaged over 500~pc beams. Galaxies show a range of mean intensity-weighted gas surface density, with M74 and M51 showing higher $\Sigma$ than the lower mass Local Group galaxies (LMC, M33, M31). The Antennae again show the highest mean $\Sigma$, and also the widest range of values.}
\label{fig:mass_hist}
\end{figure}

\begin{figure*}
\plottwo{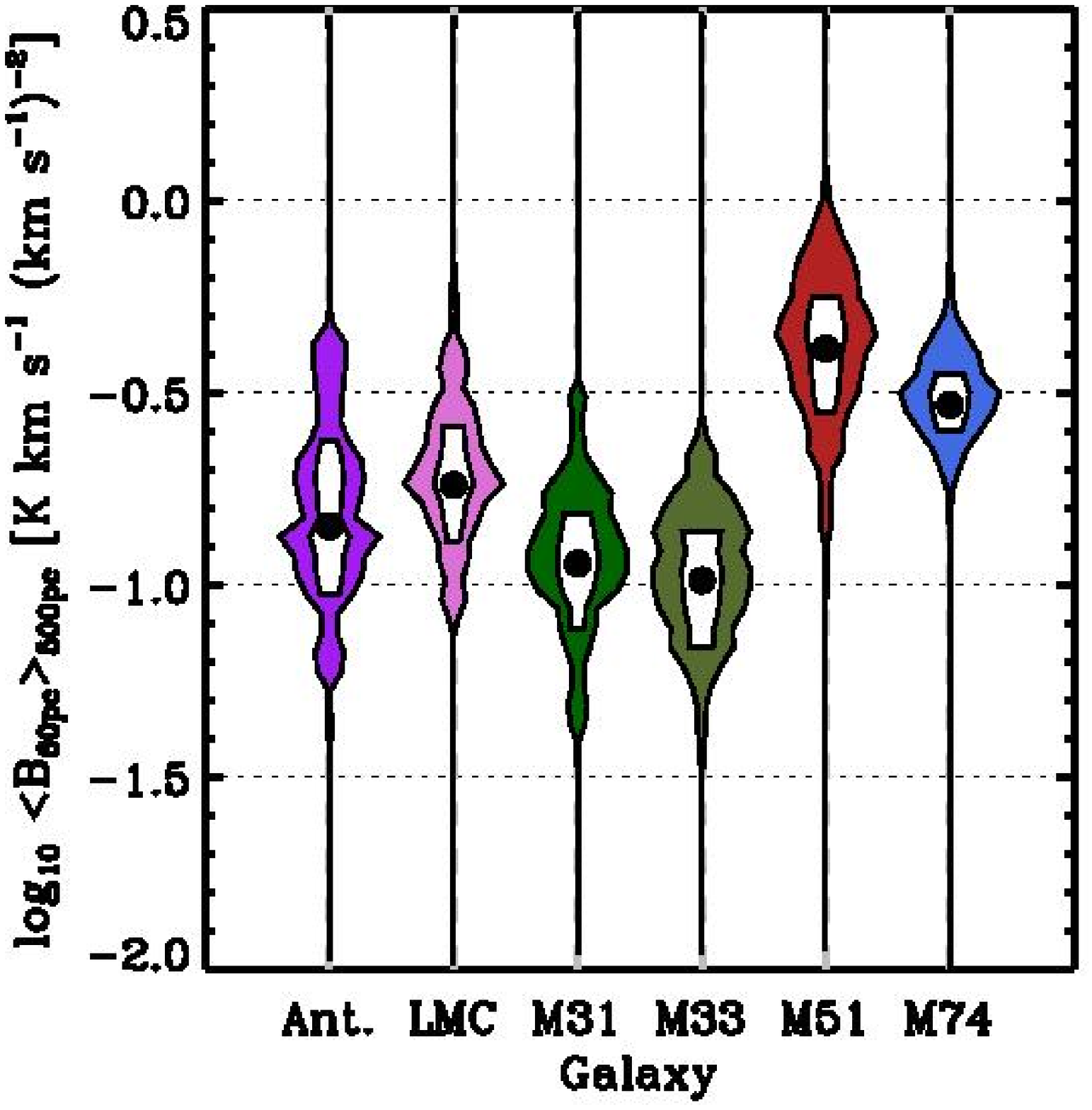}{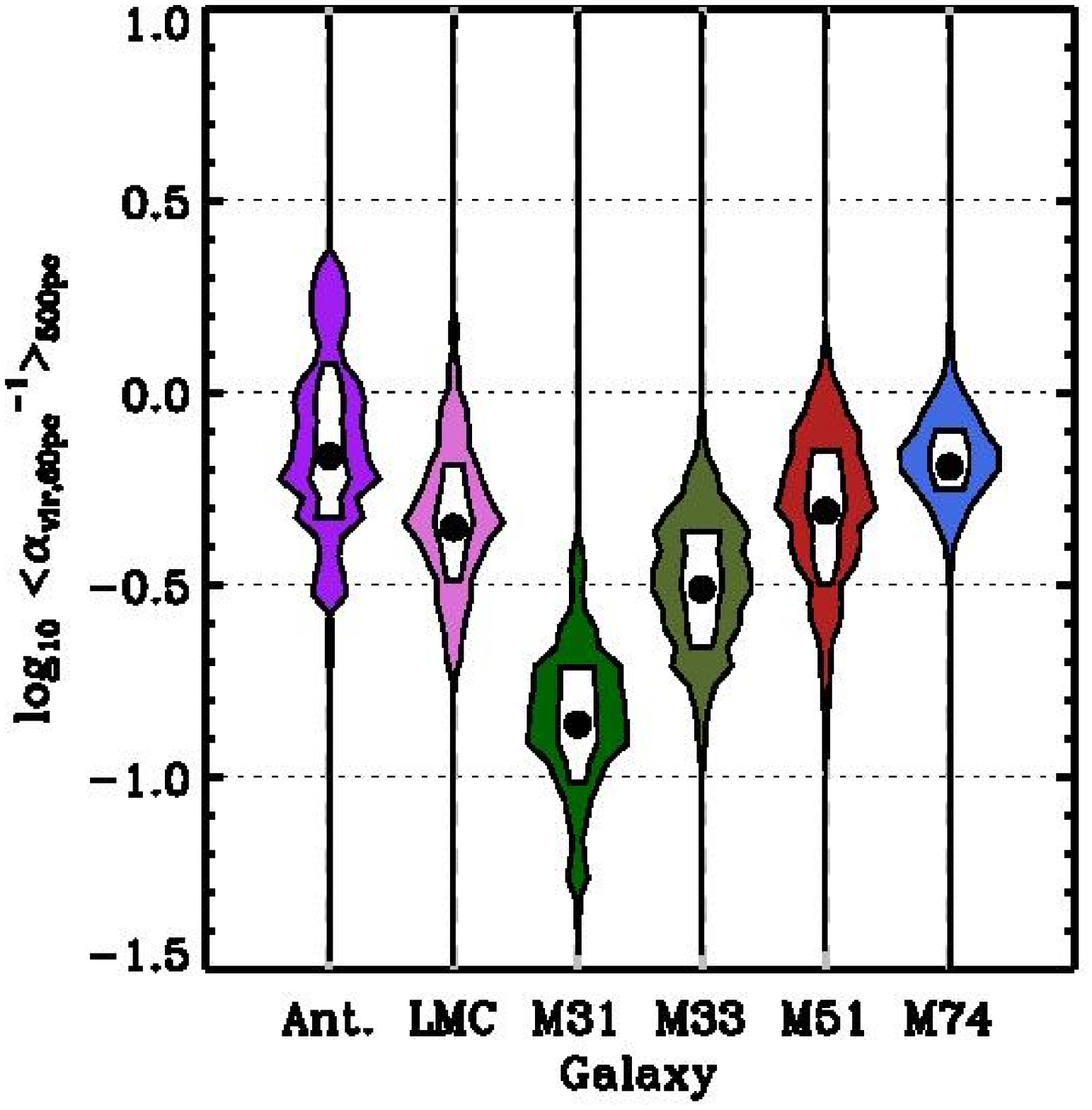}
\caption{Distribution of ({\em left}) $B \equiv I / \sigma^2$, an observational tracer of the self-gravity of the gas,  for our six data sets at 60~pc resolution averaged over 500~pc beams. Galaxies show a spread in the $B$ ratio with M74 and M51 showing higher $B$ values (i.e., more bound) than the Local Group systems. The Antennae show the widest range of $B$ values. ({\em right}) Inverse virial parameter, $\alpha_{\rm vir}^{-1}$, implied by our observed $B$. Here, too, higher values are more bound, with $\log_{10} \alpha_{\rm vir}^{-1} = 0.0$ ($\alpha_{\rm vir} = 1$) expected for virialized clouds and $\log_{10} \alpha_{\rm vir}^{-1} \approx -0.3$ ($\alpha_{\rm vir} = 2$) being the approximate boundary for boundedness.}
\label{fig:b_hist}
\end{figure*}

{\em Mass Surface Density:} With a conversion factor, the integrated intensities in Table \ref{tab:results} and Figure \ref{fig:props_hist} correspond to a mass surface density, $\Sigma$. Figure \ref{fig:mass_hist} and Table \ref{tab:results} report these values for our adopted conversion factors. The relative behavior among galaxies remains mostly the same as for $I$, except that M31 now has a lower $\Sigma$ than both M33 and the LMC as a result of our adopted conversion factors. 

For physical models, the specific values of $\Sigma$ are of interest. Most regions in the LMC, M31, and M33 have low intensity-weighted surface densities at $\theta=60$~pc, $\Sigma \approx 10{-}30$ \mbox{M$_\odot$~pc$^{-2}$}. This is comparable to the {\sc Hi} mass surface densities in these galaxies, implying a mixture of approximately equal amounts of {\sc Hi} and H$_2$ over a 60~pc box containing most molecular gas. For $\Sigma = 15$ \mbox{M$_\odot$~pc$^{-2}$}, the mass within a $60$~pc beam is $M \approx 6 \times 10^4$ M$_\odot$, roughly equivalent to a typical Milky Way molecular cloud in the beam.

The large spiral galaxies show higher $\Sigma$, with average surface densities of $\Sigma \approx 160$ \mbox{M$_\odot$~pc$^{-2}$} in M51 and $\Sigma \approx 50$ \mbox{M$_\odot$~pc$^{-2}$} in M74. These values resemble commonly quoted extragalactic GMC surface densities and imply a much sharper contrast with $\Sigma_{\rm HI}$ on 60~pc scales \citep[see][for discussion of $\Sigma_{\rm HI}$ at high resolution]{LEROY13B} than we see in the Local Group galaxies. The molecular gas mass in a beam is ${\sim} 2 \times 10^5$~M$_\odot$ for $\Sigma = 50$ \mbox{M$_\odot$~pc$^{-2}$} and ${\sim} 7 \times 10^5$~M$_\odot$ for $\Sigma = 160$ \mbox{M$_\odot$~pc$^{-2}$}, both typical of a massive Milky Way GMC.

The Antennae show both the largest $\Sigma$ and a large contrast between the median over all regions, $\Sigma \approx 650$ \mbox{M$_\odot$~pc$^{-2}$}, and the median weighted by flux, $\Sigma \approx 4,000$ \mbox{M$_\odot$~pc$^{-2}$} (Table \ref{tab:weighting}). As noted above, the light from the Antennae is dominated by a handful of regions with high $\Sigma$ at the interface between the colliding galaxies. The clouds in this zone have enormous line widths and surface densities \citep{WEI12}, comparable to those found in other extreme starbursts \citep{LEROY15A}.

The quantity $\Sigma$ that we discuss here is cleanly defined, but we note that it differs -- at least in theory -- from the definition of $\Sigma$ used in cloud property studies. Here we use a single, fixed averaging scale, $\theta = 60$~pc, and compare measurements within and among galaxies at that scale. Cloud property studies attempt to define a physical scale of interest object-by-object, although in practice this is usually an {\em ad hoc}, observational definition, rather than physical one (algorithms are designed to match by-eye structure finding results). The physical scale of cloud-property measurements does tend to resemble the beam scale. As a result, the two definitions of $\Sigma$ match in practice for a given data set, but they are not required to do so.

{\em Mass Volume Density and Free-fall Time:} Volume mass density, $\rho$, and number density, $n_{\rm H2}$, relate closely to surface density and are key parameters for most theories that explain the formation of stars from molecular clouds. Estimating $\rho$ requires knowledge of the line-of-sight depth through the gas. In a cloud view, this is usually achieved assuming spherical symmetry, so that the measured radius in the plane of the sky also corresponds to the cloud depth. In our approach, one must assume a line-of-sight depth. Here, we adopt a fiducial $l \approx 60$~pc, the size of a large molecular cloud and a reasonable approximation to the thickness of a cold gas disk; it is also our beam size, but although this makes the calculation symmetric there is no reason to expect a match between $l$ and the beam. We treat the gas as having a tophat density distribution along the line-of-sight and estimate the H$_2$ number density, $n_{\rm H2}$, following Equation \ref{eq:density}.

The $\Sigma \approx 10{-}30$ \mbox{M$_\odot$~pc$^{-2}$} in the Local Group targets implies $n_{\rm H2} \approx 2.5{-}7.5$~cm$^{-3}$ for $l=60$~pc; $\Sigma \approx 50$ \mbox{M$_\odot$~pc$^{-2}$} in M74 yields $n_{\rm H2} \approx 12$~cm$^{-3}$; $\Sigma \approx 160$ \mbox{M$_\odot$~pc$^{-2}$} in M51 corresponds to $n_{\rm H2} \approx 40$~cm$^{-3}$; and the $\Sigma \approx 650{-}4000$ \mbox{M$_\odot$~pc$^{-2}$} in the Antennae yields  $n_{\rm H2} \approx 150{-}1000$~cm$^{-3}$. As with $\Sigma$, these are densities measured within a fixed-size averaging box (with an assumed value for the third dimension). Given that these densities are often lower than those required to excite the CO emission that we observe, there must clearly be substantial sub-resolution clumping.

With our assumed depth, the high surface densities in the Antennae correspond to free-fall times of ${\langle \tau_{\rm ff,60pc} \rangle}_{\rm 500pc} \approx 1{-}3$~Myr (for $\Sigma = 650{-}4000$~M$_\odot$~pc$^{-2}$). The moderate surface densities in M51 and M74 imply ${\langle \tau_{\rm ff,60pc} \rangle}_{\rm 500pc} \approx 5{-}10$~Myr. Meanwhile in the Local Group systems, the free-fall time implied the average density over $60$~pc is quite long, $\tau_{\rm ff} \approx 10{-}20$~Myr. All of these numbers come with the caveat that they apply to the average density over $60$~pc scales. Gas clumped on scales smaller than our resolution will have a shorter $\tau_{\rm ff}$ locally.

{\em Line Width and Mach Number:} The average line width varies from ${\langle \sigma_{\rm 60pc} \rangle}_{\rm 500pc} \approx 3.5$ \mbox{km~s$^{-1}$} in the LMC to ${\sim} 4{-}5$ \mbox{km~s$^{-1}$} in M31, M33, and M74, to ${\sim} 9$~km~s$^{-1}$ in M51 and ${\sim} 18$ \mbox{km~s$^{-1}$} in the Antennae. Though other sources may contribute significantly to these line widths \citep[see][for an example in M51]{MEIDT13}, we estimate the turbulent Mach number, $\mathcal{M}$, implied for these dispersions and $T=25$~K. For this temperature, all of the line widths are highly supersonic, with three dimensional Mach numbers $\mathcal{M} \sim 15{-}20$ in the LMC, M31, M33, and M74, to $\mathcal{M} \approx 40$ in M51 and ${\sim}80$ the Antennae, though this last value certainly overestimates $\mathcal{M}$ somewhat.

Although these Mach numbers are high, the line widths that we measure are actually lower than previous measurements on larger scales \citep[e.g.,][]{TAMBURRO09,WILSON11,CALDUPRIMO13}. Applying stacking on $\sim$kpc scales, \citet{CALDUPRIMO13} found a characteristic line width $\sigma \approx 12$ \mbox{km~s$^{-1}$} for a modest sample of nearby spiral galaxies (including M74 and M51). Our result of smaller line widths at small ($60$~pc) scales agrees with the findings by \citet{CALDUPRIMO16} for M31, where they find a narrow line width component of $\sigma \approx 3.2$ \mbox{km~s$^{-1}$} on $175$~pc scales. They argue that coarser beams sample larger scales in the turbulent cascade, although the inclusion of streaming or other bulk gas motions might also contribute.

{\em Gravitational Boundedness:} Surface density and line width combine to yield an estimate of gravitational boundedness. Beams with high surface density will have stronger self-gravity, and beams with high line width have high kinetic energy. The ratio of the two is traced by the observable, $B \equiv I / \sigma^2$ (Equation \ref{eq:boundedness}), which relates to the inverse of the virial parameter, $\alpha_{\rm vir}^{-1}$. We plot ${\langle B_{\rm 60pc} \rangle}_{\rm 500pc}$ and ${\langle \alpha_{\rm vir,60pc}^{-1} \rangle}_{\rm 500pc}$ in Figure \ref{fig:b_hist}. For $\alpha_{\rm vir}^{-1}$, the specific values $\alpha_{\rm vir}^{-1} = 1$ ($\log_{10} \alpha_{\rm vir} = 0.0$) and $0.5$ ($\log_{10} \alpha_{\rm vir} \approx -0.3$) have specific physical meaning: the former corresponds to virialized clouds and the latter to marginally-bound material. Our absolute scale for $\alpha_{\rm vir}^{-1}$ depends on either a simple adopted geometry or an empirical scaling from cloud catalogs, as well as the adopted conversion factor for each system. Thus, the location of our measurements between these absolute values should not be over-emphasized, but is still worth noting.

Figure \ref{fig:b_hist} shows that our galaxies are more clustered in boundedness (${\langle B_{\rm 60pc} \rangle}_{\rm 500pc}$ or ${\langle \alpha_{\rm vir,60pc}^{-1} \rangle}_{\rm 500pc}$) than in surface density. The two spirals (M51 and M74) do show a larger $B$ than the Local Group galaxies and the Antennae, but this difference is reduced when we apply our adopted conversion factors. Then, most of the data points cluster between ${\langle \alpha_{\rm vir,60pc}^{-1} \rangle}_{\rm 500pc} = 0$ and $-0.5$, that is between virialized ($\alpha_{\rm vir}^{-1}=0$) and marginally unbound ($\alpha_{\rm vir}^{-1} < 0.5 \approx -0.3$~dex). Few regions appear virialized, on average, at $\theta = 60$~pc scales, though many appear (at least marginally) gravitationally bound. We show below that the dynamical state of the gas depends on resolution, with self-gravity becoming more important at smaller spatial scales. One way to read this result, then, could be that $\theta=60$~pc is roughly the scale at which self gravity becomes important in our targets.

M31 appears as an outlier from the other galaxies in its distribution of ${\langle \alpha_{\rm vir,60pc}^{-1} \rangle}_{\rm 500pc}$, such that the gas in M31 appears unbound by self-gravity at these scales. Schruba et al. (in prep.) note this apparent high line width compared to the gas surface density. They also show that it varies across the galaxy, with gas in the inner bulge region appearing less bound than gas in M31's 10~kpc molecular ring. They hypothesize that this could be due to clouds confined by the pressure of M31's atomic-dominated ISM. Alternative explanations would be that M31 has a higher conversion factor (by about a factor of ${\sim}2$) than our adopted Milky Way $\alpha_{\rm CO} = 4.35$ \acounits\, or that the gas in M31 is more strongly clumped than the gas in our other targets.

\subsection{A Scaling Between Surface Density and Line Width}
\label{sec:sd_lw}

\begin{figure*}
\plotone{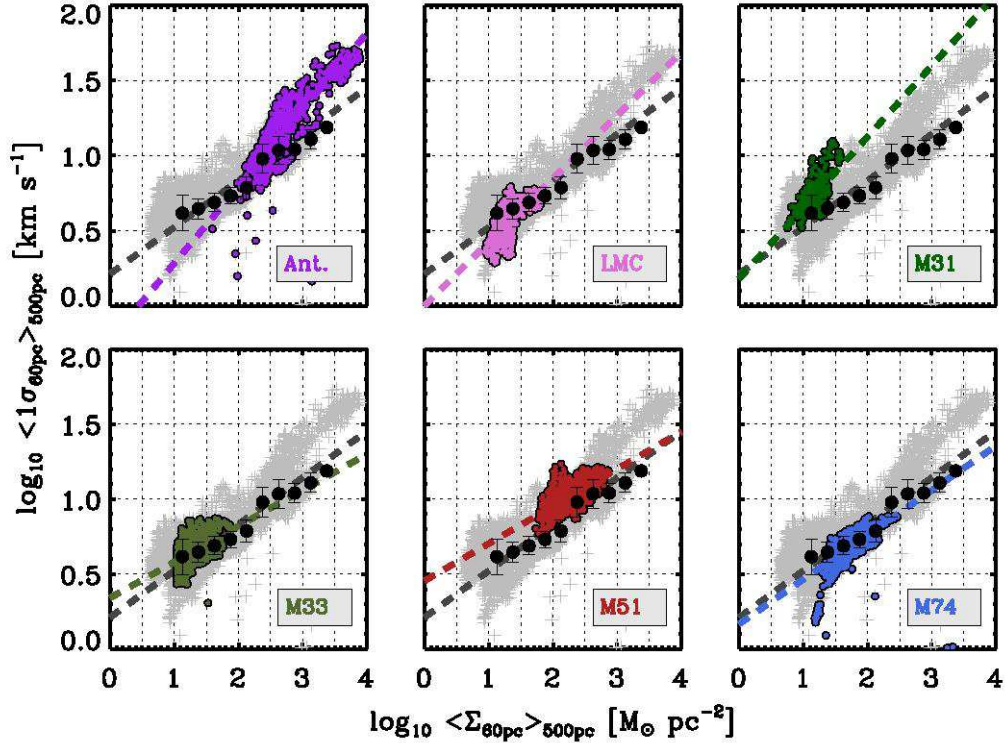}
\caption{Line width at 60~pc, $<\sigma_{\rm 60pc}>_{\rm 500pc}$, as a function of surface density, $<\Sigma_{\rm 60pc}>_{\rm 500pc}$ for our targets, both using a 500~pc averaging beam. Line width increases with surface brightness across the sample. The dark line shows Equation \ref{eq:sig_vs_sd}, which has a power law index of $\approx 0.3$; colored lines show fits to the individual data sets. Within the Antennae and M31, some regions stand out as having a high line width relative to their surface density.}
\label{fig:sig_vs_sd}
\end{figure*}

The same regions and galaxies that have a high line width also have a high mass surface density. This is reflected in both the histograms and the narrow range of ${\langle \alpha_{\rm vir,60pc}^{-1} \rangle}_{\rm 500pc}$. In fact, these two quantities correlate well across our sample. The correlation, and deviations from it, serve as a useful physical diagnostic of the dynamical state of the gas. We show this correlation for all targets, highlighting each galaxy, in Figure~\ref{fig:sig_vs_sd}. Black points show the median line width and $1\sigma$ log scatter binning all data except the Antennae by $\Sigma$. A fit to the non-Antennae data yields:

\begin{eqnarray}
\label{eq:sig_vs_sd}
&& \log_{10} \left< \sigma_{\rm 60pc} [{\rm km~s}^{-1}] \right>_{\rm 500pc} \approx \\
\nonumber && (0.31 \pm 0.04) \log_{10} \left( \frac{\left< \Sigma_{\rm 60pc} \right>_{\rm 500pc}}{50~{\rm M_\odot~pc}^{-2}} \right) +
(0.74 \pm 0.01)\,,
\end{eqnarray}

\noindent which describes the aggregate behavior of the data (the equation is a bivariate fit to the ensemble of data excluding the Antennae).

Figure \ref{fig:sig_vs_sd} and Equation \ref{eq:sig_vs_sd} show that gas with higher surface density, and so higher self-gravity, at $\theta=60$~pc also has a higher dispersion. For uniform geometry across our sample -- i.e., if clumping or size scales do not matter below 60~pc -- then we would expect $\sigma \propto \Sigma^{0.5}$ for a fixed ratio of potential to kinetic energy (see Equation \ref{eq:boundedness} with fixed $R$). Our observed scaling is shallower than $0.5$, so that gas appears somewhat more bound at higher $\Sigma_{\rm 60pc}$. This gradient is also visible in Table \ref{tab:results}, where the Antennae, M51, and M74 have higher $\alpha_{\rm vir}^{-1}$ (more bound) than the Local Group targets. In fact, this galaxy-to-galaxy variation accounts for much of the shallow slope in Equation \ref{eq:sig_vs_sd}; fits to individual galaxies, shown in color in each panel, do tend to show a steeper relationship.

High $\Sigma_{\rm 60pc}$ regions appear somewhat more bound at $60$~pc scales, but this does not mean that the Local Group targets are unbound at all scales. Sub-beam clumping means that the gas will have stronger self-gravity on smaller scales \citep[see][]{LEROY13B}. If the line width is turbulent in origin and if the outer scale of turbulence is $\gtrsim 60$~pc, then the line width, $\sigma$, will also decrease moving to smaller scales. If the dependence of surface density on size scale is stronger than that of line width, then our results may instead be interpreted as evidence that the molecular gas in Local Group targets is gravitationally bound on smaller size scales than the gas in more massive, gas-rich targets. In a picture of the ISM where GMCs are gravitationally bound, isolated spheres, this is equivalent to arguing that Figure \ref{fig:sig_vs_sd} reflects the results of a combination of filling factor and mean GMC size. Our results are not, however, consistent with variations in the filling factor of otherwise identical clouds as the sole explanation. Such a scenario would not alter the line width except via ```shadowing'' of clouds along the line-of-sight. Neither the line profiles nor the apparent surface density suggest that this effect drives the correlation that we see. As noted above, we revisit the scale dependence of this approach in a future work.

Figure \ref{fig:sig_vs_sd} can be read as a beam-wise version of the $\sigma^2/R$ vs.\ $\Sigma$ plot commonly used in Milky Way \citep{KETO86,HEYER09,FIELD11} and extragalactic \citep[see][]{LEROY15A,JOHNSON15} cloud studies to examine the dynamical state of clouds. Alternatively, when cast in terms of virial mass versus luminosity, this relation can be used to solve for $\alpha_{\rm CO}$ by assuming a dynamical state for clouds \citep[see][]{DONOVANMEYER12}. Our approach does not measure a size scale and treats each resolution element rather than each cloud as a measurement, but the encapsulated physics are similar. Regions that deviate towards high line width at a given surface density are farther from virialization, less likely to be bound, and more likely to represent gas that is part of a turbulent medium, described by an average turbulent pressure $P \approx \rho \sigma^2$ rather than $\mathit{KE} \approx \mathit{UE}$ \citep[see][]{FIELD11}.

The surface density, $\Sigma$, and line width, $\sigma$, are intrinsically correlated. Indeed, for an optically thick line like CO, an alternative view of the correlation in Figure \ref{eq:sig_vs_sd} is that a variable line width, perhaps set by  the local potential well, drives the strength of the CO line. If this is the only effect at play, this translates to a variable $\alpha_{\rm CO}$ \citep[see, e.g.,][]{MALONEY88,DOWNES98,NARAYANAN12,BOLATTO13B}. This effect certainly occurs, and helps to explain the observed contrast between ``disk'' and ``starburst'' conversion factors. However, there are several reasons to think that our observed $\sigma$-$\Sigma$ correlation is not purely explained by this effect. First, there is also a correlation between peak intensity, $I_{\rm \nu, pk, 60pc}$, and $\sigma$ (not shown) that spans approximately an order of magnitude and shows many of the same features that we see in Figure \ref{fig:sig_vs_sd}. Second, we have already accounted for our best-estimate variations of $\alpha_{\rm CO}$ in Figure \ref{fig:sig_vs_sd}. Equating line width variations to $\alpha_{\rm CO}$ variations predicts a specific behavior for $\alpha_{\rm CO}$. This behavior does not appear to be borne out by experiments using dust to infer $\alpha_{\rm CO}$ at lower resolution, although it may be somewhat at play in the centers of galaxies \citep{SANDSTROM13}. This appears particularly true in the case of M51, where multiple methods constrain $\alpha_{\rm CO}$ to a value similar to the Milky Way (Groves et al. in prep.). Beyond these arguments, our approach provides an ideal framework to test this hypothesis: by correlating dust-based $\alpha_{\rm CO}$ measurements with $\sigma_{\rm 60pc}$ one can test this hypothesis with a high degree of rigor.

Our view is that Figure \ref{fig:sig_vs_sd} reflects a degree of self-regulation in galaxies to achieve a similar, but not universal, dynamical state at $60$~pc scales. Given the values of $\alpha_{\rm vir}^{-1}$ that we observe, this dynamical state may simply be marginally bound, collapsing gas \citep[e.g.,][]{BALLESTEROSPAREDES11} or it could be virialization somewhat below our measurement scale. The shallow slope of the relation (${\sim}0.3$ rather than $0.5$) may indicate more clumped molecular gas in more quiescent regions. Variations about this relation, such as the high $\sigma$ regions seen in the inner part of M31 and the overlap region of the Antennae,  will arise due to gas in a different dynamical state. For example, the high ambient gas pressure in M31's inner region and the turbulent pressure created by the collision between the Antennae galaxies \citep[e.g.,][]{RENAUD14} may counteract the self-gravity of the gas to some degree. Though in the Antennae, in particular, local variations in $\alpha_{\rm CO}$ (which includes excitation in our formulation) and contributions to the line widths from complex geometry will also be crucial.

\subsection{Resolution and the Physical State of the Gas}
\label{sec:multiscale}

\begin{figure*}
\plotone{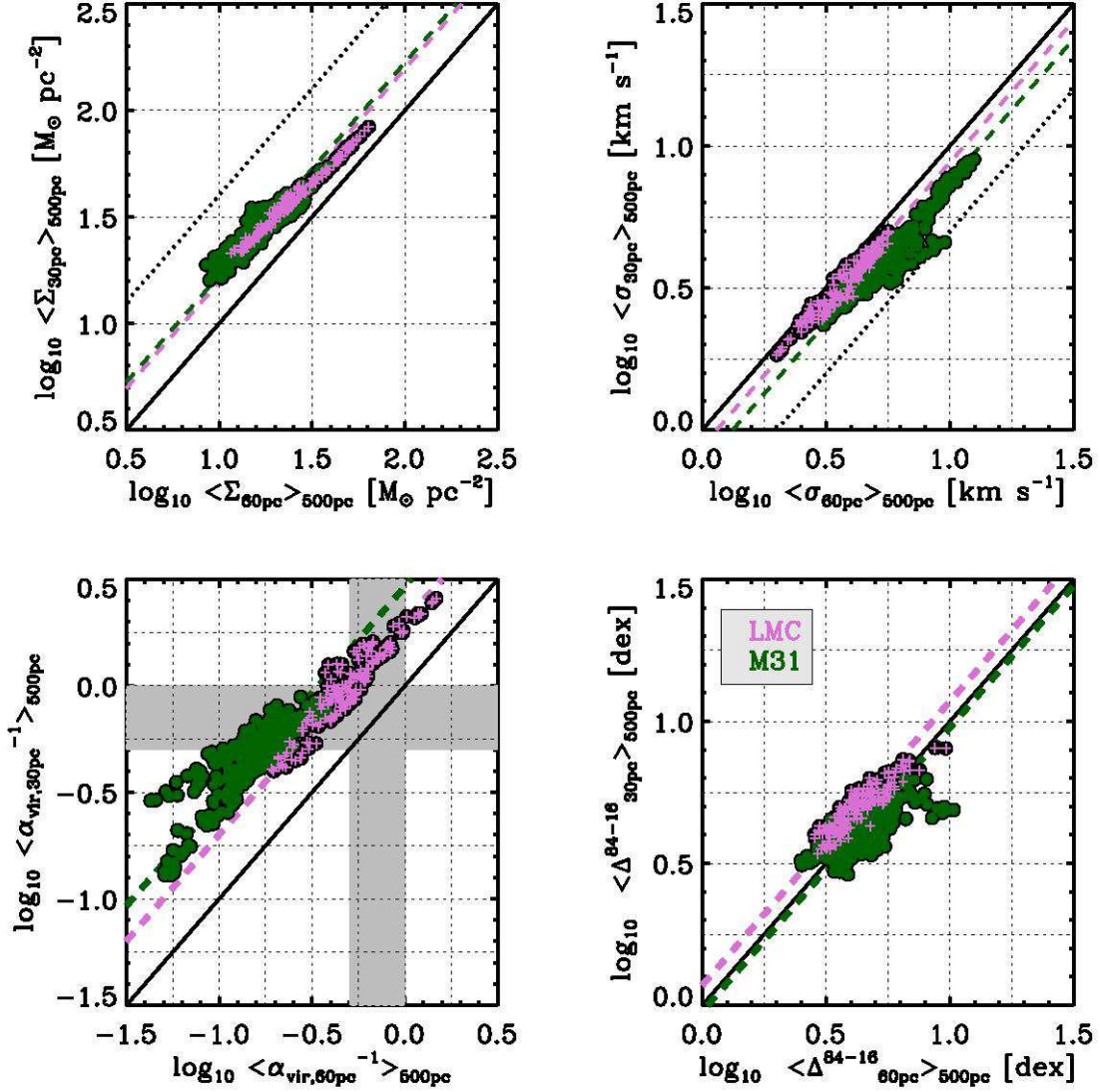}
\caption{Properties of the gas in the LMC (magenta) and M31 (green) at $\theta = 30$~pc ($y$-axis) and $\theta = 60$~pc ($x$-axis): ({\em top left}) surface density, ({\em top right}) line width, ({\em bottom left}) self-gravity, tracked by the inverse virial parameter, ({\em bottom right}) width of the intensity distribution. Solid lines show equality. Dotted lines in the top two panels show simple physical expectations: beam dilution acting on a point source for surface density and the line width-size relation for line width. In the bottom left plot, gray regions highlight the region of parameter space between marginally bound and virialized. Dashed colored lines show the median ratio relating structure between the two size scales for each galaxy. These plots show that surface density increases, line width decreases, and self-gravity increases as we consider smaller scales.}
\label{fig:multiscale}
\end{figure*}

The results throughout this section describe the average state of the cold ISM on scales of $\theta=60$~pc. However, molecular gas is structured below this scale. Indeed, by comparing the average volume densities found in the Local Group targets, $n_{\rm H2} \approx 1{-}10$~cm$^{-3}$, to the density needed to produce appreciable CO emission, ${\sim}300$~cm$^{-3}$ even with optical depth effects, one can immediately see that the gas must be highly clumped below our resolution. In our framework, a natural way to explore this smaller scale structure is to vary the resolution of the data and then re-measure the properties of the emission. Figure \ref{fig:multiscale} shows the results of this exercise for our two targets with the highest native resolution. We characterize the emission from the LMC and M31 at $\theta = 30$~pc using the same grid and $A=500$~pc averaging beam that were used for the $\theta = 60$~pc characterization. The flux recovery is worse for M31 at $\theta=30$~pc than at $\theta = 60$~pc, so we compare only points with appreciable CO flux inside our mask at both resolutions.

Figure \ref{fig:multiscale} compares the calculated surface density, line width, $\alpha_{\rm vir}^{-1}$, and the distribution width between the two resolutions. As expected, the surface density appears higher at $\theta=30$~pc than at $\theta=60$~pc, reflecting sub-resolution clumping of gas \citep[see][]{LEROY13B}. The contrast is a factor of $1.6$, which is far weaker than the factor of $4$ expected if all emission came from point sources at the center of the beam. The average line width, $\sigma$, appears $15{-}25\%$ lower at $30$~pc resolution than at $60$~pc. The sense of the change is expected from the line width-size relation \citep[][]{LARSON81,BOLATTO08,HEYER15}, but the magnitude of the change is again lower than what one would infer from the canonical $\sigma \propto r^{0.5}$. Both measurements can be explained if gas is more extended and structured on larger scales than only individual GMCs; such a situation has been observed in the clustering of clouds in M33 \citep{ROSOLOWSKY07B}.

With higher surface densities and narrower line widths, the self-gravity of gas will be higher at smaller scales. The bottom left panel of Figure \ref{fig:multiscale} shows that $\alpha_{\rm vir}^{-1}$ rises significantly moving from $\theta=60$~pc to $\theta=30$~pc, increasing by factors of ${\sim}3$ for M31 and ${\sim}2$ for the LMC. This shifts a large amount of LMC gas to an approximately virialized state, consistent with the fact that our adopted LMC CO-to-H$_2$ conversion factor was partially motivated by the virial theorem arguments made by \citet{HUGHES10} and \citet{WONG11}. Much of the M31 gas remains marginally bound or unbound, but the self-gravity of the gas is stronger at $30$~pc than at $60$~pc resolution.

The width of the intensity distribution does not appear to change dramatically with scale, so that the distribution seems to ``slide'' to higher values.

The key result from Figure \ref{fig:multiscale} is that the average properties of the molecular ISM depend on scale, though not in a trivial way. In particular, the dynamical state of gas depends sensitively on the spatial scale considered, with the narrower line widths and high surface densities found on small scales leading to stronger self-gravity.

\section{Discussion and Conclusions}
\label{sec:discuss}

We present a simple way to capture the cloud-scale properties of the interstellar medium (ISM) using high physical resolution, wide-area maps of mass-tracing spectral lines. Applying this method to wide-area, high resolution \mbox{low-$J$} CO maps of a diverse sample of six galaxies, we derive a picture of the physical state of molecular gas in galaxies at $\theta=60$~pc resolution.

\subsection{Method}

Our approach measures the line width, peak intensity, and integrated intensity at high physical resolution (``cloud scales''). We then carry out an intensity-weighted average at a larger scale to improve signal-to-noise, average over the cycling of the ISM between different evolutionary states, and measure the typical state of the ISM over a large part of the galaxy. We also measure the cumulative distribution of light as a function of intensity inside the averaging beam, parameterized by the $50^{\rm th}$ percentile value and logarithmic $68\%$ width, $\Delta^{84-16}$.

The line width ($\sigma$), integrated intensity ($I$), and their combination ($B=I/\sigma^2$) relate closely to the turbulent velocity dispersion, mass surface density ($\Sigma$), and gravitational boundedness of the ISM (parameterized via the inverse of the virial parameter $\alpha_{\rm vir}^{-1}$). For data with resolutions of ${\sim} 10{-}100$~pc, we demonstrate that this approach captures much of the same physical information that is encoded in GMC-based property studies. 

The advantage of our approach is that it avoids complex segmentation algorithms (with associated unquantified uncertainty), characterizes the whole ISM traced by the line in a simple way, and includes a natural translation to the larger scales ideal to test theories relating local properties to galactic structure or time-averaged processes like star formation. The utility of an intensity-based approach has been demonstrated in previous studies focusing on beamwise statistics by \citet{SAWADA12} and \citet{HUGHES13A}. The translation of such statistics between scales was discussed in \citet{LEROY13B}.

We apply our method to a diverse set of six galaxies with high physical resolution \mbox{low-$J$} CO maps: the Antennae Galaxies, the LMC, M31, M33, M51, and M74. We also analyze these galaxies using standard GMC property measurements, which we grid into luminosity-weighted average measurements for comparison to our beam-scale measurements.

Using these calculations, we show that common approaches to measure cloud properties almost always yield marginally resolved clouds. As a result, our adoption of the beam scale as the relevant size scale discards minimal information relative to cloud-based analysis. We also show good agreement between the gridded, luminosity-weighted results of GMC property measurements and our beam-wise approach. Although the validity of our approach does not depend on such agreement, this correspondence adds confidence that our simpler approach accesses the same physics as previous studies. We demonstrate this agreement in detail by gridding cloud property measurements, weighting by flux, across our averaging beam. The average surface density and line width measured using our method correlate well with the same quantities derived using several cloud property approaches.

A key quantity in our analysis is $B \equiv I / \sigma^2$, which traces the strength of self-gravity in the gas at our measurement scale, $B \propto \mathit{UE} / \mathit{KE}$. $B$ is an observable cognate of the quantity $\Sigma / \sigma^2$ and for a fixed size scale it relates directly to the inverse of the virial parameter or the ratio of the free-fall time to the crossing time. We demonstrate that, as expected, this anti-correlates with the virial parameter calculated from cloud property studies and note conversions from $B$ to $\alpha_{\rm vir}^{-1}$ based on both this empirical comparison and a simple geometry.

We aim to characterize the average state of all of the gas inside an averaging beam. To gauge how well one achieves this goal,  we note a useful convergence criteria: compare the flux along lines of sight in the cube for which the first moment can be measured to the total flux in the cube. At $\theta = 60$~pc resolution, the data sets in this paper recover $\gtrsim 70\%$ of the emission in the cube.

The Appendices tackle several important technical points. Our measurements incorporate a stacking methodology that helps to avoid some of the biases in sensitivity introduced by signal-to-noise based masking. We also present methods to help account for the spectral response of the data beyond only the finite channel width, which is an often-neglected topic in radio data analysis. We also outline a straightforward path to estimate statistical uncertainties in our framework using Monte Carlo calculations. Applying this method to our data, we derive fractional uncertainties in our ensemble averages of $\sim 2{-}10\%$ and note a strong covariance among statistical errors in the quantities $I$, $\sigma$, and $I_{\rm \nu, pk}$, consistent with their definition. 

\subsection{Capturing the Physics of ``Larson's Laws'' Using our Approach}

With a few exceptions, most analysis of the structure of the molecular ISM in galaxies at $\lesssim 100$~pc resolution has been carried out through the lens of ``Larson's Laws'' \citep[][among many others]{LARSON81,ROSOLOWSKY03,BOLATTO08,FUKUI10,DONOVANMEYER12,COLOMBO14A,LEROY15A}. Our framework captures most the key physics of this approach as it attempts to describe the structure of the gas. Specifically:

\begin{enumerate}
\item We measure the intensity-weighted surface density, $\Sigma_{\rm \theta pc}$, which is often a main focus of the mass-radius relation for GMCs.
\item The cumulative flux distribution as a function of integrated intensity (or surface density) captures physics closely related to the GMC mass function and cloud filling factor.
\item The parameter $B$ or $\Sigma/\sigma^2$ relates to the dynamical state of the gas and accesses information similar to that revealed by comparing virial mass to luminosity.
\item The fact that gas at ${\sim} 60$~pc scales regulates to a somewhat fixed dynamical state creates a scaling relation between line width and surface density. The position of a region relative to this scaling also serves as a diagnostic of dynamical state.
\item The line width distribution at a fixed size scale (for us, the measurement beam) relates closely to the normalization of the line width-size relation.
\end{enumerate}

Though explored only briefly in this paper, for the LMC and M31, varying the measurement scale and repeating our measurements has the potential to further refine this information, for example revealing a version of the line width-size relation and measuring the clumpiness of structure within the beam.

\subsection{Molecular Gas in Galaxies at 60~pc Resolution}

Our application to a diverse set of six galaxies yields a sketch of the molecular ISM at 60~pc spanning from Local Group dwarf spirals (the LMC and M33) to the nearest major merger (the Antennae) and including massive spiral galaxies where most stars are formed (M31, M51, M74).

Except in the Antennae, the distribution of CO flux, and so presumably mass, as a function of integrated intensity appears to be approximately described by a lognormal. This has not, to our knowledge, been cleanly noted before, but relates closely to the work of \citet{HUGHES13A}, \citet{SAWADA12}, and \citet{ROSOLOWSKY05B}. The width of this distribution appears fairly narrow $1\sigma \approx 0.3 {-} 0.4$~dex in most of our targets, but also displays notable environmental variation in M51 and the Antennae. In the Antennae galaxies, the distribution appears bimodal with the SGMCs showing a distinct distribution from emission in the rest of the overlap region.

We observe significant variations in the intensity-weighted surface density and line width within and among galaxies. To first order, the Local Group galaxies show a narrow range of conditions, while the Antennae galaxies and the spirals M74 and M51 exhibit higher surface densities and line widths. This extends a main result of PAWS shown by \cite{HUGHES13B} comparing the LMC, M33, and M51.

At 60~pc resolution, the molecular gas surface density varies from $\sim 10{-}30$ \mbox{M$_\odot$~pc$^{-2}$} in the Local Group targets, to $\sim 50$ \mbox{M$_\odot$~pc$^{-2}$} in M74, $\sim 160$ \mbox{M$_\odot$~pc$^{-2}$} in M51, and then as high as $\sim 4,000$ \mbox{M$_\odot$~pc$^{-2}$} in the Antennae. Assuming a fiducial (but not known) depth of $60$~pc, these correspond to a range of average volume densities from a few particles per cm$^{-3}$ up to $n_{\rm H2} \approx 1000$~cm$^{-3}$. In addition to a wide range of free-fall times, this implies a wide range of density contrasts with the atomic gas on 60~pc scales across our sample.

The line width, $\sigma$, varies from ${\langle \sigma_{\rm 60pc} \rangle}_{\rm 500pc} \approx 3$--$5$ \mbox{km~s$^{-1}$} in the smaller targets up to ${\sim} 20$ \mbox{km~s$^{-1}$} for the Antennae (though sometimes confused by multiple components). For a fiducial gas temperature of $T = 25$~K and assuming the line widths to be totally turbulent, these values imply Mach numbers $\mathcal{M} \approx 15{-}80$ that vary systematically among our targets.

The sense of the variations in both surface density and line width are that targets with high masses and higher overall gas surface densities tend to also have higher local line widths and surface densities. This supports a good correspondence between ambient ISM pressure and internal pressure in the molecular gas \citep[see][]{HUGHES13B}.

Because surface density and line width correlate, both rising towards higher pressure regions, the dynamical state of the gas at $60$~pc scales, parameterized by $\alpha_{\rm vir,60pc}^{-1}$, appears more constant. We see a relatively narrow range of $\alpha_{\rm vir}^{-1} \approx 0.33{-}1$ in all of our targets except M31. This means that the cold gas in our targets ranges from virialized to marginally unbound at $60$~pc scales, with most data hovering around $\alpha_{\rm vir}^{-1} \approx 0.5$, the marginally bound case. Given the tendency of gas to appear more bound by self-gravity at smaller spatial scales, we interpret our $\alpha_{\rm vir}^{-1}$ measurements to indicate that $\theta \approx 60$~pc may be near the characteristic scale where gas self-gravity becomes dominant.

The cold gas in M31 appears strikingly unbound compared to that in our other targets, reflecting moderately high line widths but relatively low surface densities. This is noted and explored by Schruba et al.\ (in prep.). It could reflect either unbound gas that is part of a turbulent medium and perhaps mixed with the dominant atomic medium. It could also reflect more highly clumped gas in M31 compared to our other targets. Or, despite previous evidence, it could reflect a higher conversion factor in M31 than we adopt here.

These conclusions demonstrate how our approach captures the physical state of the ISM in absolute terms and reveals variations among galaxies. This characterization does depend on the measurement scale. In this paper we choose to characterize the properties of the ISM on $60$~pc scales, but we emphasize that the surface density, line width, and self-gravity all depend on the measurement scale \citep{LEROY13B}. At sharper resolution, the surface density increases, the line width decreases and -- via the combination of these two effects -- self-gravity increases. We show these variations for two of our targets, the LMC and M31. The variations with scale are weaker than expected for a beam-diluted point source or the standard line width-size relation, indicating extended structure beneath our resolution. Self-gravity increases dramatically in both targets at higher resolution, such that many regions that are unbound when averaged over $60$~pc scales appear bound when viewed at $30$~pc scales.

\subsection{Future Applications}
\label{sec:future}

Our method is designed to test hypotheses that link cloud-scale conditions to large scale conditions in a galaxy and the time-averaged efficiency of processes like star formation and feedback. Though we characterize the structure of the molecular gas in this paper, we defer such an analysis (beyond galaxy-to-galaxy comparisons) to future work. Key first applications include contrasting the hydrostatic pressure to the internal pressure in the molecular gas and measuring the dependence of the star formation efficiency per free-fall time on environment. Strong theoretical expectations exist for both cases and our method is explicitly designed to test these expectations.

Section \ref{sec:multiscale} highlights another key test. Varying the measurement scale, e.g., via progressive convolution of the data, probes the physical state of the gas as a function of scale. This will access the scale at which self-gravity becomes dominant, the clumping of molecular gas, and a version of the turbulent line width-size relation.

Finally, although we have applied our calculations to only CO observations, the method can be naturally applied to high physical resolution {\sc Hi} observations \citep[e.g.,][]{BRAUN12} to probe the state of that gas \citep[e.g., see][]{GOODMAN09,LEROY13B}. It is also readily applicable to numerical simulations, which now regularly predict CO or {\sc Hi} emission from galaxies \citep[e.g.,][]{NARAYANAN12,SMITH14}. Such applications offer the potential for large gain in both directions: observations offer a powerful benchmark for simulations, while simulations should help to calibrate the translation of our observables to physical quantities.

\acknowledgments 

We gratefully acknowledge the ALMA, IRAM, CARMA, and MOPRA facilities and thank the M33 and MAGMA teams for making their data public. This paper makes use of the following ALMA data: ADS/JAO.ALMA\#2011.0.00876.S, 2012.00650. ALMA is a partnership of ESO (representing its member states), NSF (USA) and NINS (Japan), together with NRC (Canada), NSC and ASIAA (Taiwan), and KASI (Republic of Korea), in cooperation with the Republic of Chile. The Joint ALMA Observatory is operated by ESO, AUI/NRAO and NAOJ. The National Radio Astronomy Observatory is a facility of the National Science Foundation operated under cooperative agreement by Associated Universities, Inc.  Based on observations carried out with the IRAM 30-m telescope. IRAM is supported by INSU/CNRS (France), MPG (Germany) and IGN (Spain). 

AE acknowledges partial support from grants PFB \#06, FONDECYT \#1130458 and ACT \#1101. AH acknowledges support from the Centre National d'Etudes Spatiales (CNES).  AU acknowledges support from Spanish MINECO grants AYA2012-32295 and FIS2012-32096.  JMDK gratefully acknowledges financial support in the form of a Gliese Fellowship and an Emmy Noether Research Group from the Deutsche Forschungsgemeinschaft (DFG), grant number KR4801/1-1. JP thanks the CNRS/INSU programme PCMI for support. MQ and SEM acknowledge funding from the Deutsche Forschungsgemeinschaft (DFG) via grants SCHI~536/7-2, SCHI~536/5-1, and SCHI~536/7-1 as part of the priority program SPP~1573 ``ISM-SPP: Physics of the Interstellar Medium''. MQ, SEM, and ES acknowledge financial support to the DAGAL network from the People Programme (Marie Curie Actions) of the European Union's Seventh Framework Programme FP7/2007- 2013/ under REA grant agreement number PITN-GA-2011-289313. ER is supported by a Discovery Grant from NSERC of Canada.


\begin{appendix}

\section{A. Methods}

This appendix steps through methodological details that, while crucial to the implementation of our calculations, may only be of interest to a small subset of readers. In Section \ref{sec:stack}, we detail the spectral stacking technique that we use to circumvent biases due to limited sensitivity. In Section \ref{sec:uncertainties}, we describe our Monte Carlo approach to calculate uncertainties. Section \ref{sec:specresponse} discusses how to account for the finite spectral resolution of radio data; this includes a treatment of channel-to-channel correlation, an effect that is usually neglected in analysis of radio data.

\subsection{Incorporating Spectral Stacking}
\label{sec:stack}

\begin{figure*}
\plottwo{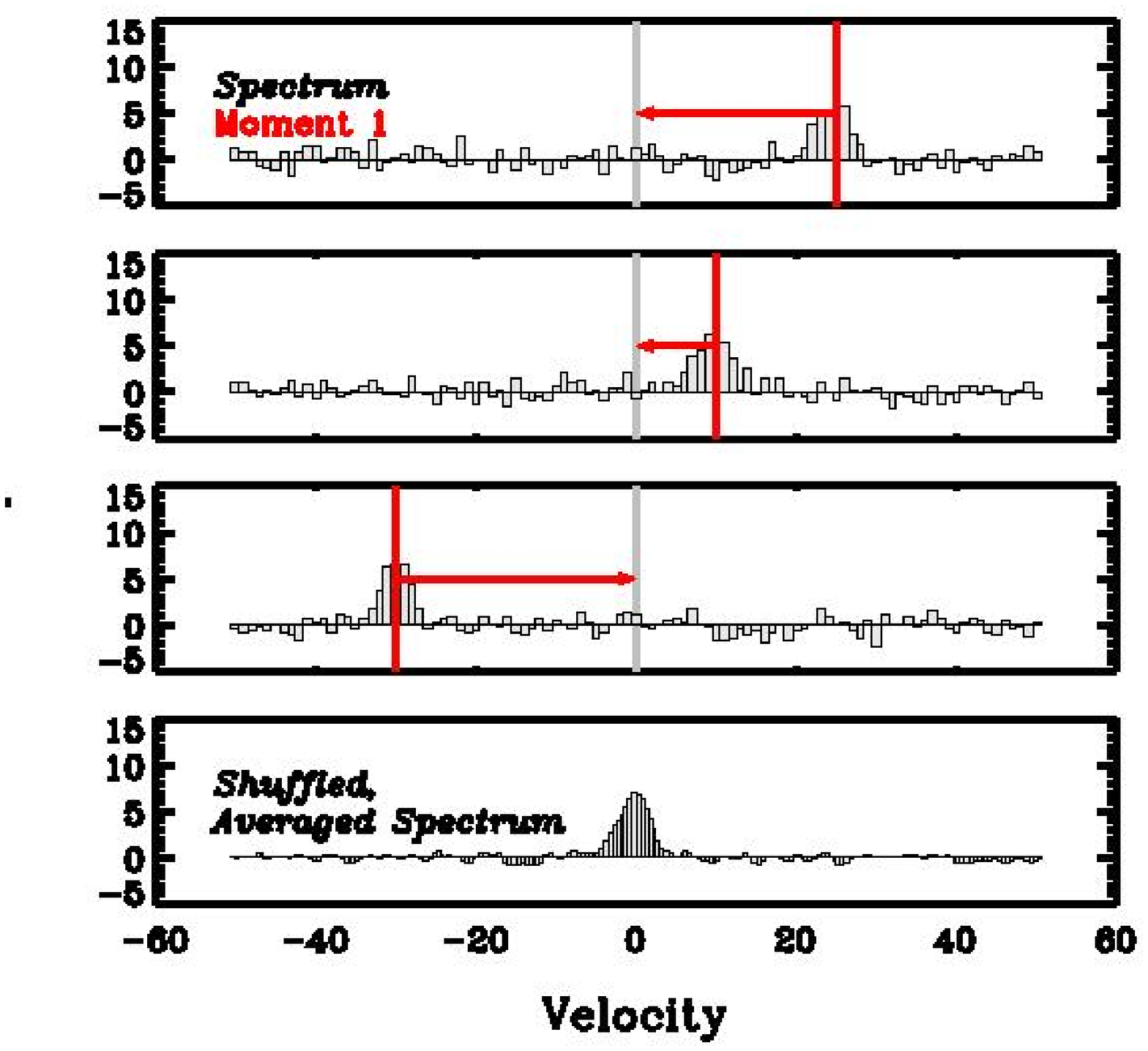}{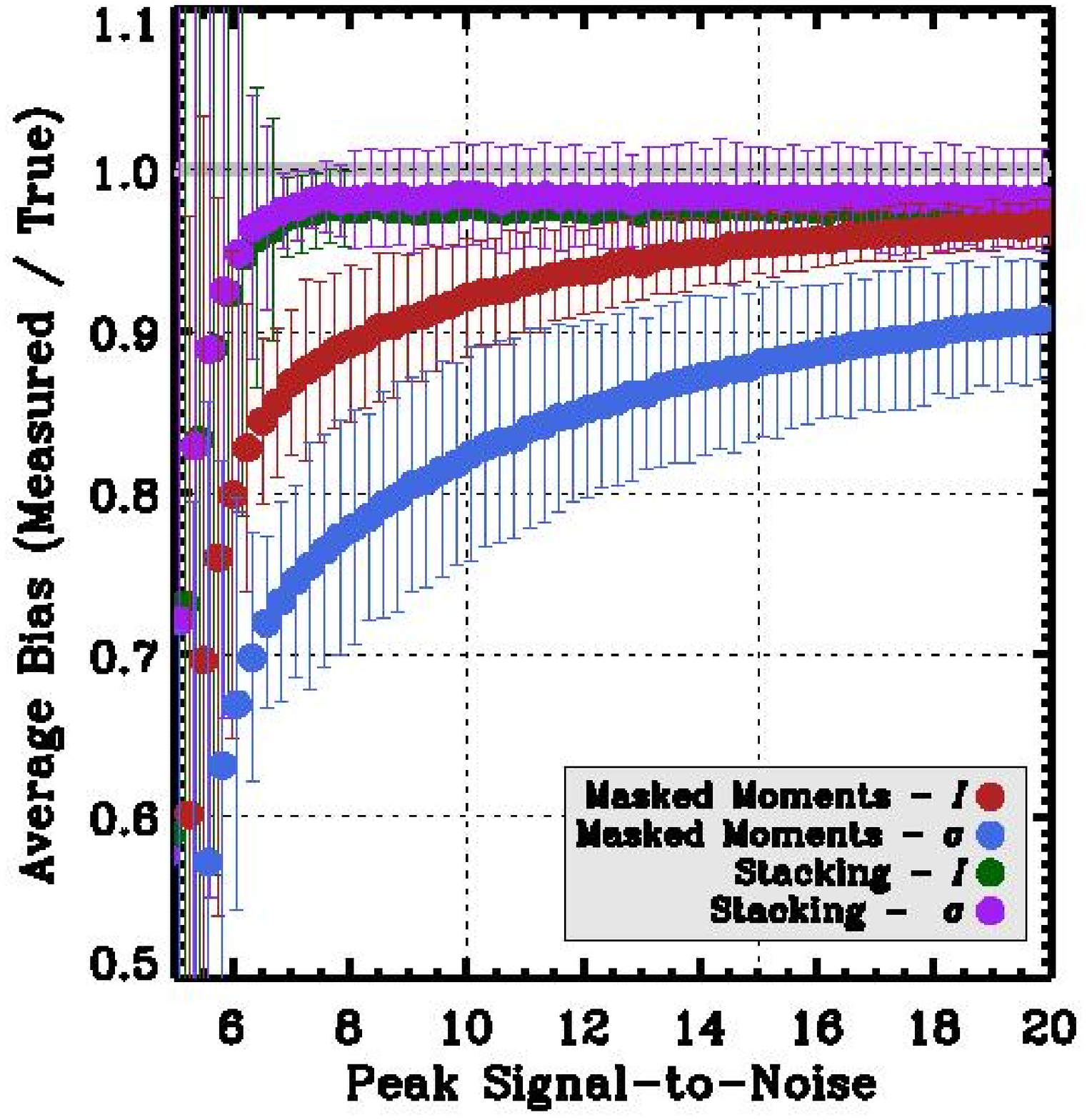}
\caption{({\em left}) Illustration of our stacking (``shuffling'') approach. In this example, we combine three lower signal-to-noise spectra (upper panels) into a shuffled, stacked spectrum (bottom panel). First, we estimate the intensity-weighted mean velocity from the first moment (the red bar). Then, we shift each spectrum to a new velocity axis for which this mean velocity is defined as $v=0$. Finally, we co-add the shuffled spectra, weighting each by its integrated intensity (see text). ({\em right}) The bias, defined as measured over true value, in the recovery of integrated intensity and line width measurements from a Monte Carlo simulation. Results for moments calculated from masked data are indicated in red and blue. Results from a fit to a stack of $25$ spectra co-added after removing their mean velocity (from the first moment) are indicated in green and purple. Individual moment measurements are biased, and these would persist into any average of moment measurements. Thus, the curves can be fairly compared. The error bars show the range of bias for lines with $1\sigma$ line width 2--10 times the channel width.}
\label{fig:bias}
\end{figure*}

We consider two methods to calculate the peak intensity, integrated intensity, and line width at the measurement scale. First, we simply calculate the first three moments of the masked data cube at the measurement scale, yielding the integrated intensity, mean velocity, and second moment. This is a standard approach but it can yield biased results in the case of low signal-to-noise data. Masking is necessary to pick out regions of interest from the mostly empty data cube. This enables the use of the second moment, which diverges in the presence of noise, to estimate the line width. However, the masking suppresses faint, extended emission around the bright peaks. This introduces a bias towards low values into the integrated intensity and second moment measurements. \citet{ROSOLOWSKY06} discuss these biases and propose corrections for them, which are implemented in the CPROPS software. In the current version of CPROPS, clouds are treated as three dimensional Gaussians to calculate a first-order correction to the measured moments based on the ratio of peak intensity to masking threshold for each cloud. Line profile fitting or the use of more robust statistics like the equivalent width (Equation \ref{eq:ew}) may also mitigate these effects.

The right panel of Figure \ref{fig:bias} shows the bias in the integrated intensity and second moment measurements for an individual spectrum. The model spectrum is a Gaussian line profile with noise added and then masked following our standard prescription (an initial mask identifying at least two consecutive channels with $\mathrm{S/N} = $, expanded out to adjacent channels with $\mathrm{S/N} > 2$). Red and blue lines show the bias in line properties measured using moments as a function of peak signal-to-noise\footnote{The error bars indicate the spread in results for different line widths; the significant width of these bars show that the bias depends on both channel width and signal-to-noise \citep[see][]{HEYER01,ROSOLOWSKY06}}. The bias is strongest for the second moment near the signal-to-noise threshold adopted for the mask. Bias in the line width persists even to high signal-to-noise, however, with moment-based measurements of $\sigma$ biased low by $\sim 10{-}30$\%. Integrated intensity measurements, which depend less on recovering emission in the line wings of the line, perform better. These are biased by $\lesssim 10\%$ at about $\mathrm{S/N} \approx 8$.

The green and purple lines in the right panel show results using stacking instead of beam-by-beam moments. The procedure, illustrated in the left panel, is to measure the mean velocity of a spectrum from its first moment. Then the spectrum is interpolated to a new velocity axis with this local first moment defined to be the new $v=0$. These spectra are summed to produce a stacked, high signal-to-noise spectrum. The resulting spectrum is tractable to fitting or statistical analysis, which yield less biased measurements because of the higher signal-to-noise. The use of the first moment to do the shuffling does introduce some statistical uncertainty. However, this is similar to the use of the first moment to compute the second in the moment method. 

In the right panel of Figure \ref{fig:bias}, the green and purple lines show results stacking $25$ independent spectra, each of which has the indicated peak signal-to-noise. The Figure \ref{fig:bias} shows that the stacking results, at least for our simulated noisy Gaussian, recover the line width and integrated intensity with minimal bias. In contrast, the bias seen in the blue and red lines would affect each moment measurement and so persists into the measured average of many moment measurements.

In practice, our stacking methodology resembles that of \citet{SCHRUBA11}, but differs in a few important details. First, we use the first moment of the line itself to stack the data \citep[similar to][]{STILP13} instead of a velocity estimated {\em a priori} from another brighter line; e.g., \citet{SCHRUBA11} used the velocity centroid of the {\sc Hi} emission to stack CO data. This introduces some statistical uncertainty, but does not appear to strongly bias the measurement. The larger difference is that we implement our weighted averaging scheme to aggregate the shuffled data into a stacked spectrum. The procedure is:

\begin{enumerate}
\item Shuffle the original, measurement-scale data cube using the local first moment.
\item Weight each spectrum in the original data cube by the local integrated intensity (zeroth moment).
\item Convolve the weighted cube to the averaging scale.
\item Divide the convolved, weighted cube by the sum of weights, which is the convolved integrated intensity map.
\end{enumerate}

\noindent The result is an integrated-intensity weighted, self-shuffled data cube that is stacked on scales of the averaging beam. This is our preferred methodology for obtaining line width measurements in this paper, because it allows a more detailed characterization of the line profile. However, note that there is information lost in this approach: analysis of the measurement-scale distribution of line widths width still requires beam-wise calculations.

\subsubsection{Comparing Stacking and Moments in Our Data}

\begin{figure}
\plottwo{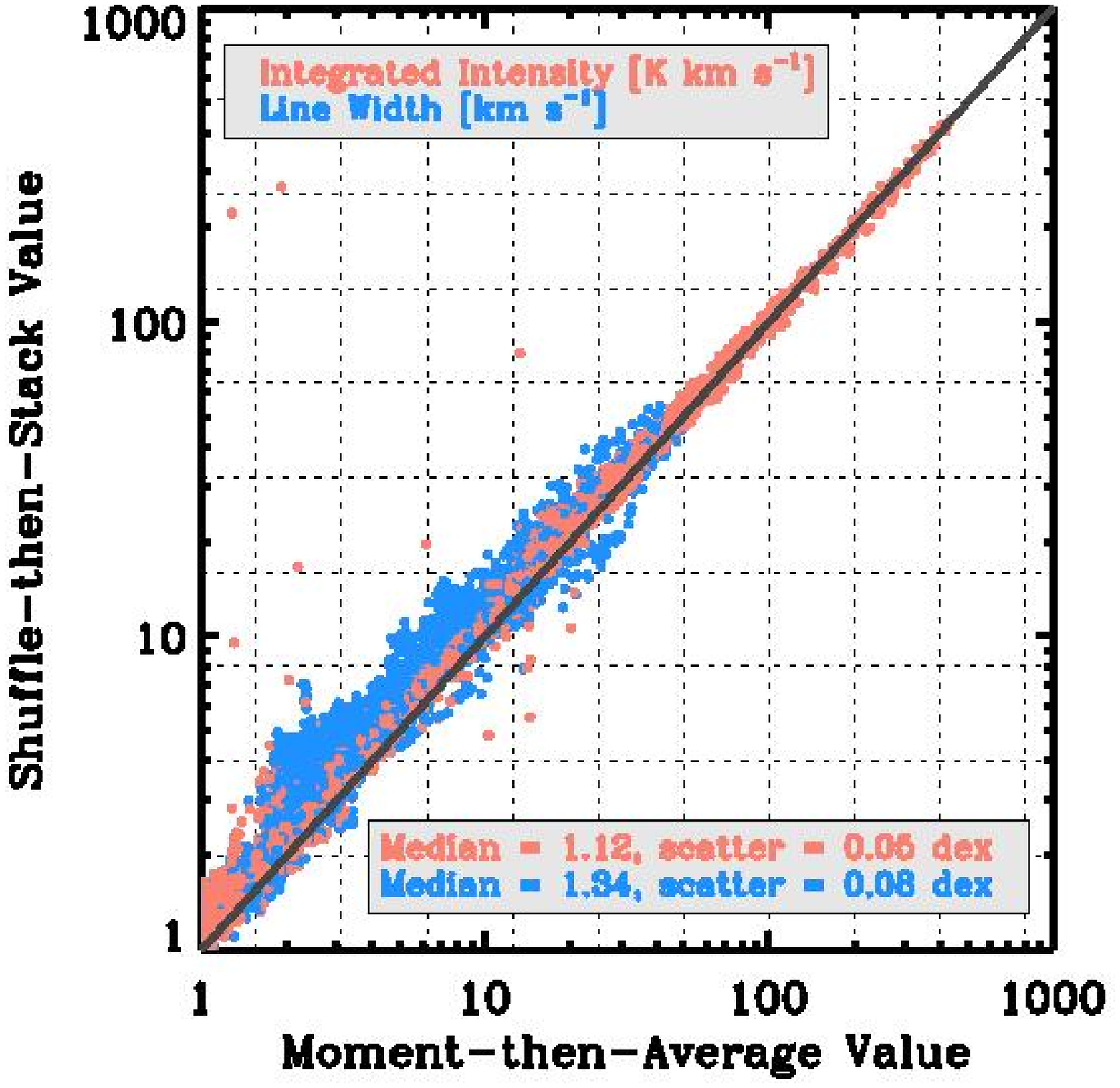}{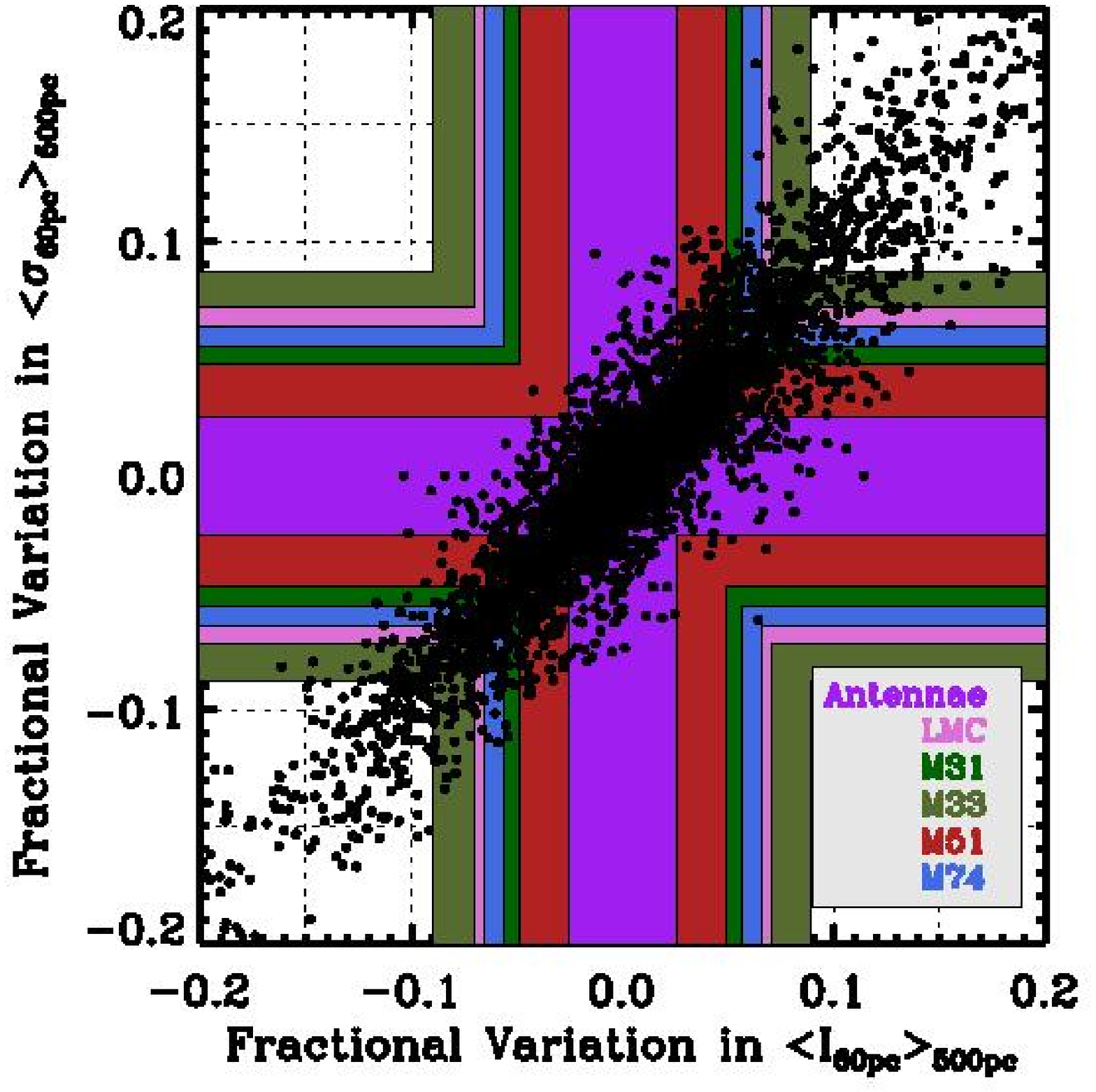}
\caption{({\em left}) Line width (blue) and integrated intensity (red) from intensity-weighted stacking ($y$-axis) and moment methods ($x$-axis) for our six data sets at 60~pc resolution. The two methods agree well, but the stacked results recover larger values on average, especially for the line width. This reflects the sensitivity bias using moment methods, illustrated in Figure \ref{fig:bias}. Methods exist to correct the moments for clipping and sensitivity effects \citep[see][]{ROSOLOWSKY06}, but here we prefer to use the stacking approach, which also yields results that are more tractable to detailed line profile analysis. ({\em right}) Fractional scatter in line width and integrated intensity for our six targets. Colored regions show the rms fractional scatter (defined as scatter in the measurement over the median value) for each quantity and target (we plot the rms scatter times $-1$, too, so that the lines bound the expected $1\sigma$ range). Black dots show results for individual Monte Carlo realizations. These show the strong correlation between the statistical error for $I$ and $\sigma$. We do not expect statistical errors to heavily affect the results in the main text, given the modest values of the uncertainty, but the ability to calculate rigorous statistical errors and covariance among the errors in another advantage of our approach.}
\label{fig:stack_vs_moments}
\end{figure}

The left panel of Figure \ref{fig:stack_vs_moments} compares results for the line width (blue) and integrated intensity (red) from this stacking approach ($y$-axis) to moment-based ($x$-axis) results for our six data sets. The figure reports the median ratio and the scatter in the ratio derived from the two approaches. 

Results from the two methods track one another well over an order of magnitude in line width and three orders of magnitude in integrated intensity. We do measure an offset, which has the sense expected from our discussion of sensitivity biases in the previous section: the stacked results recover higher values than the moments, with a larger discrepancy for the line width than for the integrated intensity. The sensitivity bias of moment-based analysis has been recognized before, with previous works addressing the issue via curve-of-growth corrections or leveraging the peak-to-edge ratio and an assumed Gaussian geometry \citep[see][]{ROSOLOWSKY06}. Here, we advocate the simpler solution of shuffling and weighted stacking. This recovers the true values for a model Gaussian line well, as long as enough emission is detected to estimate the intensity-weighted mean velocity for stacking.

\subsection{Calculation of Uncertainties}
\label{sec:uncertainties}

We adopt a Monte Carlo approach to estimate uncertainties. For a given data set at a given measurement resolution, we identify bright signal and estimate the noise from signal-free regions (including spatial variations, if present). We then add a new realization of normally distributed noise to the masked data. This creates a simulated data set in which the masked data are the true value by construction. We realize a succession of such data sets. For each, we measure the intensity-weighted quantities of interest described above. The scatter among the measurements of this mock data set provides a reasonable estimate of the statistical uncertainties in the quantities of interest. Moreover, the Monte Carlo treatment naturally yields an estimate of the covariance in the uncertainties among the quantities of interest. This is important given the intrinsic correlation, for example, between $I$ and $\sigma$. 

This approach embeds an approximation: that the masked real data (which includes noise) is the true intensity distribution. Therefore the uncertainties that we derive should strictly be interpreted as the simulated uncertainties for a model intensity distribution equal to the real data. In this paper, we neglect this difference and consider our simulated uncertainties to be a good representation of the true ones.

We consider normally distributed noise that is spatially correlated by the nominal beam of the telescope. In principle, one could generalize this approach to a truly rigorous treatment by beginning the Monte Carlo treatment in the $u-v$ data and raw single dish data. At this point, it is not clear that adding this degree of rigor is needed. The uncertainty in the translation from observed emission properties to physical quantities instead appears to represent a much larger obstacle to rigorous hypothesis testing.

We note that radio and millimeter-wave (mm-wave) astronomy still lacks a universally accepted treatment of the spectral response in a data set. By ``spectral response'', we mean the spectral shape of a delta function emission feature, i.e., the spectral ``beam'' of the data, similar to  the ``line spread function'' for optical data. Most analyses assume independent successive channels. However most real correlators and data processing pathways introduce non-zero channel-to-channel correlation in a given spectrum. This effect needs to be quantified and treated in a field-standard way for rigorous uncertainty calculations in the cube domain to be easily tractable. We describe our approach, which is approximate but an improvement over no treatment, in Section~\ref{sec:specresponse}.

Our approach to simulate noise with appropriate intrinsic correlation is:

\begin{enumerate}
\item Measure the amplitude of the noise from the signal-free region of the real data cube.
\item Measure the channel-to-channel correlation in each cube via a linear correlation coefficient.
\item Generate a cube of normally distributed random data with unity amplitude.
\item Convolve each plane of the random, normally distributed data with the beam of the real data cube.
\item Convolve each spectrum in the random, normally distributed data with a kernel designed to produce the appropriate channel-to-channel correlation.
\item Renormalize the amplitude of the simulated noise cube so that it has the same rms scatter about zero as the real data cube with appropriate correlations.
\end{enumerate}

\noindent Then we add the simulated noise to the masked real data cube. Repeating this process, we take the scatter and covariance in the measured results to represent realistic uncertainties for our real measurements.

After averaging among large areas over high quality data, systematic rather than statistical uncertainties often dominate the error budget. Calibration uncertainties are easy to treat; for mm-wave data these are usually $\sim 10{-}15\%$ and reasonably simulated as a lognormal distribution that applies once to each data set (i.e., they are multiplicative and more or less 100\% covariant across a single cube). Uncertainties in image reconstruction, baseline subtraction, and other aspects of calibration remain harder to treat quantitatively; the combination of a Monte Carlo treatment and automated image reconstruction algorithms offers an appealing route forward, but one that lies beyond the scope of this paper.

\subsubsection{Uncertainties and Covariance in Our Data}

\begin{deluxetable*}{lcccccccccc}[t!]
\tabletypesize{\scriptsize}
\tablecaption{Uncertainties in Ensemble Averages From Our Data \label{tab:unc}}
\tablewidth{0pt}
\tablehead{
\colhead{Galaxy} & 
\multicolumn{3}{c}{Median uncertainty in ...}  &
\multicolumn{3}{c}{Fractional uncertainty in ...}  &
\multicolumn{3}{c}{Covariance in uncertainty ...}  \\
\colhead{} &
\colhead{$I_{\rm \nu, pk}$} &
\colhead{$I$} &
\colhead{$\sigma$} &
\colhead{$I_{\rm \nu, pk}$} &
\colhead{$I$} &
\colhead{$\sigma$} &
\colhead{$I_{\rm \nu, pk}$ and $I$} &
\colhead{$I_{\rm \nu, pk}$ and $\sigma$} &
\colhead{$I$ and $\sigma$} \\
\colhead{} &
\colhead{[K]} &
\colhead{[K~km~s$^{-1}$]} &
\colhead{[km~s$^{-1}$]} &
\colhead{} &
\colhead{} &
\colhead{} &
\colhead{} &
\colhead{} &
\colhead{} 
}
\startdata
Antennae & 0.007 &  0.47 &  0.39 & 0.014 &  0.02 &  0.02 &  0.27 & -0.28 &  0.68 \\
LMC & 0.001 &  0.02 &  0.22 & 0.029 &  0.07 &  0.07 &  0.25 & -0.19 &  0.78 \\
M31 & 0.002 &  0.06 &  0.24 & 0.019 &  0.06 &  0.05 &  0.22 & -0.13 &  0.86 \\
M33 & 0.001 &  0.04 &  0.33 & 0.026 &  0.09 &  0.09 &  0.22 & -0.06 &  0.81 \\
M51 & 0.015 &  0.92 &  0.43 & 0.018 &  0.05 &  0.05 &  0.21 & -0.12 &  0.82 \\
M74 & 0.005 &  0.13 &  0.23 & 0.026 &  0.06 &  0.06 &  0.25 & -0.14 &  0.83 

\enddata
\tablecomments{Based on 100 Monte Carlo realizations of the data, as described in the text.}
\label{tab:unc}
\end{deluxetable*}

Table \ref{tab:unc} and the right panel of Figure \ref{fig:stack_vs_moments} present the statistical uncertainties in our data based on a Monte Carlo calculation. We report both absolute and fractional uncertainties. The fractional uncertainties range from $\sim 1{-}10\%$. The low values should not be surprising given that our calculations aggregate data with good signal-to-noise. The magnitude of the uncertainties places the statistical uncertainties in the same range as systematic uncertainties in flux calibration ($\sim 5{-}15\%$). Although unquantified, uncertainties due to limited $u-v$ coverage and image reconstruction algorithms (combination of single dish and interferometer, error beam treatment, etc.) likely contribute uncertainties at the same level.

The uncertainties on $I$, $\sigma$, and $I_{\rm \nu, pk}$ are not independent. The right panel in Figure \ref{fig:stack_vs_moments} plots contours of the fractional offset from the median value in line width ($y$) and integrated intensity ($x$) for each point and Monte Carlo realization. The two are strongly correlated, as one would expect from the definition of $I$. As a result, a statistical fluctuation leading to a large line width is also likely to yield a large integrated intensity. Conversely, because our definition of line width uses the peak intensity to define the equivalent width, a scatter to high $I_{\rm \nu, pk}$ leads to a lower $\sigma$, though the effect is more modest. The last three columns of Table \ref{tab:unc} report the correlation between offsets from the median value across our set of Monte Carlo iterations. Though the statistical uncertainties in our current analysis are modest, these correlated uncertainties can be important to the interpretation of scaling relations in more marginal cases. 

\subsection{Treatment of Spectral Response} 
\label{sec:specresponse}

\begin{figure*}
\plottwo{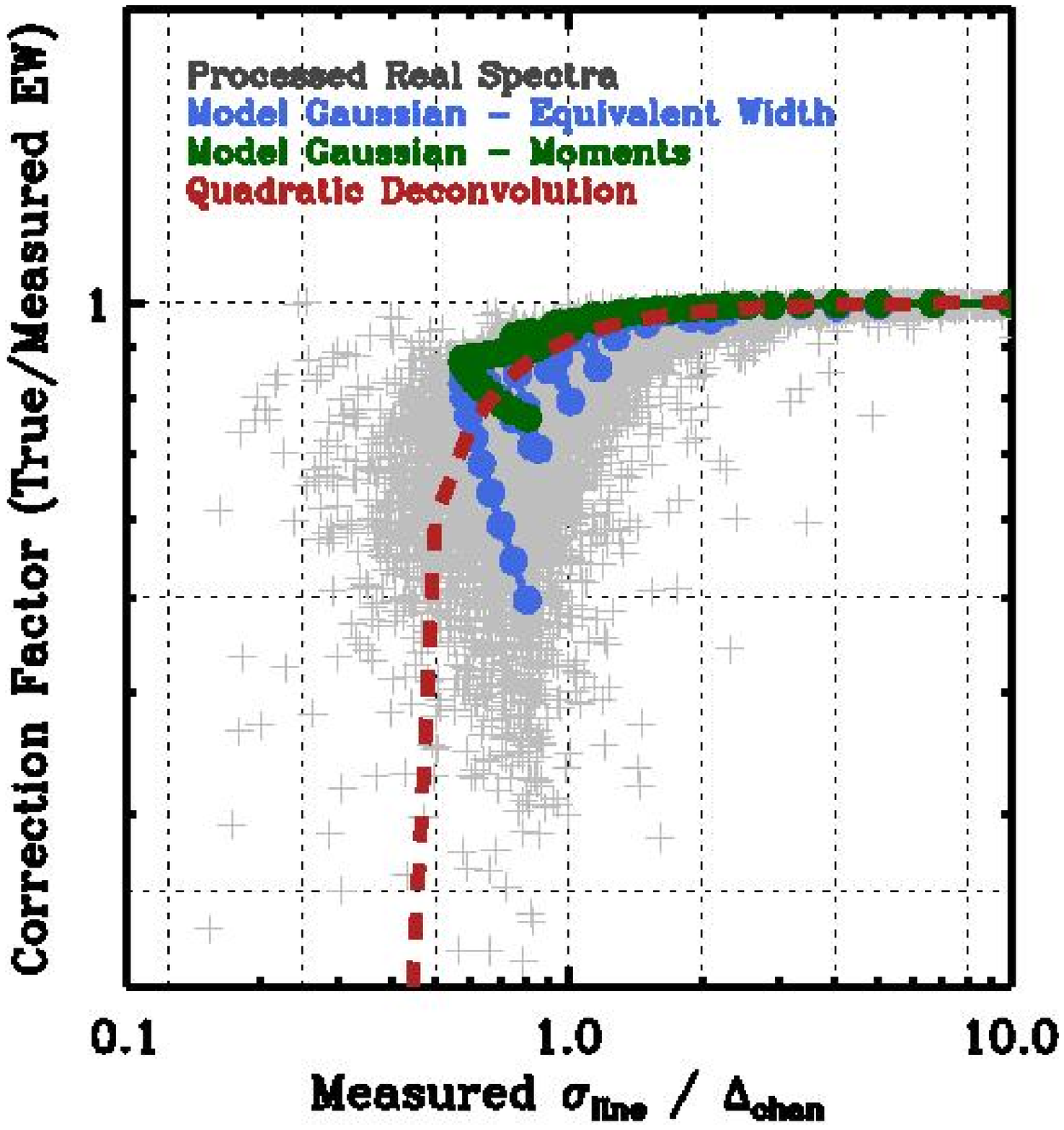}{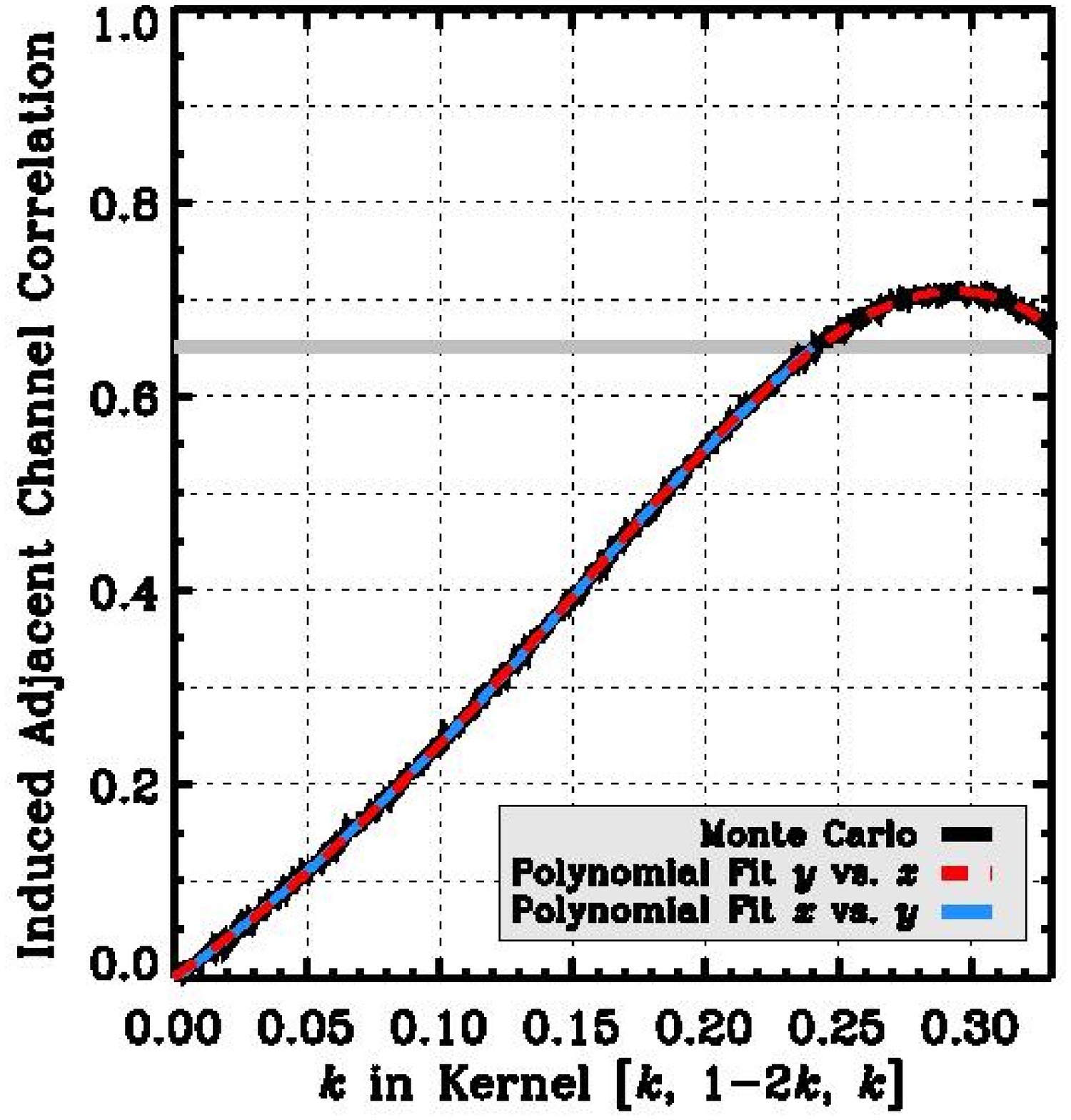}
\caption{({\em left}) Bias in the measured line width due to finite channel width, estimated from real spectra (gray points), model Gaussian lines (blue and green lines), and a simple quadratic prescription (red). The $y$-axis shows the correction factor that must be applied to the measured line width to match the known, true input. The $x$-axis shows the ratio between the measured line width, expressed as the rms dispersion $\sigma$, and the channel width. Corrections become large and unstable when the measured line width approaches a channel width. We adopt this as our cutoff, below which we consider a linewidth measurement to be an upper limit, $\sigma < \Delta v_{\rm chan}$. ({\em right}) The linear correlation coefficient, $r$, between noise in successive channels as a diagnostic of the spectral response beyond a channel width. Our simple treatment models the spectral response as a three-element normalized kernel of the form $[k, 1-2k, k]$ (so that $k=0.25$ is the Hann kernel). The plot shows the induced channel-to-channel noise correlation for different values of $k$. We use a polynomial fit to values $r < 0.65$ and values of $r$ measured from signal-free regions of the real data to model a spectral response beyond the channel width in our Monte Carlo treatment.}
\label{fig:specresponse}
\end{figure*}

The spectral response affects the measured line width, so that the measured line profile is the convolution of the true line profile shape with the channel profile. CPROPS \citep{ROSOLOWSKY06} uses subtraction in quadrature, approximating the channel as a Gaussian with the same equivalent width (Equation \ref{eq:ew}) as the tophat channel. In the case of known channel width, channel-to-channel coupling ($k$), and a good model for the line shape (e.g., a Gaussian), an alternative would be to include the spectral response in forward modeling the line. 

We adopt the deconvolution-in-quadrature approach, modified to account for channel-to-channel correlation (see below), but only in the regime where corrections are modest. We identify this regime by modeling the effects of finite channel width on a large set of real spectra. From our spectral stacking of six galaxies below, we have many high signal-to-noise line profiles from a diverse set of nearby galaxies. We take the observed, uncorrected line profiles, resampled to a very fine grid, as a reasonable family of templates (note that this is an approximation, similar to treating the data as the true value in the Monte Carlo calculation). We channelize these spectra and re-measure the line width of the data, varying the channel width from small to large values. We also carry out the same calculation for a series of model Gaussian line profiles. From these channelized, model lines we measure the new, degraded line width using both moment methods and the equivalent width. We compare this to the same statistic obtained for the line before channelization. The result is a correction factor that would need to be applied to the measured line width after channelization to recover the true line width before channelization. We plot this factor as a function of the channel width divided by the measured line width in Figure \ref{fig:bias}.

The figure shows that corrections become unstable and substantial below $\Delta v_{\rm chan} \approx \sigma$, where $\sigma$ is the measured line width before any correction. Corrections are small for lines better resolved than this. We therefore adopt this as a practical cutoff for measuring the line width. We suggest to treat narrower lines as having an upper limit $\sigma < \Delta v_{\rm chan}$. Equation \ref{eq:deconv} gives our formula for deconvolution in quadrature while accounting for any channel-to-channel correlation, which broadens the spectral response.

Note that with precise knowledge of the functional form of the line profile, the spectral response, and sufficient signal-to-noise, one could derive the width of spectral features much narrower than the channel width. The spectral response of most current mm-wave facilities does tend to be well-characterized as part of the development of the instrument \citep[e.g., the spectrometers at the IRAM 30-m have a well-characterized noise-equivalent bandwidth, e.g.,][]{KLEIN12}. However, given the diversity of observed line shapes in galaxies, our view is that the functional form of the astrophysical line profile is not sufficiently well-known for a forward modeling approach to work far below the resolution of the instrument.

\subsubsection{Channel-to-Channel Correlation}

Real mm-wave data seldom have perfectly independent spectral channels, reflecting the response of telescope backends, the common practice of Hanning smoothing, and frequent interpolation during data processing. Accounting for this correlation is important to model uncertainties and to measure the width of marginally resolved spectral lines. We adopt an {\em ad hoc} approach to this problem, modeling the spectral response outside an individual channel with a normalized three-element kernel similar to a Hann kernel but with variable magnitude, i.e., 

\begin{equation}
\label{eq:kern}
\left[ k, 1-2k, k \right]~.
\end{equation} 

Convolution with such a kernel will introduce a channel-to-channel correlation in the noise in a data set. This is measurable via the linear correlation coefficient, $r$, between noise in successive spectral channels. That is, in a signal free part of the data cube, we measure the correlation of the intensity in channel $n$ with the intensity at the same spatial position but channel $n+1$ (or equivalently $n-1$). For independent spectral channels in a signal-free region of the cube, we expect $r = 0$.

Figure \ref{fig:specresponse} shows the magnitude of channel-to-channel correlation, $r$, induced by a kernel specified according to Equation \ref{eq:kern}. Below $r \approx 0.65$, a measured $r$ can be inverted to yield $k$. This kernel can then be used to treat uncertainties or estimate line widths. Based on the numerical calculations shown in Figure \ref{fig:specresponse}, the following yields $k$ with $\sim 1\%$ accuracy:

\begin{equation}
k \approx 0.0 + 0.47 r - 0.23 r^2 - 0.16 r^3 + 0.43 r^4~.
\end{equation}

Our experience with real mm-wave data (including those used below) is that $r \approx 0.15$ to $0.4$ for real data. This should be expected from the common practice of Hanning smoothing and then downsampling the data by a factor of $2$, which yields $r \approx 0.15$. Complex instrumental profiles, additional interpolation and smoothing of the data can add further complications.

From this approach, we have an estimate of the spectral response that we have bootstrapped from data that we can use during the Monte Carlo treatment or for line width analysis. The adopted kernel is easily extensible beyond adjacent channels; however, our exploration of the data considered in this paper does not suggest strong correlations between channel $n$ and $n \pm 2$. Therefore we treat the spectral response as a three element kernel (Equation \ref{eq:kern}).

Channel-to-channel correlation implies a broader spectral response than only the channel width, which should be accounted for in the estimate of the astrophysical line width. Following \citet{ROSOLOWSKY06}, we account for broadening by deconvolving the measured line width by the equivalent width of the spectral response, expressed as the rms of an equivalent-area Gaussian. They considered the channel width, $\Delta v_{\rm chan}$. We extend this to an effective spectral response width

\begin{equation}
\label{eq:effchan}
\sigma_{\rm response} = \frac{\Delta v_{\rm chan}}{\sqrt{2 \pi}} \times \left( 1.0 + 1.1 8k + 10.4 k^2 \right)
\end{equation}

\noindent where the term in parentheses (which is unity for $k=0$) accounts for the broadening of the response by a kernel specified by Equation \ref{eq:kern} in the range $k \in [0., 0.25]$. We then correct the measured line width via

\begin{equation}
\label{eq:deconv}
\sigma_{\rm true} = \sqrt{\sigma_{\rm measured}^2 - \sigma_{\rm response}^2}
\end{equation}

Note that this correction is required even for $\sigma$ estimated from the equivalent width because the spectral response ``dilutes'' the peak intensity, $I_{\rm peak}$, in Equation \ref{eq:ew}. To first order (within ${\sim} 10\%$), the dilution yields the same apparent broadening of the line as Equation \ref{eq:effchan}.

\section{B. Atlas of CO Structural Properties}

Though the main text focuses on statistical distributions, our analysis returns spatial information, suitable for cross-correlation with environmental metrics. That is, our analysis involves the construction of maps of average cloud-scale properties. We defer a detailed analysis of the spatial distribution of molecular gas properties within galaxies to future work, but we present these data as maps in Figures \ref{fig:atlas_ant} -- \ref{fig:atlas_m74}. These maps motivate our statements that the inner region of M31 differs from the star-forming ring, the LMC ridge from the main body of the galaxy, and the Antennae SGMCs from the rest of the galaxy. Radial variations are also evident in M33 and M51.

\begin{figure*}
\plottwo{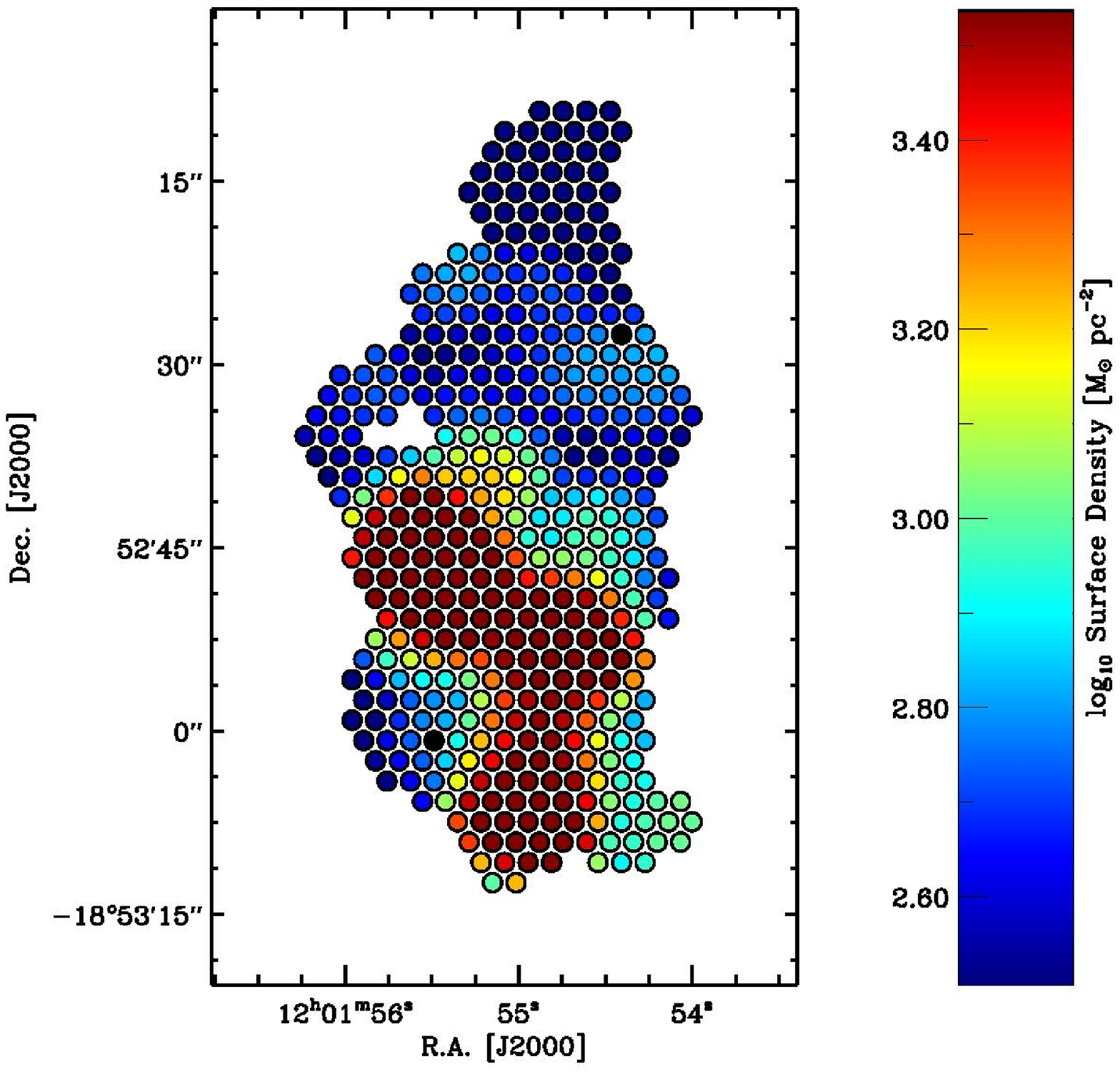}{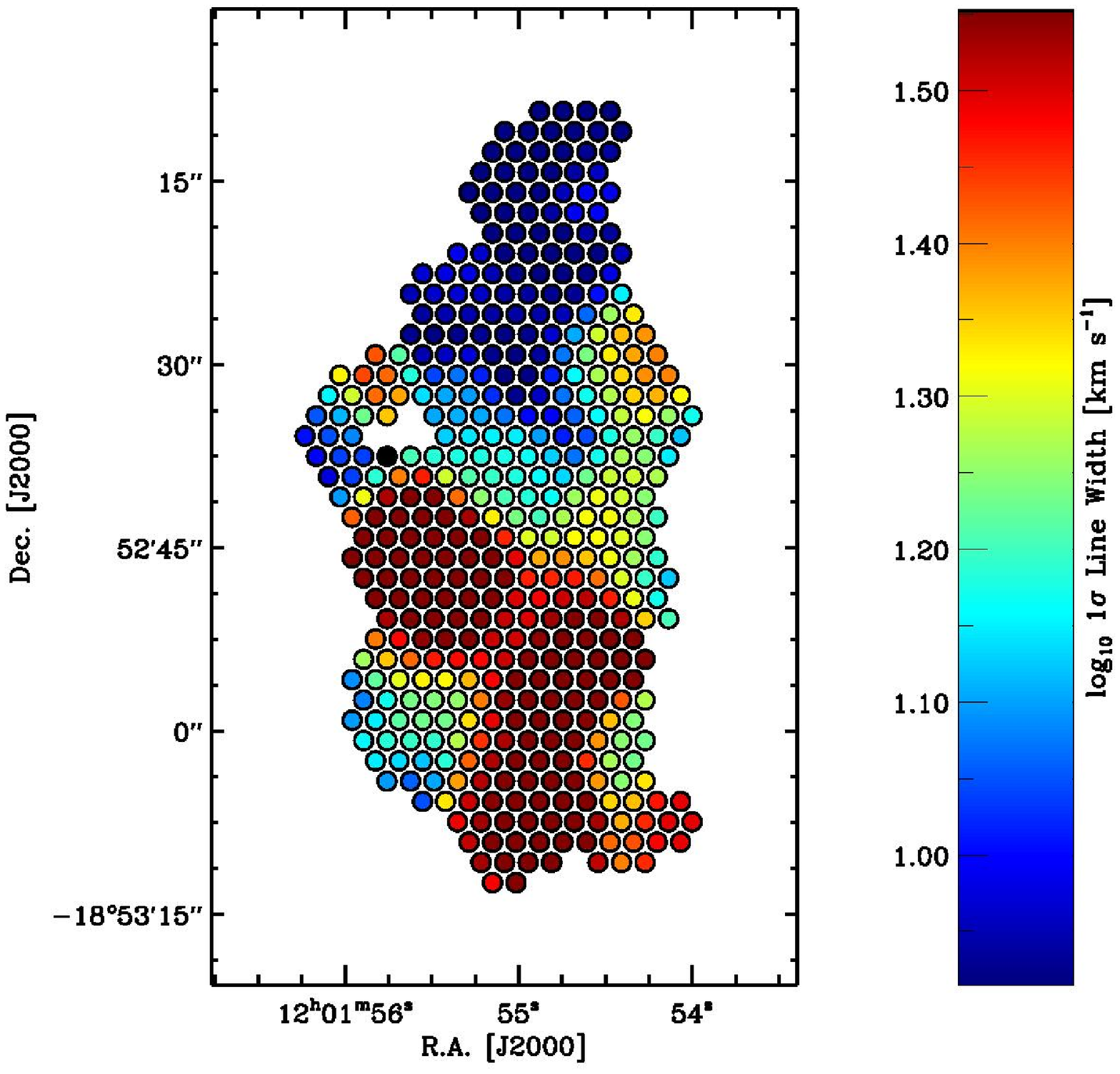}
\plottwo{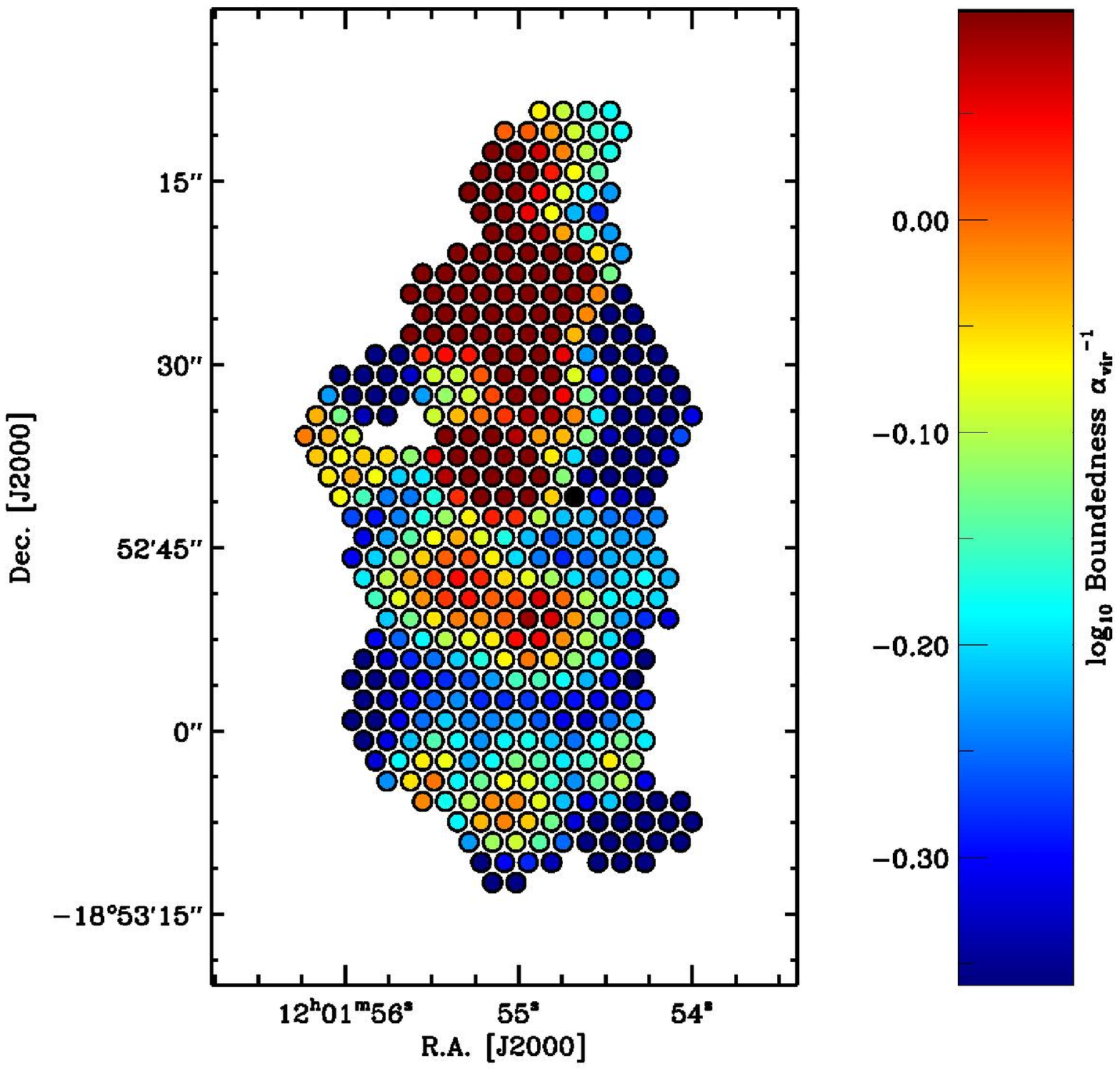}{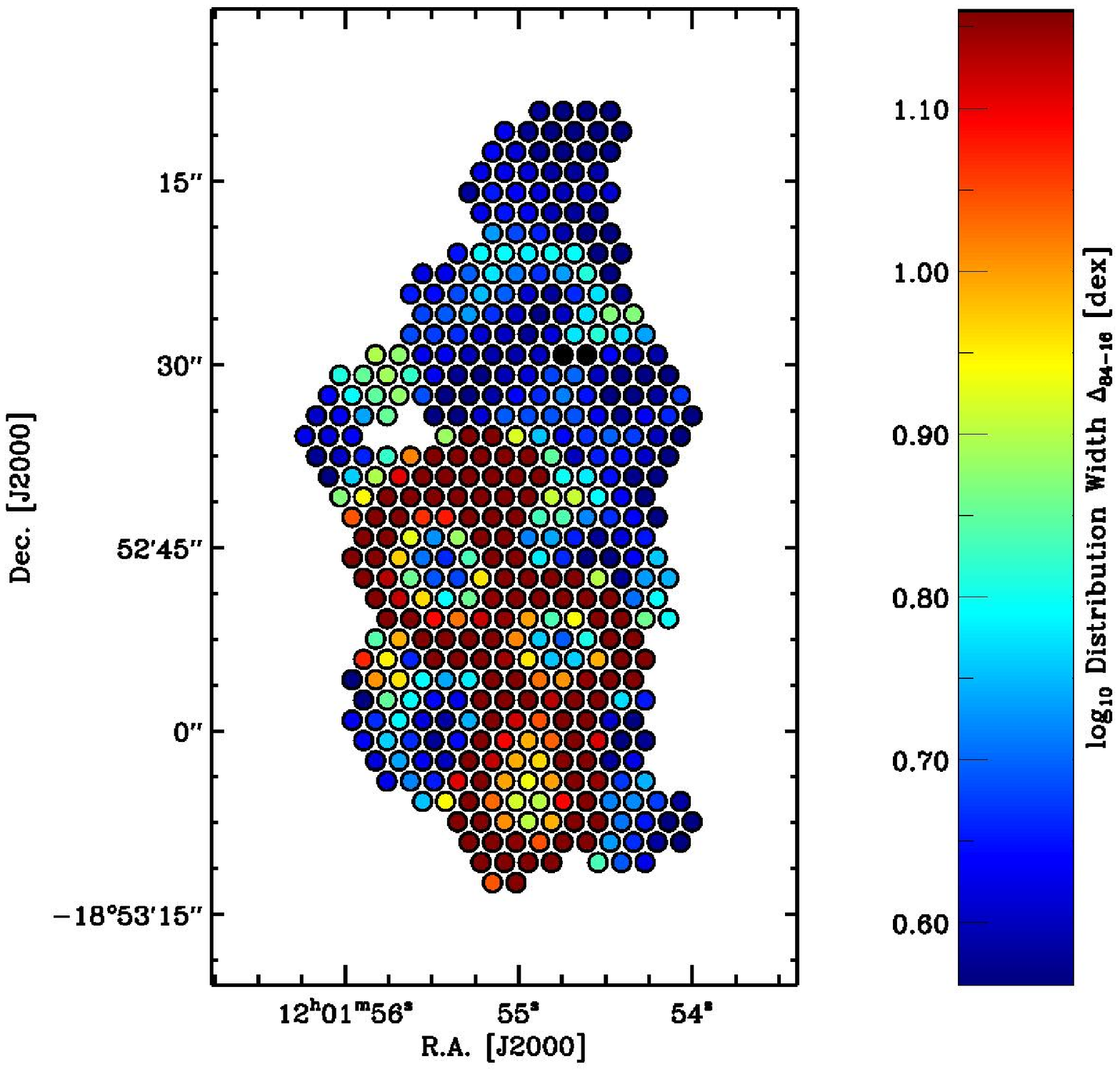}
\caption{Atlas images of the intensity-weighted average  cloud-scale surface density (top left), line width (top right) and boundedness (bottom left) of molecular gas in the Antennae. The averages are measured within 500~pc apertures. The logarithmic $68\%$ width, $\Delta^{84-16}$, of the CO integrated intensity distribution within each 500~pc region is also shown (bottom right).}
\label{fig:atlas_ant}
\end{figure*}

\begin{figure*}
\plottwo{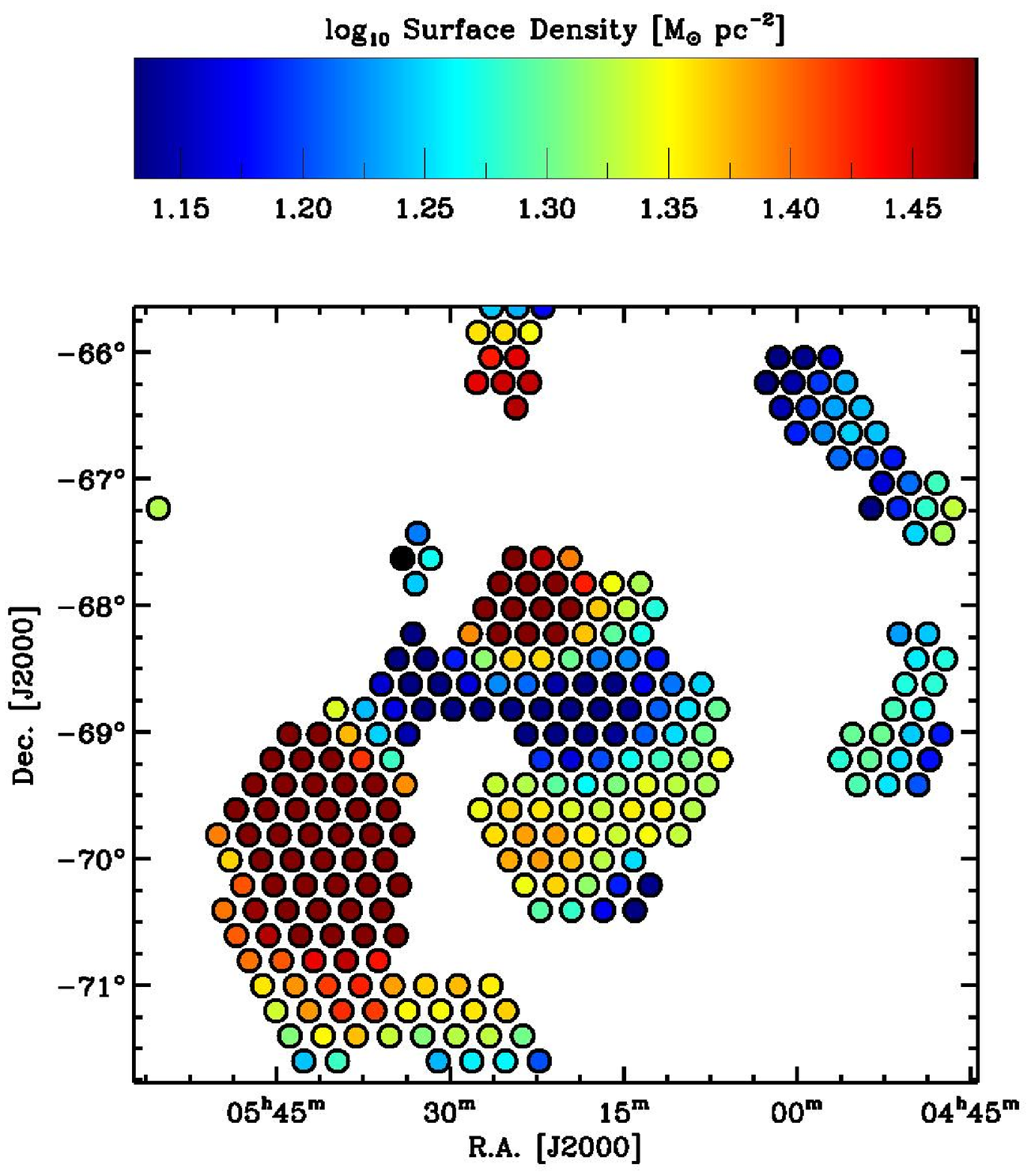}{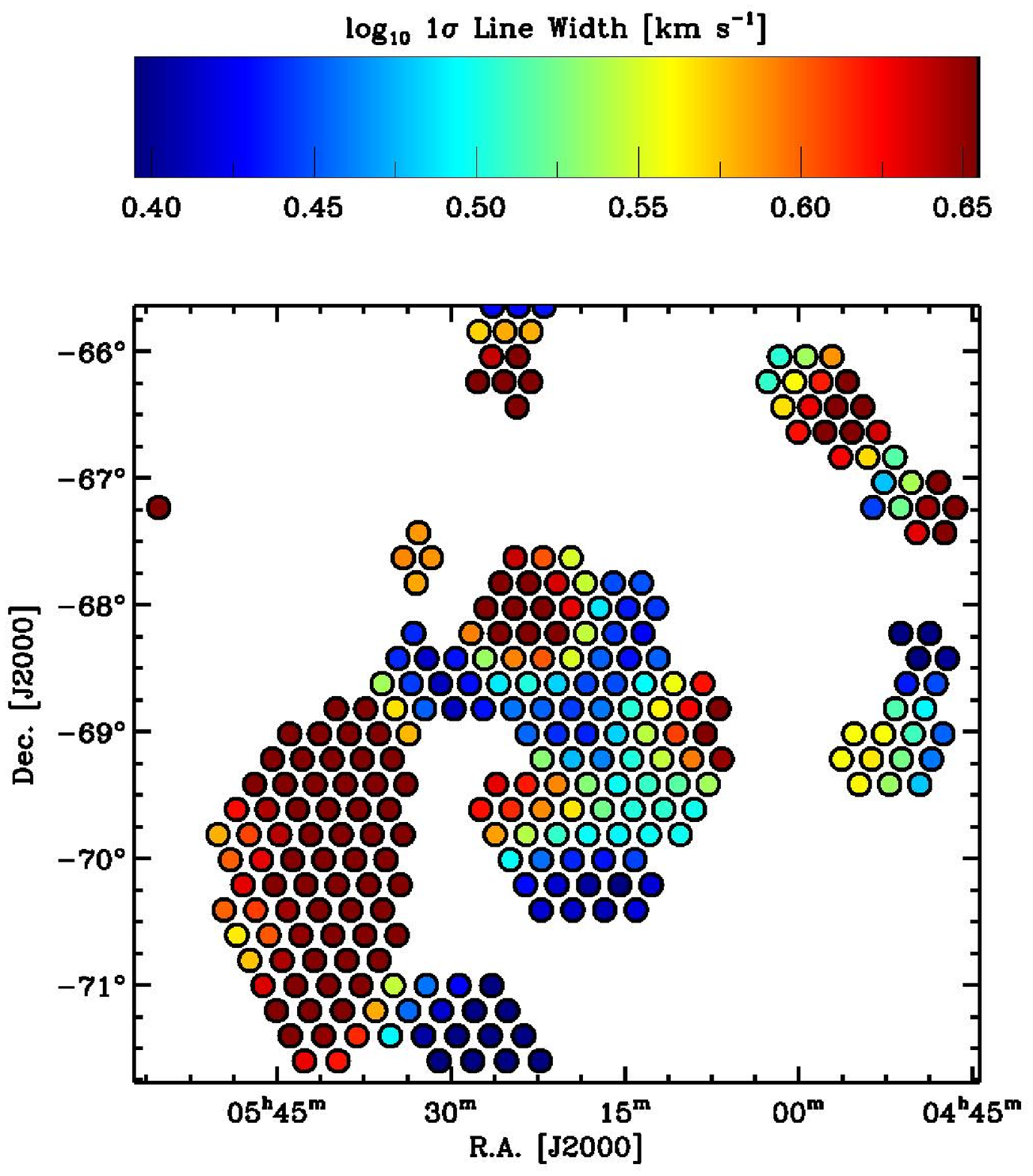}
\plottwo{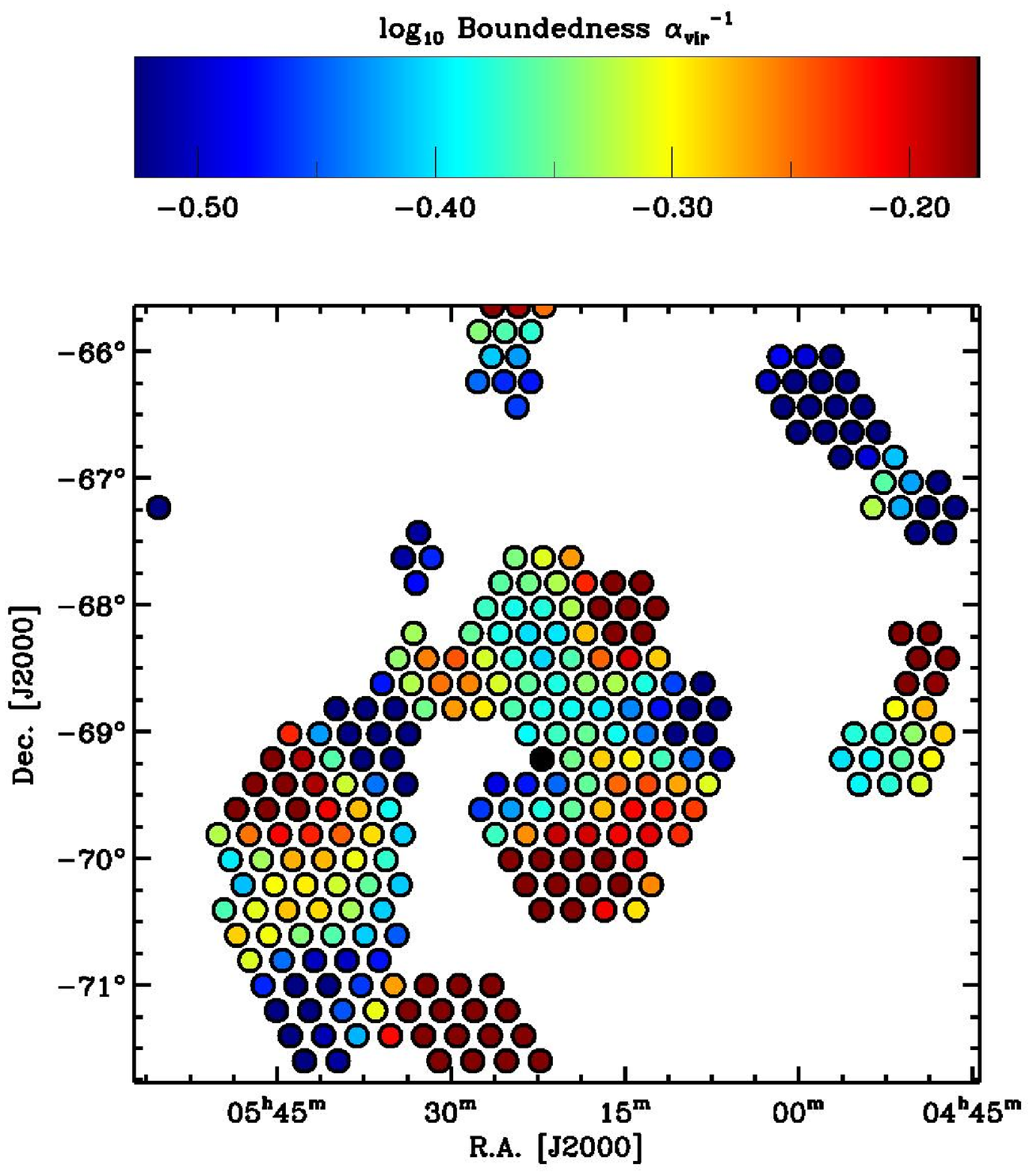}{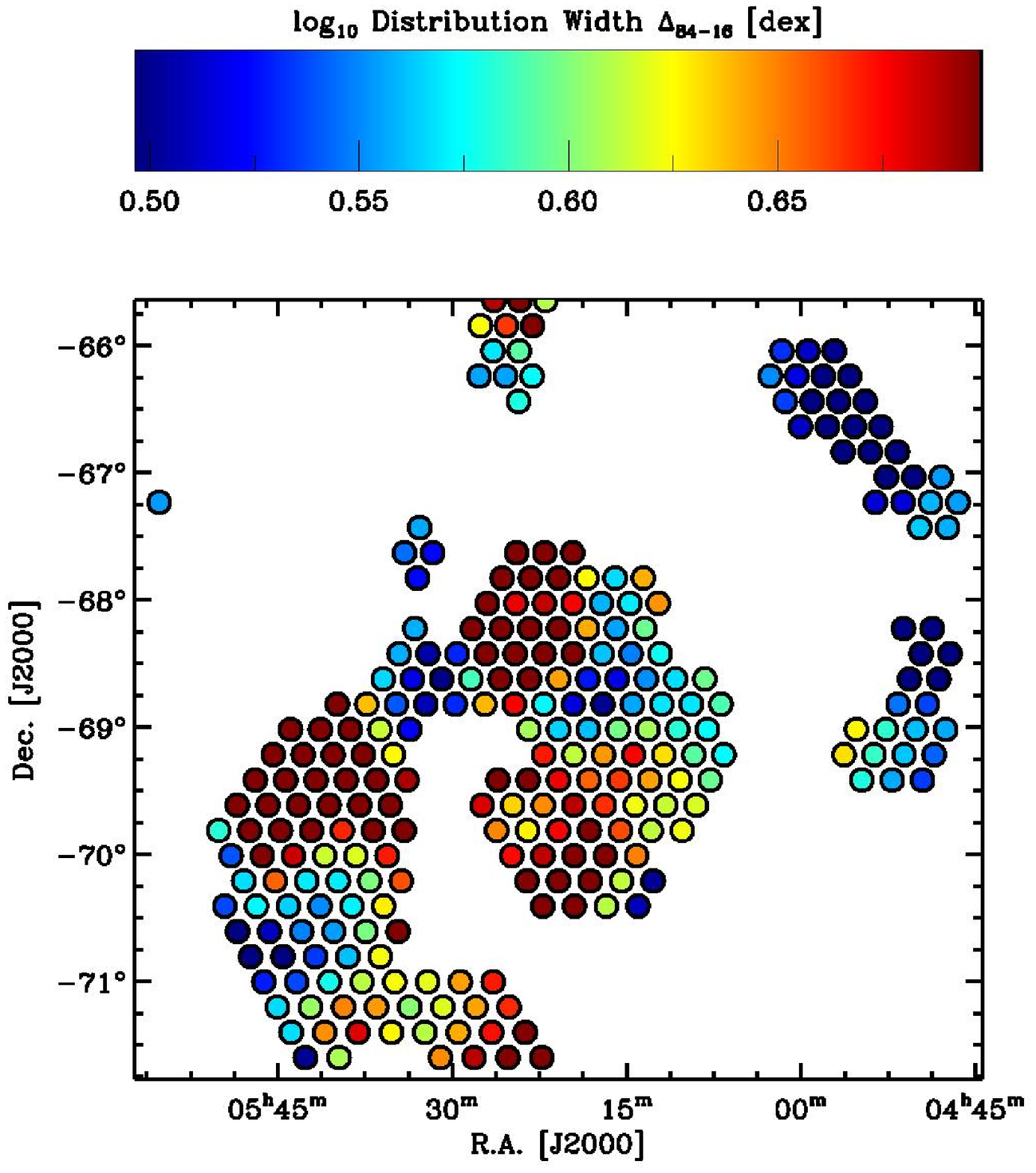}
\caption{Same as Fig.~\ref{fig:atlas_ant}, but for the Large Magellanic Cloud.}
\label{fig:atlas_lmc}
\end{figure*}

\begin{figure*}
\plottwo{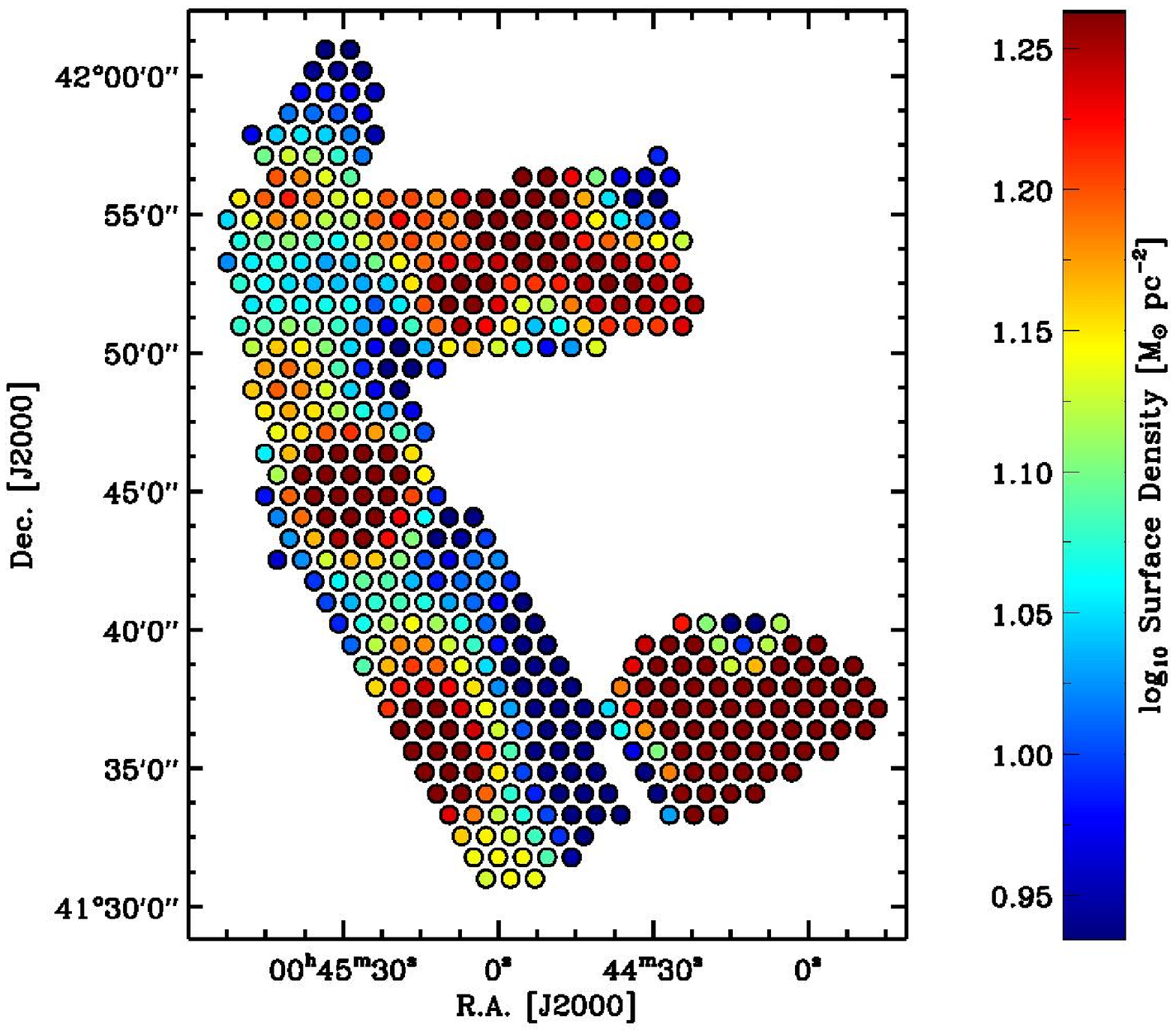}{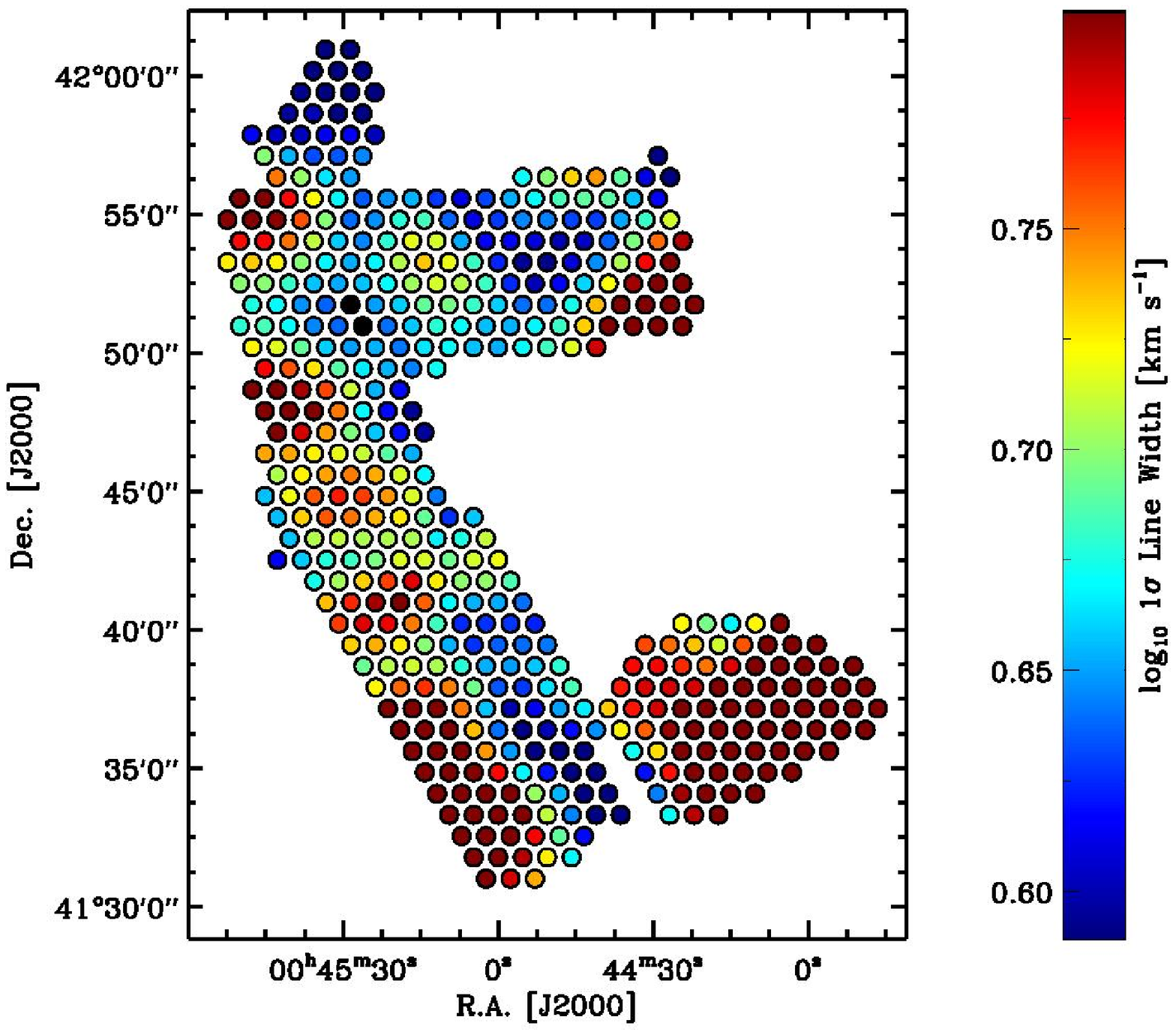}
\plottwo{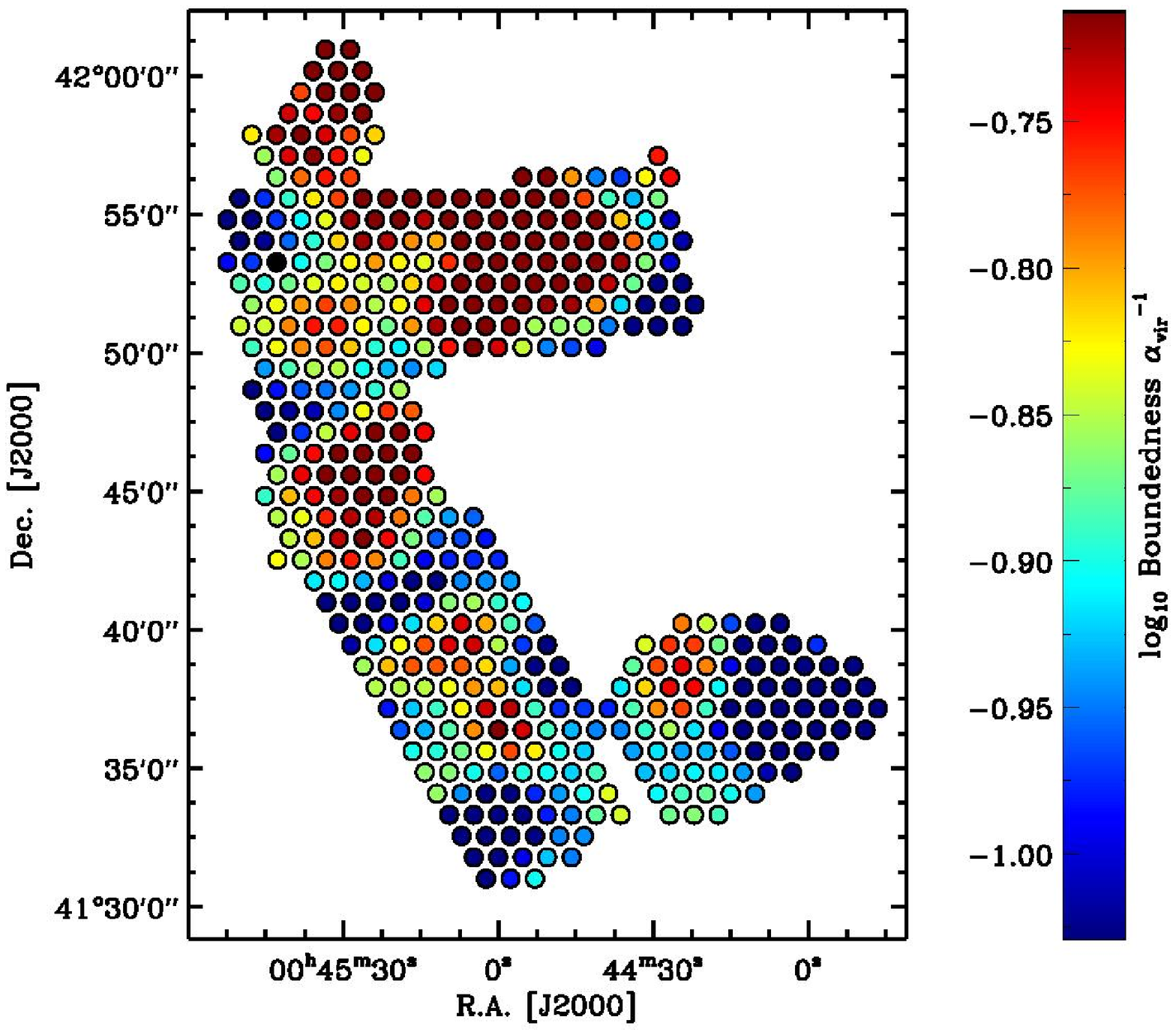}{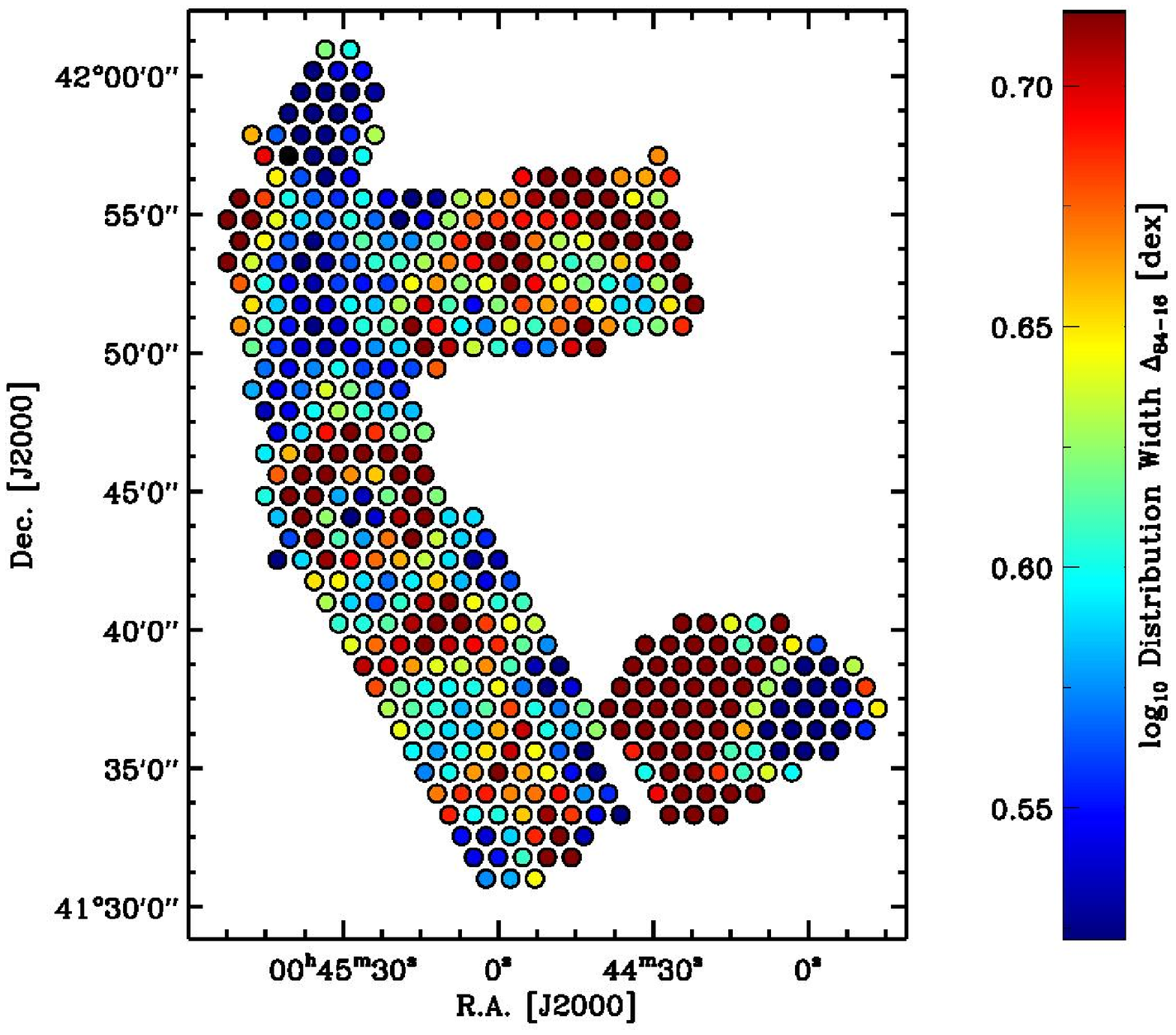}
\caption{Same as Fig.~\ref{fig:atlas_ant}, but for M31.}
\label{fig:atlas_m31}
\end{figure*}

\begin{figure*}
\plottwo{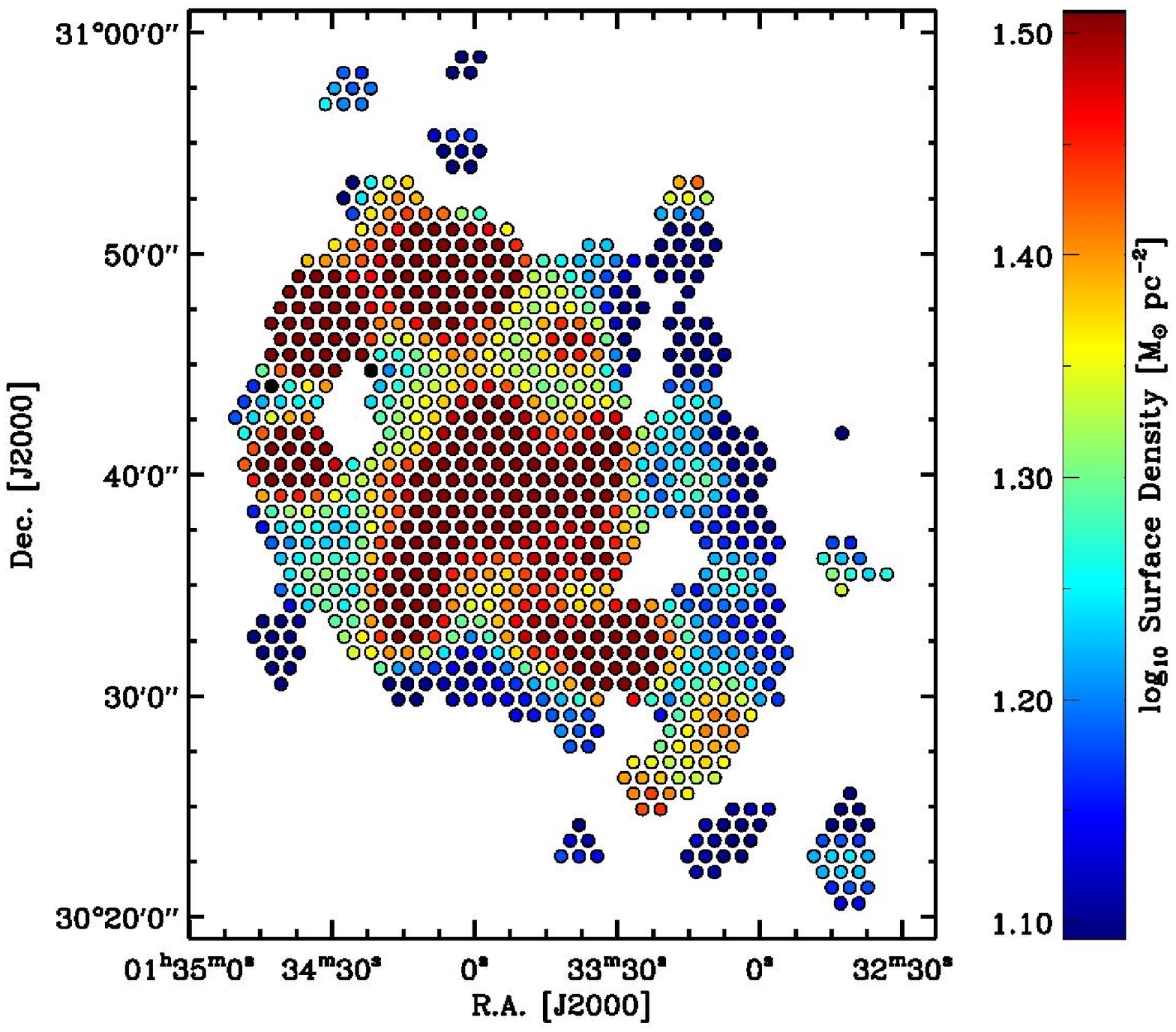}{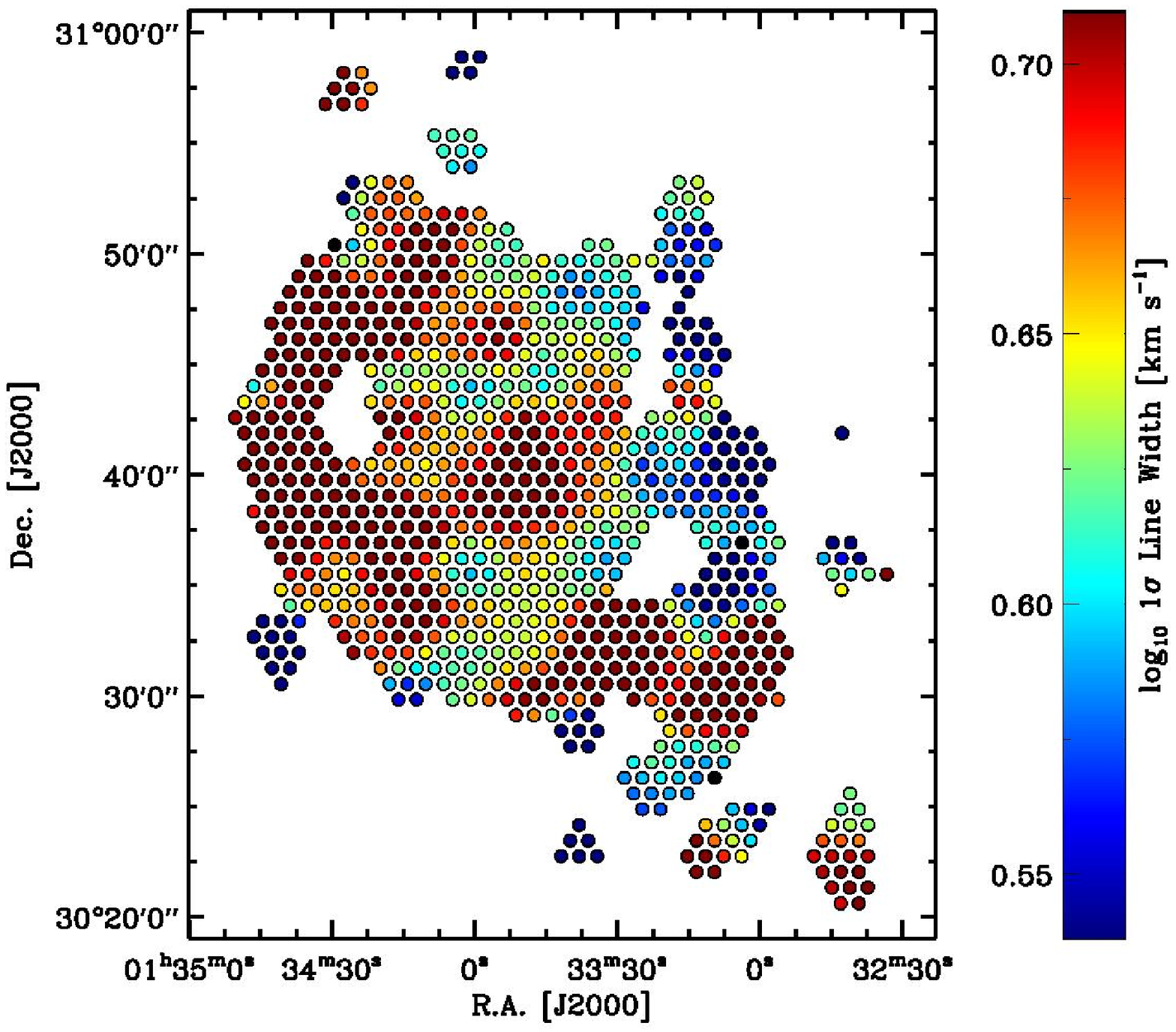}
\plottwo{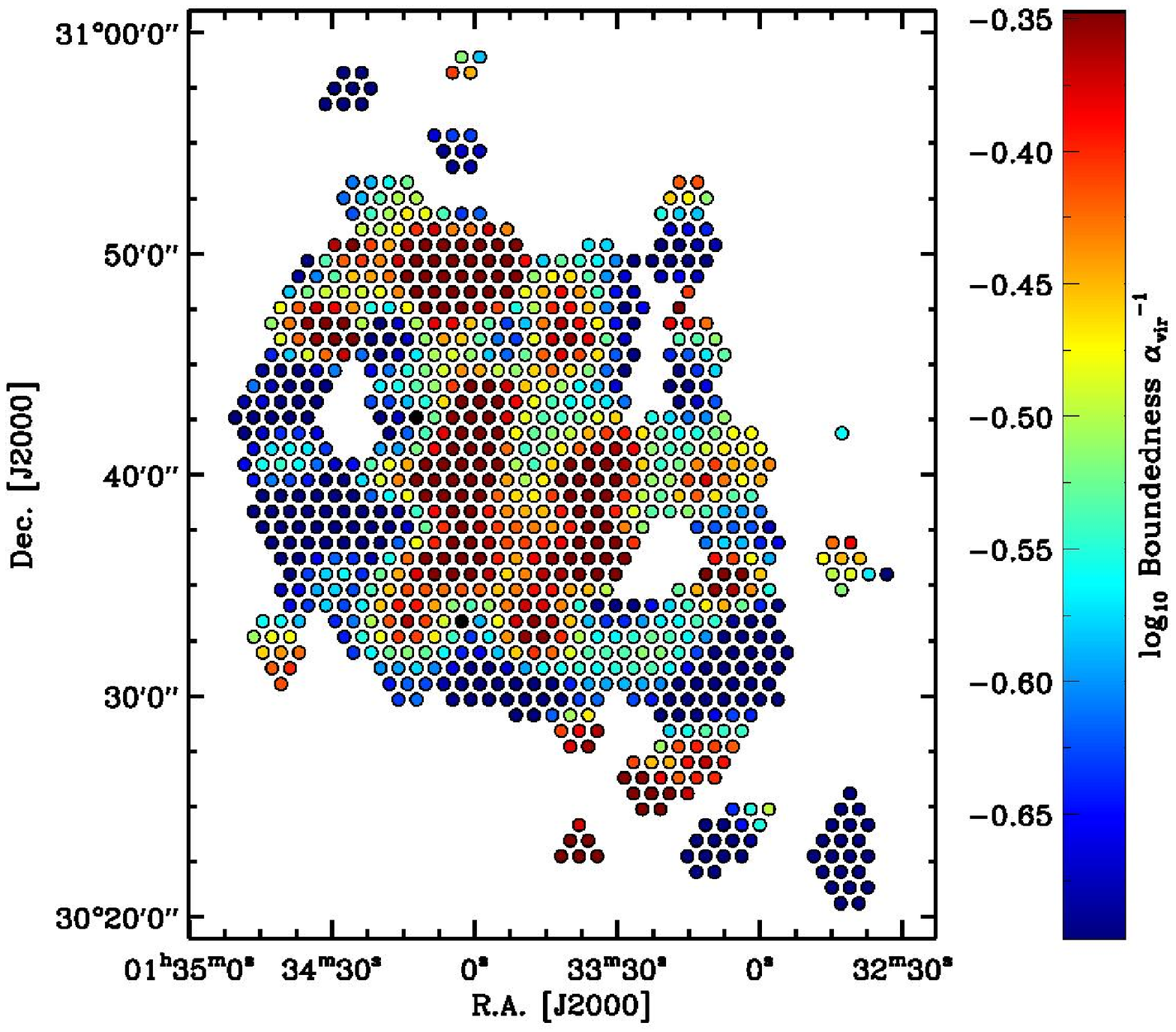}{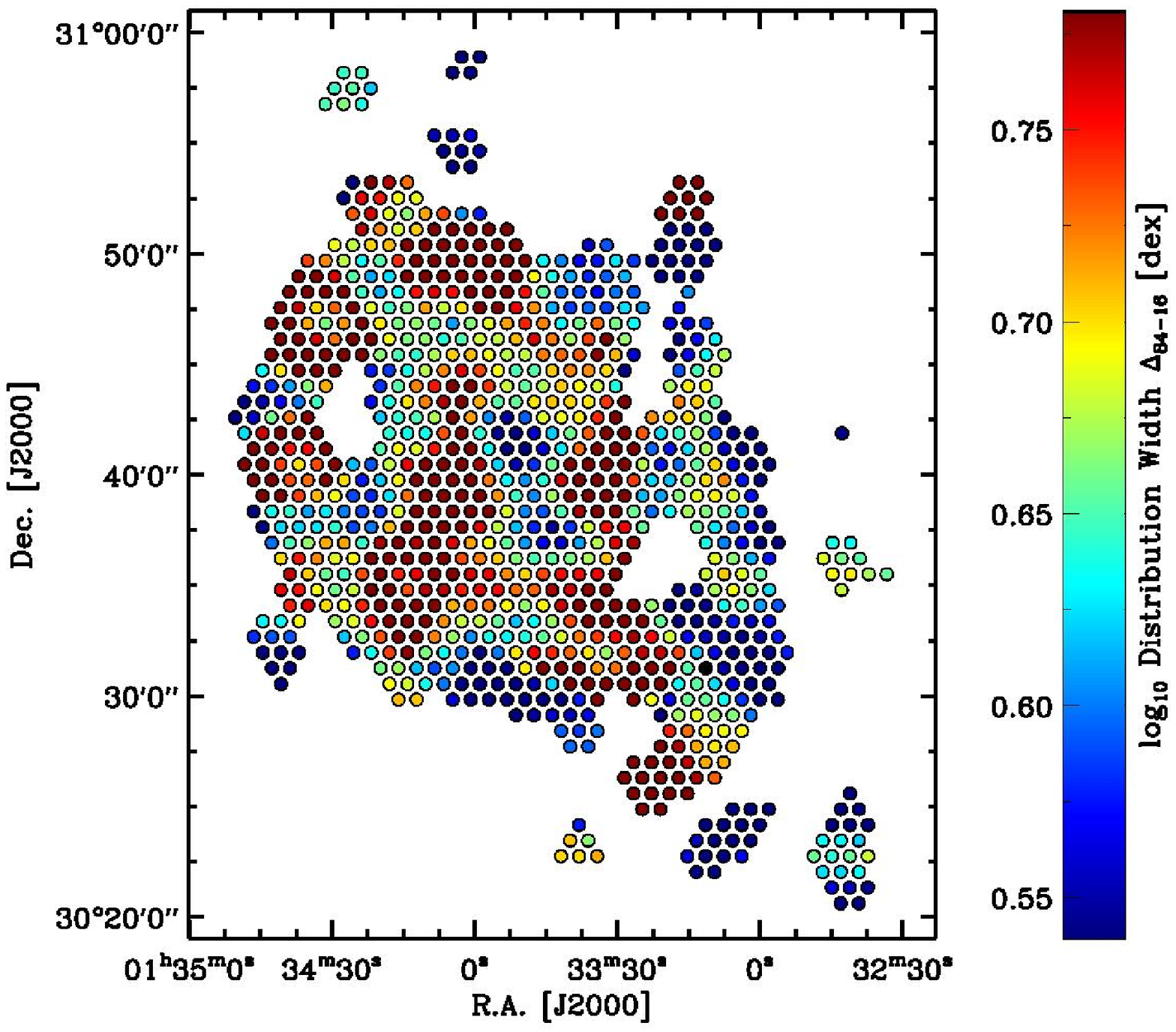}
\caption{Same as Fig.~\ref{fig:atlas_ant}, but for M33.}
\label{fig:atlas_m33}
\end{figure*}

\begin{figure*}
\plottwo{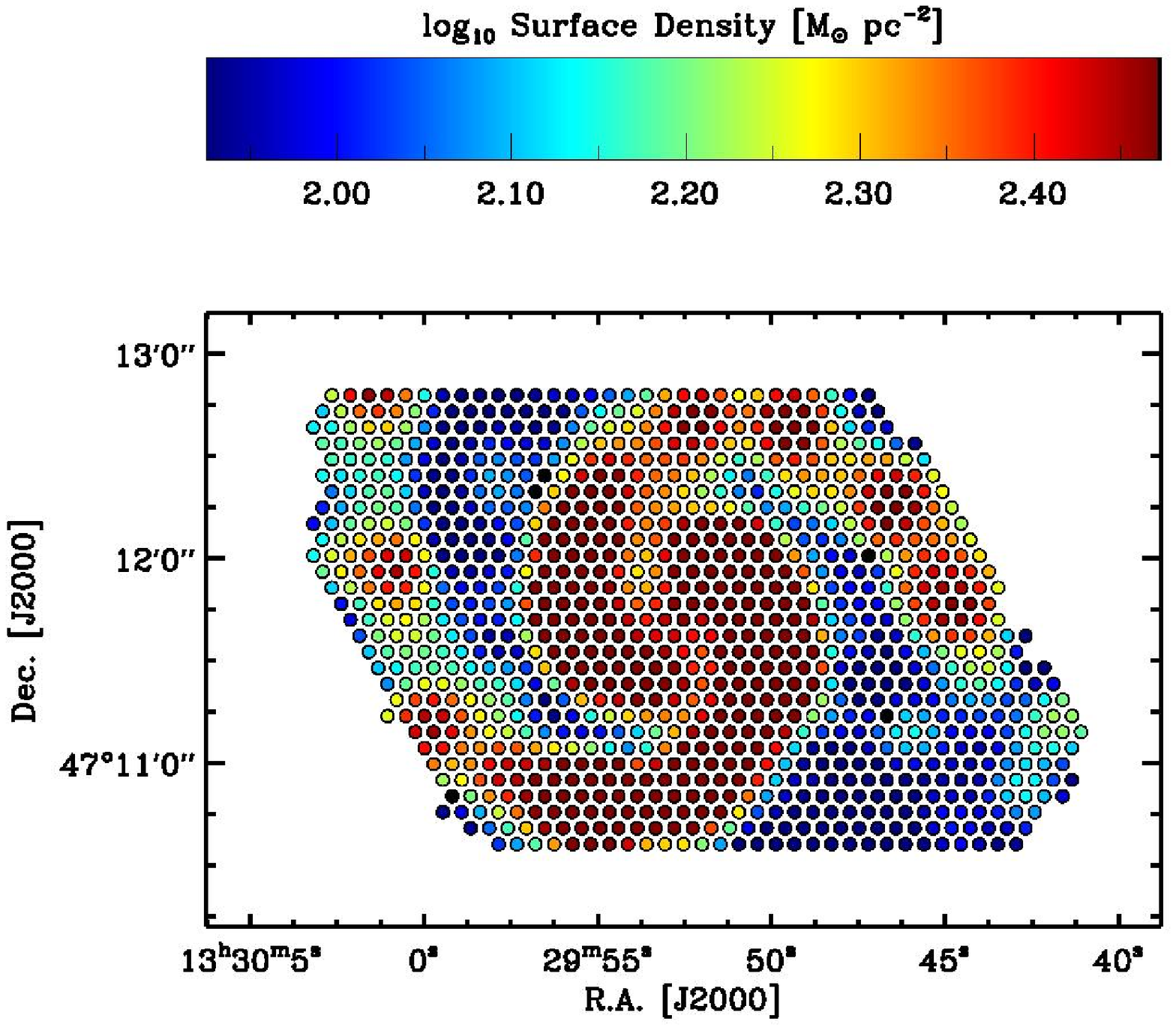}{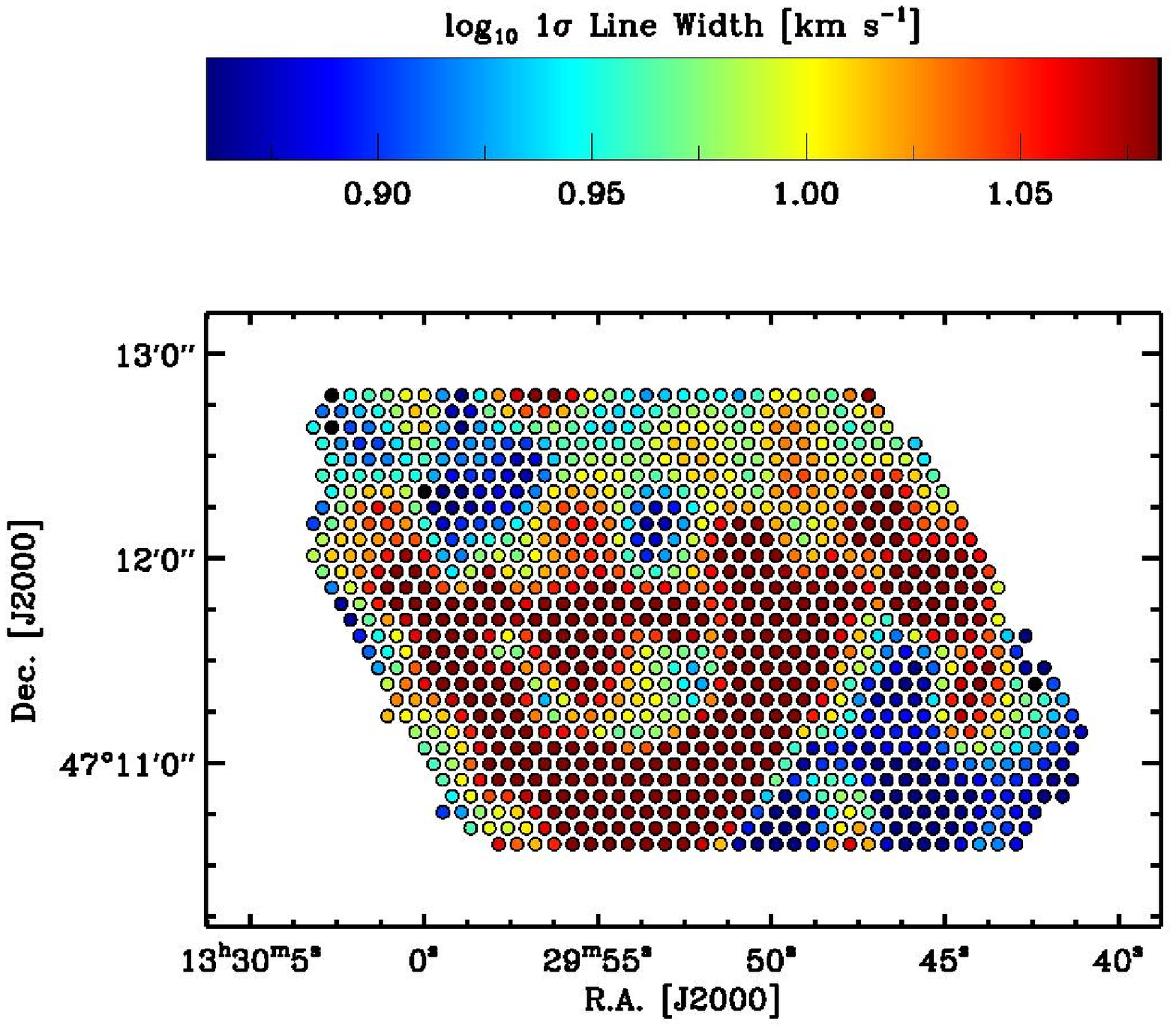}
\plottwo{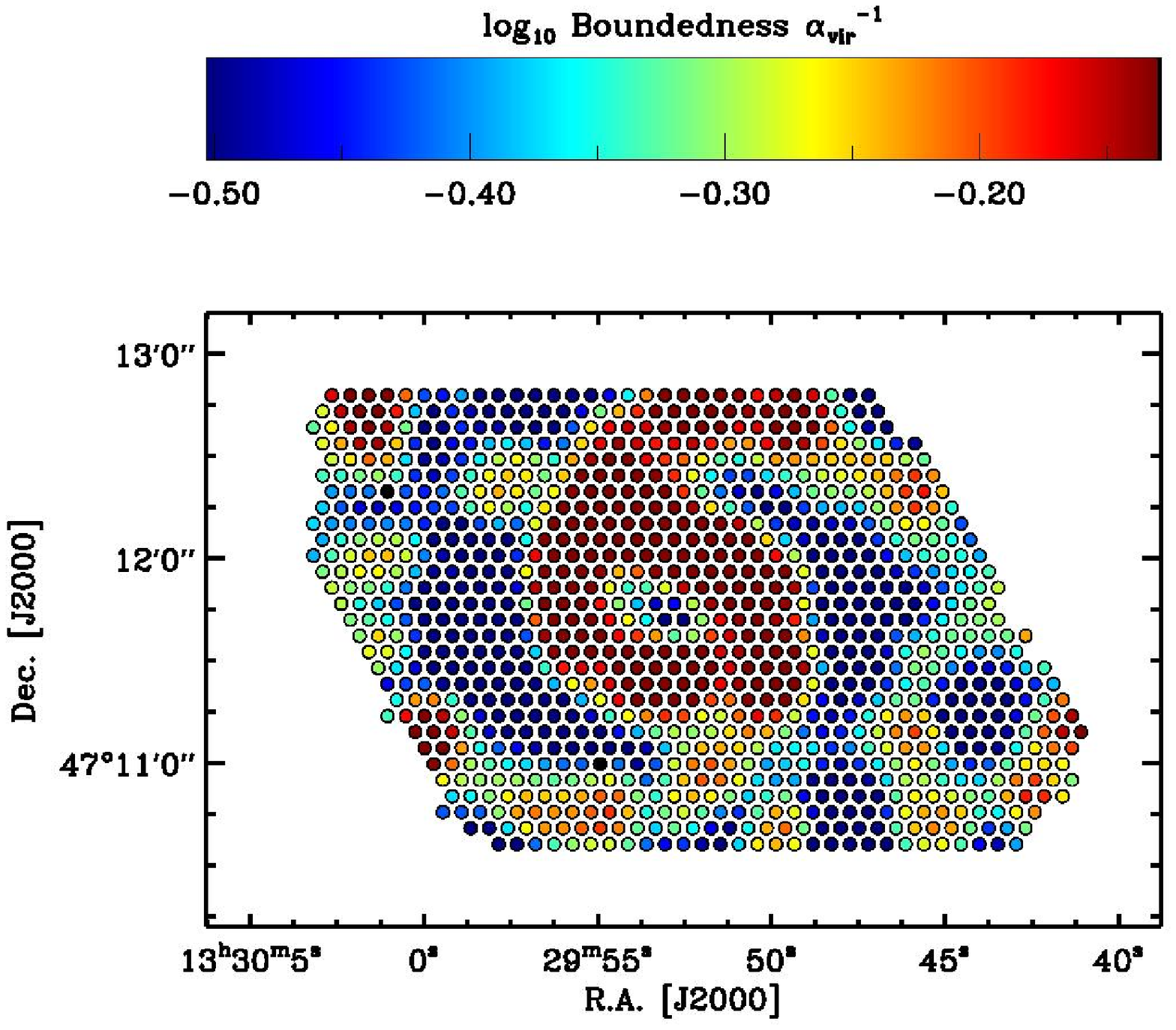}{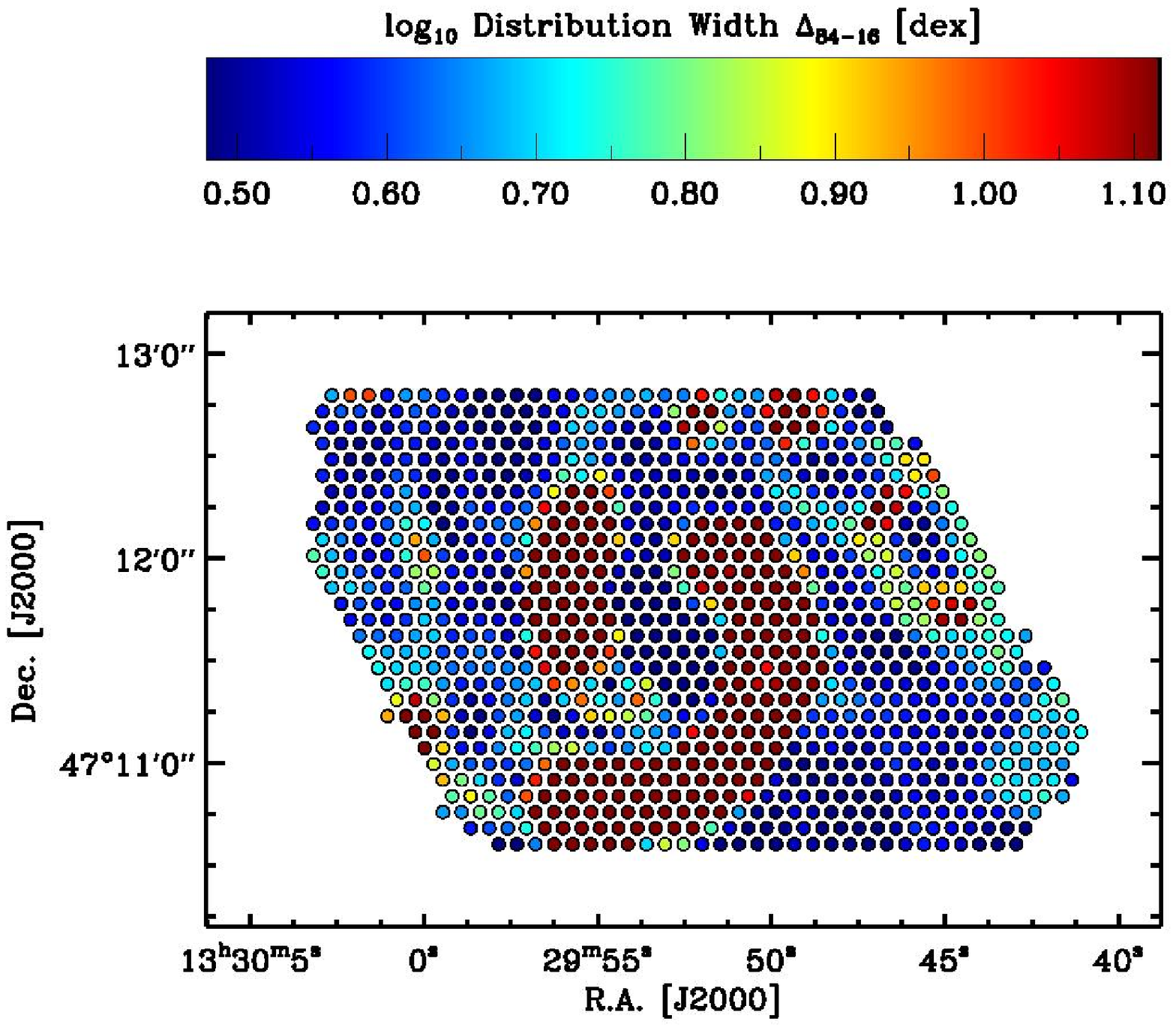}
\caption{Same as Fig.~\ref{fig:atlas_ant}, but for M51.}
\label{fig:atlas_m51}
\end{figure*}

\begin{figure*}
\plottwo{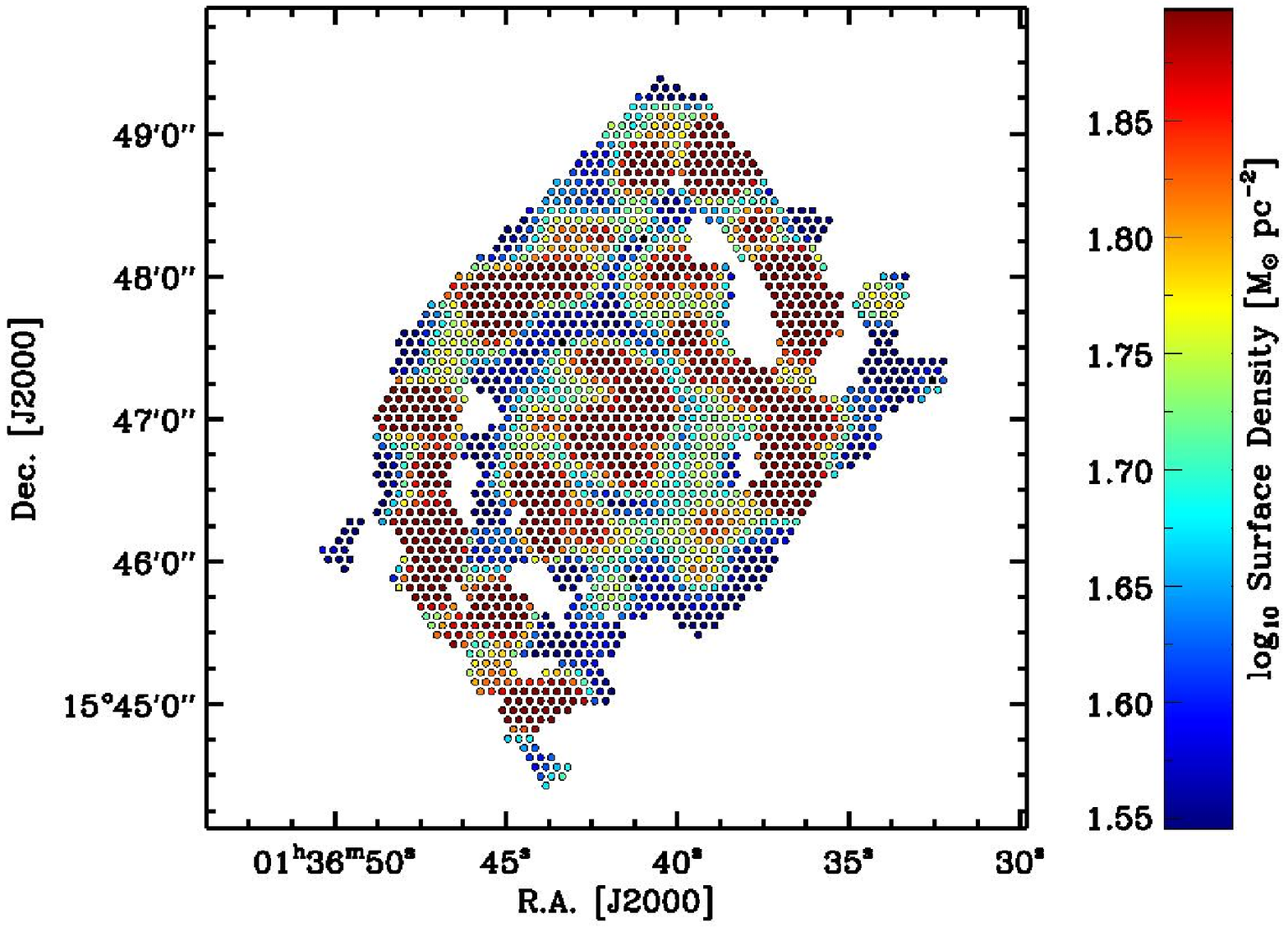}{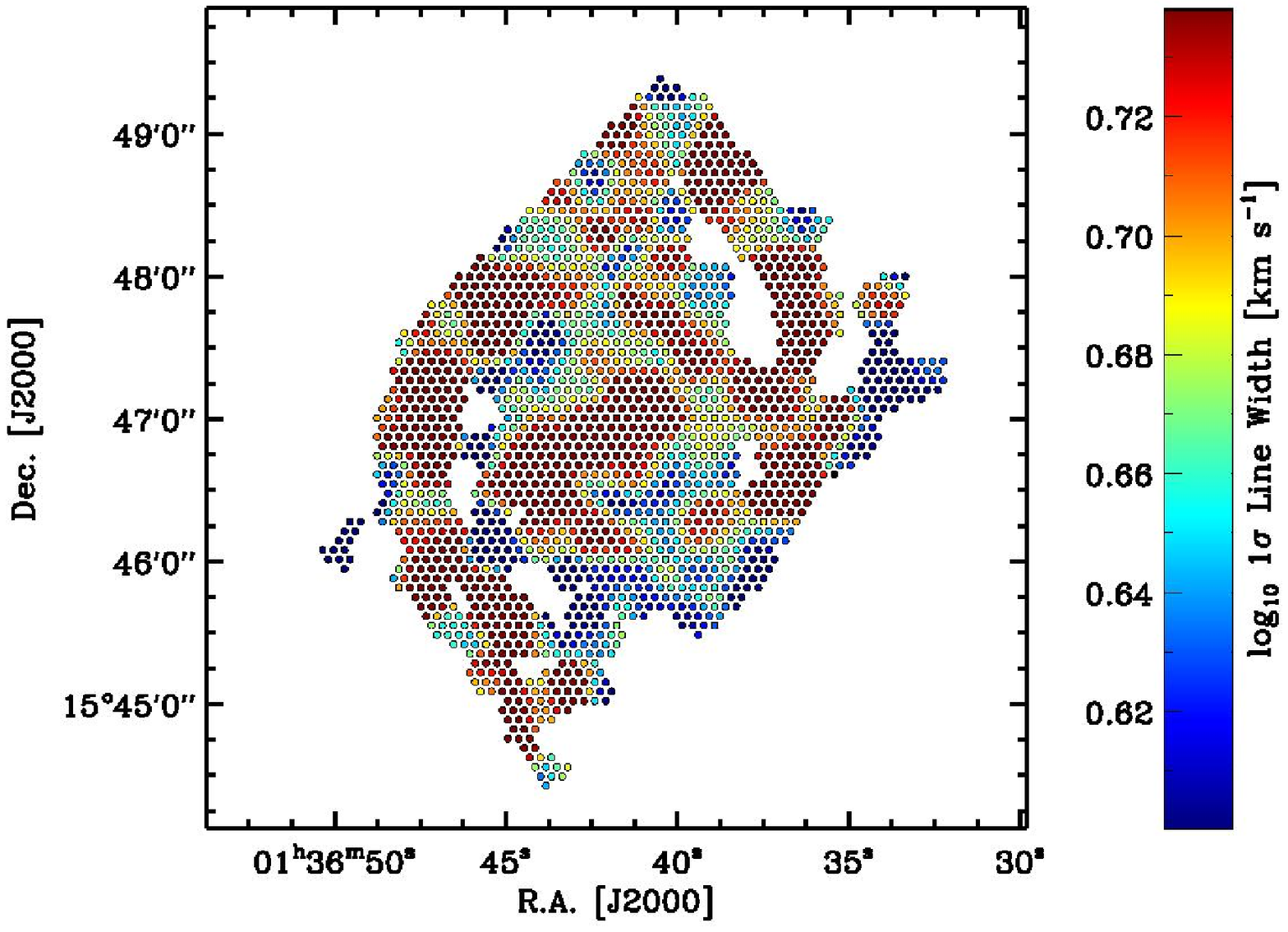}
\plottwo{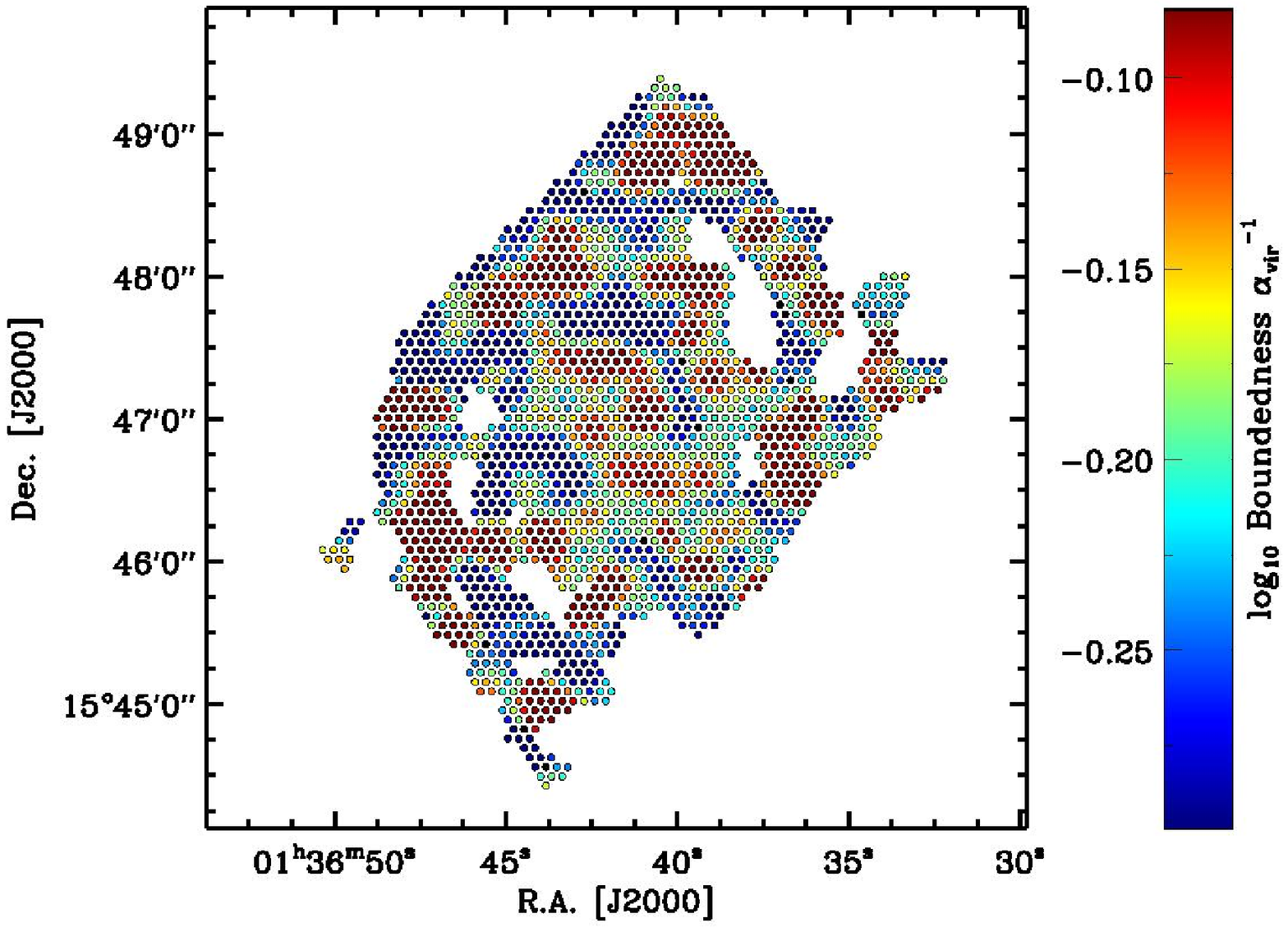}{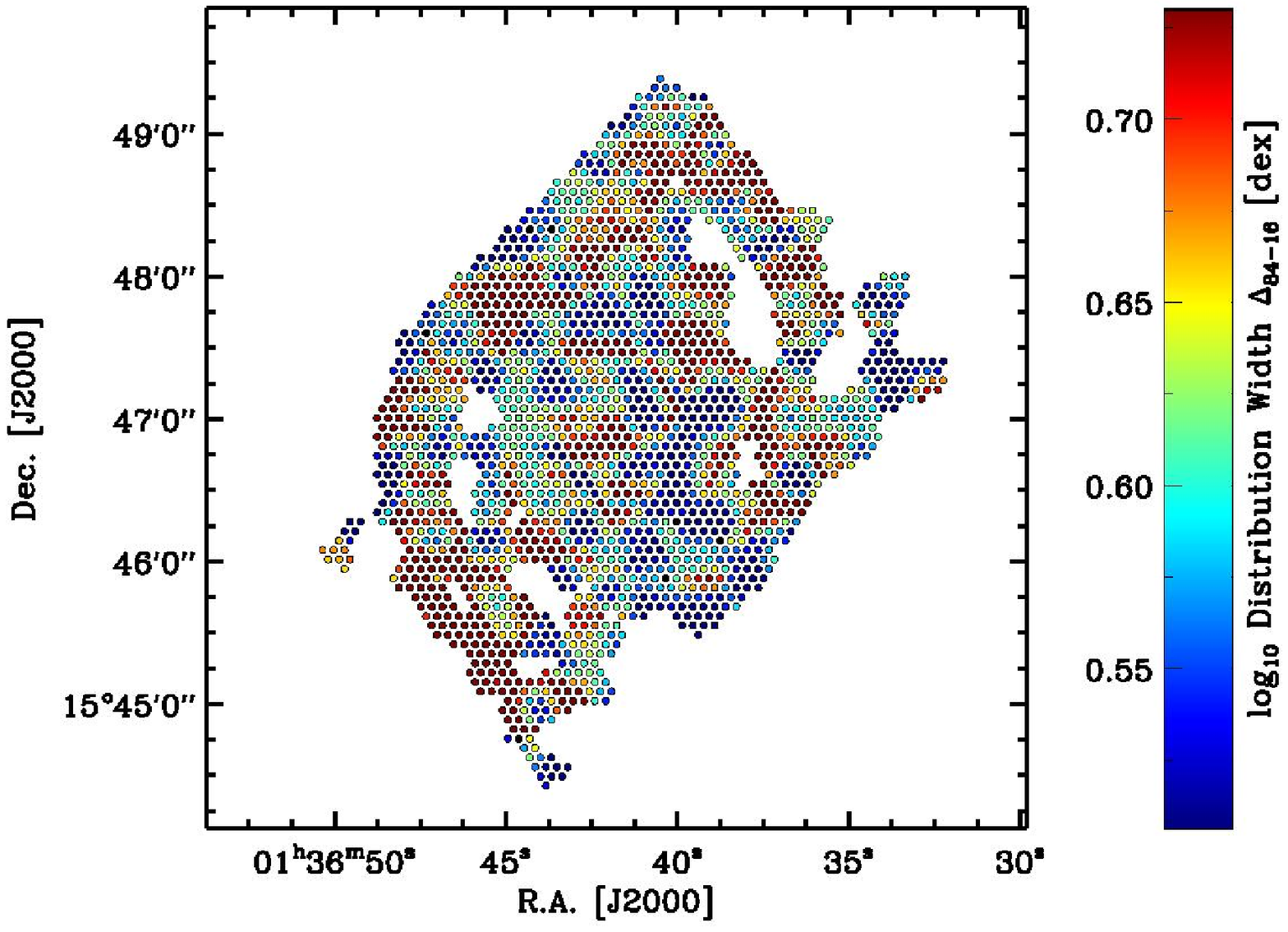}
\caption{Same as Fig.~\ref{fig:atlas_ant}, but for M74}
\label{fig:atlas_m74}
\end{figure*}

\end{appendix}

\end{document}